# Hamiltonian Cycle Problem is in P

Aimin Hou
Dongguan University of Technology
Da Xue Road 1#, Songshan Lake, Dongguan, Guangdong, China
Email: houam@dgut.edu.cn

**Abstract**

In this paper we present the first deterministic polynomial time algorithm for detecting the existence of Hamiltonian cycles and finding a Hamiltonian cycle in general graphs. Our algorithm can also solve the Hamiltonian path problem in traceable graphs. Besides, our algorithm can deal with the finite general graphs including undirected, directed, and mixed.

To achieve the results presented in this paper we construct the related path hologram transformed from the original graph and calculate the path set of every vertex in the path hologram with greedy strategy. The path hologram is a multi-segment graph with the vertex <*u, k*> where *u* is a vertex in the original graph and *k* is the segment level of *u* in the path hologram. The path set of the vertex <*u, k*> is a collection of segment sets consisting of all the vertices located on the same segment level among all the longest basic paths from the initial vertex ***S*** or any vertex on segment level *j*, where 1≤*j*≤*k*-1, to the vertex <*u, k*> in the path hologram. The core skill of our method is the "consecutive" deleting-replenishing operations recursively on the left/right action field of "invalid" vertices, respectively. This skill will ensure that a Hamiltonian cycle can be found in deterministic polynomial time at each step in Backward Searches and all invalid path fragments cannot be visited.

The space complexity of our algorithm **PHG-BP** is $O(n^4)$. With the parallel computing of *n*×*n* computers, the space complexity can be improved to $O(n^2)$. The time complexity of our algorithm **PHG-BP** are theoretically $O(n^8 \times d^3)$ on average and $O(n^9 \times d^3)$ in the worst case respectively, where *d* is the maximum degree of vertex. With the parallel computing of *n*×*n* computers, the time complexity can be improved to $O(n^5)$ on average and $O(n^6)$ in the worst case.

## 1  Introduction

Since the computational complexity theory had been put forward, problems can be classified into easy to be solved and difficult to be solved. The former belonging to the class **P** can be solved by deterministic polynomial time algorithms, while the latter in class **NP** can be solved by non-deterministic polynomial time algorithms. There exists a subset **NPC** problem of **NP**. Up to now, deterministic polynomial time algorithm for **NPC** problems, including many **NP** problems, has not been found for a general graph in the worst case.

As one known, the Hamiltonian Cycle Problem (**HCP** for short) is one of the famous **NPC** problems. Here we summarize some results on closely related topics of **HCP** algorithms to date in Section 1.1.

### 1.1  Hamiltonian Algorithms

Despite the fact that the Hamiltonian Cycle Problem is **NPC**, algorithms of a probabilistic nature and algorithms for special classes of graphs have been developed. Representative literatures, but not limited to these, in the line of research include Sufficient Conditions (see, [Dir52], [Ore60], [CE72], [TH12], [Wel17], [AGK21]), Random Graphs (see, [Pós76], [FF83], [Coo91], [FKT99], [BCFF00], [CFPP18], [Tur20]), Special Graphs (see, [RW92], [Won95], [HC11], [KB16], [ZP19], [BKK20], [KPS20]), Forbidden Subgraphs (see, [Zha88], [BDK00], [Sha20], [Sha21]), and Parameterized Algorithms (see, [Mon85], [KL94], [AYZ94], [CY02], [BMT02], [Gab07]). The algorithms based on conditions cannot deal with Hamiltonian graphs which do not satisfy the conditions. The algorithms for Random Graphs are all probabilistic algorithms. Although most random graphs, which are generated by edge probabilistic distribution, can be solved in polynomial time in the sense of probability, there are always particular instances that make this kind of algorithm degenerate into exponential time algorithm. The algorithms for Special Graphs and Forbidden Subgraphs have polynomial time complexity. Due to the particular requirements on vertex degree of special graphs, the cost of searching space for permutation of path selection or basic cycle selection is greatly reduced. However, the disadvantage of these algorithms is that the same strategy cannot be used to deal with general graphs. The parameterized algorithms mainly deal with the finding of the basic path or basic cycle of a given length. The above algorithms mainly use these methods including path extension rotation, edge cutting, edge coloring, disjoint cycles decomposition, or vertex cutting set, etc.

Recently, Björklund [FOCS 2010] used a new reduction inspired by the algebraic sieving method for *k*-Path to design a Monte Carlo algorithm for Hamiltonicity detection in an *n*-vertex undirected graph running in $O^*(1.657^n)$ time. Cygan, Kratsch, and Nederlof [STOC 2013] used the permutation matrices within *Ht* induced from the sets of matchings *Xt* to derive a $1.888^n \times n^{O(1)}$ time Monte Carlo algorithm that solves the Hamiltonicity problem in directed bipartite graphs. Björklund and Husfeldt [FOCS 2013] used a new combinatorial formula for the number of Hamiltonian cycles based on modulo a positive integer to derive a deterministic algorithm that given any directed graph on *n* vertices computes the parity of its number of Hamiltonian cycles in $O(1.619^n)$ time and polynomial space. Haah, Hastings, Kothari, and Low [FOCS 2018] used $O(nT\text{polylog}(nT/\varepsilon))$ gates with depth $O(T\text{polylog}(nT/\varepsilon))$ of tree to design the first simulation algorithm that achieves the cost of quasilinear in *nT* and poly-logarithmic in *nT*/ε.

Although many outstanding achievements have been obtained about the **HCP**, a deterministic polynomial time algorithm for general graphs in the worst case has not been found up to date.

## 1.2   Our Results

In this paper, we present the first deterministic polynomial time algorithm for detecting the existence of Hamiltonian cycles and finding a Hamiltonian cycle in general graphs. Its proof of correctness is relied on the following theorems which are proven in Appendix.

**Definition 1.** Given a finite connected undirected graph ***G***=(***V***, ***E***), a path hologram ***H***=<$V_H$, $E_H$, ***S***, ***D***, ***L***> is constructed by the following steps:

  1. Any a vertex of $V_H$ is expressed as the form of <*u*, *k*>, where *u*∈***V*** and *k* is the segment level on which *u* is located in a path, 0≤*k*≤*L*=|***V***|.

  2. Select any a vertex *v* of ***V*** to be as a starting point. Generate the initial vertex ***S***=<*v*, 0> and the final vertex ***D***=<*v*, *L*> of $V_H$, where ***S*** and ***D*** are located on segment level 0 and segment level *L*, respectively.

  3. For any other vertex *w* of ***V*** except *v*, generate *L*-1 vertices <*w*, *k*> of $V_H$ respectively, where 1≤*k*≤*L*-1.

  4. For each undirected edge *e*=*vw* of ***E***, where the vertex *v* selected in step 2 is the endpoint of the edge *e*, generate two directed edges (<*v*, 0>, <*w*, 1>) and (<*w*, *L*-1>, <*v*, *L*>) of $E_H$ respectively.

5. For any other undirected edge $e=uw$ of $E$, which does not contain the vertex $v$ selected in step 2 as its endpoint, generate $2(L-2)$ directed edges $(<u, k>, <w, k+1>)$ and $(<w, k>, <u, k+1>)$ of $E_H$ respectively, where $1 \leq k \leq L-2$.

Thus, if an undirected graph $G=(V, E)$ has the order $n=|V|$ and the size $e=|E|$, in the path hologram $H=<V_H, E_H, S, D, L>$ constructed by **Definition 1** we have $|V_H|=(n-1)^2+2$ and $|E_H|=2\deg(v)+2(n-2)(e-\deg(v))$, where the vertex $v$ selected in step 2 is the starting point.

**Definition 2.** Given a path hologram $H=<V_H, E_H, S, D, L>$, assume that the initial vertex $S=<u_0, 0>$ and the final vertex $D=<u_0, L>$. If there exists a path $P$ from $S$ to $D$, without loss of generality assume that $P=S-<u_1, 1>-<u_2, 2>-\ldots-<u_{L-1}, L-1>-D$, such that all vertices $u_0, u_1, u_2,\ldots,u_{L-1}$ are distinct and located on different segment levels, then the path $P$ is called a basic path from $S$ to $D$ in $H$.

**Definition 3.** Given a path hologram $H=<V_H, E_H, S, D, L>$, the path set of the vertex $<u, k>$ in $H$, denoted by $PS[<u, k>]$, is a collection of segment sets consisting of all the vertices located on the same segment level among all the longest basic paths from the initial vertex $S$ or any vertex on segment level $j$, where $1 \leq j \leq k-1$, to the vertex $<u, k>$ in $H$ if these paths exist. Otherwise, $PS[<u, k>]=\{\{u\}\}$.

That is, $PS[<u, k>]=\{PS_{<u,k>}[i] \mid 0 \leq i \leq k\}$. Here $PS_{<u,k>}[i]$ is the vertex set consisting of all vertices located on segment level $i$ among all the longest basic paths from the initial vertex $S$ or any vertex on segment level $j$, where $1 \leq j \leq k-1$, to the vertex $<u, k>$ in $H$ and called the $i$-th segment set.

**Definition 4.** Given a path hologram $H=<V_H, E_H, S, D, L>$, assume that the vertex $v \in PS_{<u,k>}[i]$, the left action filed of the vertex $<v, i>$ is defined as $N^-(v) \cap PS_{<u,k>}[i-1]$ and denoted by $AF^-(<v, i>)$. Analogously, the right action filed of the vertex $<v, i>$ is defined as $N^+(v) \cap PS_{<u,k>}[i+1]$ and denoted by $AF^+(<v, i>)$.

Notice that $N^-(v)=N^+(v)=N(v)$ for any vertex $v$ in an undirected graph $G$, where $N(v)$ is the neighbourhood of $v$.

**Theorem 1.** A finite connected undirected graph $G=(V, E)$ has a Hamiltonian cycle if and only if the corresponding path hologram $H=<V_H, E_H, S, D, L>$ has a basic path from the initial vertex $S$ to the final vertex $D$.

**Theorem 2.** A finite connected undirected graph $G$ is a Hamiltonian graph if and only if the **PHG-BP** algorithm must save the basic path from the initial vertex $S$ to the final vertex $D$ in the path set of the vertex $D$ in the corresponding path hologram $H$.

**Theorem 3.** If there exists a basic path from the initial vertex $S$ or a vertex on segment level $j$, where $1 \leq j \leq k-1$, to the vertex $<u, k>$, then the **CM** operator, i.e., Consistency Maintain operator, ensures that all the valid basic paths saved in the path set $PS[<u, k>]$ can be found in Backward Searches, but all the invalid path fragments, which are broken and remained in $PS[<u, k>]$, cannot be visited in Backward Searches.

**Theorem 4.** $PS[<u, k>]$ has no basic paths from the initial vertex $S$ to the vertex $<u, k>$ if and only if some segment set is empty.

**Theorem 5.** The **CM** operator is complete.

**Theorem 6.** The **CM** operator can be calculated in polynomial time.

**Theorem 7.** The **LPM** operator, i.e., Longest Path Merging operator, satisfies the optimal principle of greedy strategy.

**Theorem 8.** If the original graph $G$ is Hamiltonian, then any a Hamiltonian cycle can be found by the **FHC** operator in deterministic polynomial time.

The following defines three operation symbols $\cup_{max}, \cap_{min}, \otimes$. Here $A, B$ and $PS[<u, k>]$ are sets.

$$A\cup_{max}B = \begin{cases} \{A[i] \cup B[i]\} & where\ |A|=|B|,\ 0\leq i \leq |A| \\ A & where\ |A|>|B| \\ B & where\ |A|<|B| \end{cases} \quad (1)$$

$$A\cap_{min}B=\{A[i]\cap B[i]\} \quad where\ 0\leq i\leq \min\{|A|,|B|\} \quad (2)$$

$$PS[<u,k>]=PS[<v,k-1>]\otimes\{\{u\}\}=PS[<v,k-1>]\cup\{\{u\}\} \quad (3)$$

where $PS_{<u,k>}[k]=\{u\}$, $PS_{<u,k>}[i]=PS_{<v,k-1>}[i]$ for $0\leq i\leq k-1$, and $(<v,k-1>,<u,k>)\in E_H$

## 1.3 Paper Organization

In Section 2 we give the preliminaries of the graph theory. In Section 3 we present our algorithm and its pseudocode, and analyze its complexity. We give some instances in Section 4 and discuss our method in Section 5. The proof of our theorems is given in Appendix.

## 2 Preliminaries

Throughout this paper, we rely only on the classic notation and terminology of the graph theory. A graph **G** is a non-empty set of objects called vertices together with a (possibly empty) set of pairs of distinct vertices of **G** called edges. *V*(**G**) and *E*(**G**) denote the vertex set and the edge set of **G**, respectively. Edge can be classified into directed edge (or arc) or undirected edge according to whether it is an ordered pair of vertices, denoted by (*u*, *v*), or an unordered pair of vertices, denoted by *uv*. An undirected edge *uv* can be replaced by two directed edges with opposite arrow directions, i.e., (*u*, *v*) and (*v*, *u*). A graph is called a directed graph, or digraph for short, if all edges are directed. A graph is called an undirected graph if all edges are undirected. If a graph has both directed and undirected edges, it is called a mixed graph. A finite graph with *n* vertices is called a graph of order *n*. In this paper, we discuss only finite connected graphs.

In an undirected graph **G**, if *e*=*uv* is an undirected edge, then *u* and *v* are adjacent (or neighbours). The degree of the vertex *u*, denoted by deg(*u*), is the number of edges incident with *u*. The set *N*(*u*) of neighbours of a vertex *u* is called the neighbourhood of *u*. Similarly, in a directed graph **G**, if *e*=(*u*, *v*) is a directed edge, then *u* is the starting vertex and *v* the end vertex. The out-degree of the vertex *u*, denoted by $deg^+(u)$, is the number of edges with *u* as their starting vertex, while the in-degree of the vertex *u*, denoted by $deg^-(u)$, is the number of edges with *u* as their end vertex. The out-neighbourhood of *u*, denoted by $N^+(u)$, is the set of the end vertices of the directed edges with *u* as their starting vertex, while the in-neighbourhood of *u*, denoted by $N^-(u)$, is the set of the starting vertices of the directed edges with *u* as their end vertex.

Let *u* and *v* be vertices of a graph **G**. A *u*-*v* path in **G** is a subgraph *P*⊂**G** with *V*(*P*)={*u*=$u_0$, $u_1$,…, $u_{k-1}$, $u_k$=*v*} and *E*(*P*)={$u_{i-1}u_i$ or ($u_{i-1}$, $u_i$) | 1≤*i*≤*k*}. The vertices *u* and *v* are called endpoints of *P*. The number *k* is the length of *P*. If the endpoints *u* and *v* are identical, *P* is called a cycle (or circuit). A Hamiltonian cycle or path of **G** is a cycle or path that contains every vertex of **G** exactly once.

## 3 Algorithm and Its Complexity

In this Section, we present the pseudocode of our algorithm and some related operators in detail. The **PHG-BP** algorithm, with greedy strategy, can tell us whether or not there exists a basic path from the initial vertex **S** to the final vertex **D** in a path hologram and find it if it exists. It is straightforward to calculate the time complexity and space complexity of our algorithm with the standard analysis.

## 3.1 PHG-BP Algorithm

The principle of **PHG-BP** algorithm can be expressed as follows. We start from the initial vertex **S** and recursively extend the path by a vertex $<u, k>$ segment by segment until reach the final vertex **D**. When we plan to generate the path set $PS[<u, k>]$ of the vertex $<u, k>$, each parent vertex $<v, k-1>$ of $<u, k>$ is considered according to each directed edge $(<v, k-1>, <u, k>) \in E_H$. If the vertex $u$ joining the $PS[<v, k-1>]$ does not cause a conflict, that is, $u \notin PS_{<v,k-1>}[i]$, where $0 \leq i < k-1$ and $PS_{<v,k-1>}[i] \in PS[<v, k-1>]$, then a basic path from the initial vertex **S** or a vertex on segment level $j$, where $1 \leq j \leq k-1$, to the vertex $<u, k>$ through the parent vertex $<v, k-1>$ can be obtained, i.e., $PS[<u, k>]=PS[<v, k-1>] \otimes \{\{u\}\}$, where $\otimes$, see the **formula (3)**, stands for "join" operation on two sets $PS[<v, k-1>]$ and $\{\{u\}\}$. If a conflict occurs, that is, there exists some $i$ ($0<i<k-1$) such that $u \in PS_{<v, k-1>}[i]$, then the path from the initial vertex **S** or a vertex on segment level $j$, where $1 \leq j \leq k-1$, to the vertex $<v, k-1>$ through the vertex $<u, i>$ is invalid and should be deleted from $PS[<v, k-1>]$. After all such invalid paths have been deleted, all valid paths stored in the remainder of $PS[<v, k-1>]$ can be used to construct the path from the initial vertex **S** or a vertex on segment level $j$, where $1 \leq j \leq k-1$, to the vertex $<u, k>$ through the parent vertex $<v, k-1>$ if the remainder is not $\{\varnothing\}$. At this time, $PS[<u, k>]$ can be obtained by joining the vertex $u$ in the new temporary path set $PStemp[<v, k-1>]$, i.e., $PS[<u, k>]= PStemp[<v, k-1>] \otimes \{\{u\}\}$. Further, those "broken" invalid path fragments stored in $PStemp[<v, k-1>]$ will not be visited in future Backward Searches if they exist. Next, we will compute the path set union with greedy strategy, see the **formula (1)**, by merging the sets on the same segment level of all the same longest paths from the initial vertex **S** or a vertex on segment level $j$, where $1 \leq j \leq k-1$, to the vertex $<u, k>$, i.e., $PS[<u, k>]=\cup_{max} \{PStemp[<v, k-1>] \otimes \{\{u\}\} \mid \forall (<v, k-1>, <u, k>) \in E_H$ and $PStemp[<v, k-1>] \neq \{\varnothing\}$ and $PStemp[<v, k-1>]$ has the same longest length$\}$. That is, after the merging operation has been calculated, $PS[<u, k>]$ has saved all the valid paths, which have the same longest length, from the initial vertex **S** or a vertex on segment level $j$, where $1 \leq j \leq k-1$, to the vertex $<u, k>$. Such path expansion process continues until no vertex can be joined to the tail of the paths. Finally, if there exists a basic path of length $n=|V(G)|$ from the initial vertex **S** to the final vertex **D** in $PS[D]$, then the original graph **G** is Hamiltonian, where $|PS[D]|=n+1$; otherwise **G** is nonHamiltonian.

In the case that a Hamiltonian cycle existed in the original graph, a non-redundant valid parent vertex in the Hamiltonian cycle can be backward searched based on the intersection of its path set and the path set of its parent vertex recursively at each step under the condition that the resulting path set is not $\{\varnothing\}$. The calculation at each step of the Backward Searching from the final vertex **D** to the initial vertex **S** is deterministic in polynomial time.

**Algorithm: PHG-BP**
BEGIN
for (each segment level $i$)
    for (each vertex $<u, i>$ on each segment level $i$)
        for (each parent vertex $<v, i-1>$ of $<u, i>$)
            { call **CM** to calculate the final $PStemp[<v, i-1>]$ in order to obtain $PStemp[<u, i>]$;
              $PS[<u, i>]=PS[<u, i>] \cup_{max} PStemp[<u, i>]$; } //greedy strategy
for (each parent vertex $<v, n-1>$ of $<1, n>$)
    { $PStemp[<1, n>]=PS[<v, n-1>] \otimes \{\{1\}\}$;
      $PS[<1, n>]=PS[<1, n>] \cup_{max} PStemp[<1, n>]$; }

if (length(*PS*[<1, *n*>])= *n*+1)
   then   {   HC=**FHC**(*PS*[ ], ***D***), print(HC); }    //*PS*[ ] is the array of path sets of all vertices
   else print("nonHamiltonian Graph");
END

---

**Pseudocode of Algorithm: PHG-BP**   //Path HoloGram - Basic Path (V1.0)
Input: path hologram ***H***=<$V_H$, $E_H$, ***S***, ***D***, ***L***>, initial vertex ***S***=<1, 0>, final vertex ***D***=<1, *n*>
Output: return a Hamiltonian cycle if it exists, otherwise return "nonHamiltonian Graph"

BEGIN
1. for (*u*=1 to *n* step +1)    //consider each vertex
2.   for (*i*=1 to *n* step +1)    //consider each segment level
3.     *PS*[<*u*, *i*>]={{*u*}};    //initialize
4. *PS*[<1, 0>]={{1}};
5. for (*i*=1 to *n*-1 step +1) {
6.   for (*u*=2 to *n* step +1) {
7.     for (every start vertex *v* of edge (<*v*, *i*-1>, <*u*, *i*>)∈$E_H$) {
8.       PStemp[<*u*, *i*>]=**CM**(*PS*[ ], <*v*, *i*-1>, <*u*, *i*>);
9.       *PS*[<*u*, *i*>]=**LPM**(*PS*[<*u*, *i*>], PStemp[<*u*, *i*>]);    //greedy strategy
10.     }  //end for every *v*
11.   }  //end for *u*
12. }  //end for *i*
13. for (every start vertex *v* of edge (<*v*, *n*-1>, <1, *n*>)∈$E_H$) {
14.   PStemp[<1, *n*>]=*PS*[<*v*, *n*-1>]⊗{{1}};
15.   *PS*[<1, *n*>]=**LPM**(*PS*[<1, *n*>], PStemp[<1, *n*>]);
16. }  //end for every *v*
17. if (length(*PS*[<1, *n*>])= *n*+1) {
18.   HC=**FHC**(*PS*[ ], ***D***);    //*PS*[ ] is the array of path sets of all vertices
19.   print(HC);
20. }
21. else print("nonHamiltonian Graph");
END

   In step 5, the segment levels of the path hologram can be handled in time *O*(*n*-1) steps. In step 6, the vertices on each segment level can be handled in time *O*(*n*-1) steps. However, with the parallel computing, we can process *n*-1 vertices at *n*-1 computers simultaneously. In this case, the step 6 can be performed in time *O*(1) steps. In step 7, we consider the parent vertices of a vertex in time *O*(*d*) steps, where *d* is the maximum degree of vertex. However, with the parallel computing, we can process *d* parent vertices at *d* computers simultaneously. In this case, the step 7 can be performed in time *O*(1) steps. In step 8, the **CM** operator can be performed in time $O(n^6 \times d^2)$ steps on average and $O(n^7 \times d^2)$ steps in the worst case. Thus the total time complexity of step 5 is $O(n^8 \times d^3)$ on average and $O(n^9 \times d^3)$ in the worst case. With the parallel computing of *n*×*n* computers, the total time complexity of step 5 can be improved to $O(n^5)$ on average and $O(n^6)$ in the worst case.

   The path set of a vertex needs $O(n^2)$ space and the path hologram has $2+(n-1)^2$ vertices. Thus the space complexity of the **PHG-BP** algorithm is $O(n^4)$. With the parallel computing of *n* computers, the space complexity of the **PHG-BP** algorithm can be improved to $O(n^3)$.

## 3.2 CM Operator

As discussed above, for every vertex $<u, k>$ in the path hologram $H$, when the path set $PS[<u, k>]$ will be generated, each parent vertex $<v, k-1>$ of $<u, k>$ should be considered to judge whether a conflict will occur if the vertex $u$ joins the $PS[<v, k-1>]$. That is whether $u \in PS_{<v,k-1>}[i]$ or not, where $0 \leq i < k-1$ and $PS_{<v,k-1>}[i] \in PS[<v, k-1>]$.

If $u \notin PS_{<v,k-1>}[i]$ for each $i$, where $0 \leq i < k-1$, then a conflict will not occur and $PS[<u, k>]$ can be obtained by $PS[<v, k-1>] \otimes \{\{u\}\}$. That is, there exists a basic path from the initial vertex $S$ or a vertex on segment level $j$, where $1 \leq j \leq k-1$, to the vertex $<u, k>$ through the vertex $<v, k-1>$ in $H$.

If $u \in PS_{<v,k-1>}[i]$ for some $i$, where $0 \leq i < k-1$, then a conflict will occur. It means that the path, on which $u$ reappeared, from the initial vertex $S$ or a vertex on segment level $j$, where $1 \leq j \leq k-1$, to the vertex $<u, k>$ through the vertex $<v, k-1>$ is not basic and is an invalid path, and should be abandoned. After all such operations of abandonment have been accomplished, all valid basic paths stored in the remainder of the $PS[<v, k-1>]$ can be used to construct the path from the initial vertex $S$ or a vertex on segment level $j$, where $1 \leq j \leq k-1$, to the vertex $<u, k>$ through the vertex $<v, k-1>$ if the remainder is not $\{\varnothing\}$. On the contrary, if the remainder of the $PS[<v, k-1>]$ is $\{\varnothing\}$, then there is not a basic path from the initial vertex $S$ or a vertex on segment level $j$, where $1 \leq j \leq k-1$, to the vertex $<u, k>$ through the vertex $<v, k-1>$ and $PS[<v, k-1>]$ is not considered. In order to properly implement the operations of abandonment, every duplicated vertex of $u$ in the $PS[<v, k-1>]$ should be deleted first. If the vertices in the left/right action filed of the duplicated vertex of $u$ are also located on the invalid path, then they should be deleted again; otherwise they will appear on another valid path and should be retained.

Further, if the vertex $w$ in the left/right action filed of the duplicated vertex of $u$ has been deleted, the vertices in the left/right action filed of the vertex $w$ should be considered in order to determine to delete or retain again. Such "consecutive" vertex deleting-replenishing operations should be carried out recursively on the left action field and/or the right action field of these "invalid" vertices, which have been deleted, respectively until another valid path is arrived, i.e., no vertex can be deleted.

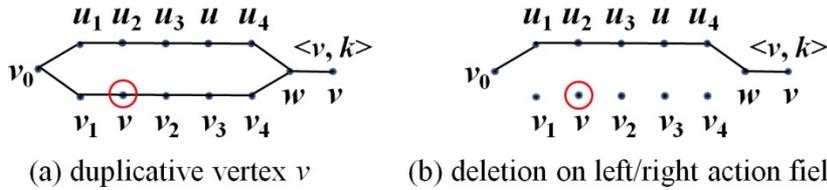

(a) duplicative vertex $v$      (b) deletion on left/right action fields

**Figure 1** Analysis of deleting "duplicative vertex" and its left/right action fields

For example, consider the situation of the **Figure 1**. When we consider the vertex $<v, k>$, we find that the vertex $v$ appears on its parent vertex path set $PS[<w, k-1>]$. So that we should delete these vertices $v_1$, $v$, $v_2$, $v_3$, and $v_4$. Here, $v_1 \in AF^-(<v, k-5>)$, $v_2 \in AF^+(<v, k-5>)$, $v_3 \in AF^+(<v_2, k-4>)$, and so on. When we want to delete the vertex $v_0 \in AF^-(<v_1, k-6>)$, $v_0$ should be replenished due to the reason of $v_0 \in AF^-(<u_1, k-6>)$. So is the vertex $w$. Here, the deleted path fragment $P_{del} = v_1\text{-}v\text{-}v_2\text{-}v_3\text{-}v_4$, the parallel path fragment $\bar{P} = u_1\text{-}u_2\text{-}u_3\text{-}u\text{-}u_4$, $P_{del}$ and $\bar{P}$ have a common left endpoint $v_0$ and a common right endpoint $w$. The same "parallel path cluster" means that $\{P_{del}; \bar{P}\}$.

**Operator: CM**
BEGIN
for (each segment set $PStemp_{<v,k-1>}[j]$ of $PStemp[<v, k-1>]$)

    if ($u \in \text{PStemp}_{<v,k-1>}[j]$)    //handle all the duplicated vertices of $u$
       then   {  delete $u$ from $\text{PStemp}_{<v,k-1>}[j]$;
            "consecutive" vertex deleting-replenishing operations on the left action field;
            "consecutive" vertex deleting-replenishing operations on the right action field;
            if ($\emptyset \in \text{PStemp}[<v, k\text{-}1>]$)   then   { $\text{PStemp}[<v, k\text{-}1>]=\{\emptyset\}$, stop;}
        }
while (there exists a singleton set $\{z\}$ in $\text{PStemp}[<v, k\text{-}1>]$)
   for (each segment set $\text{PStemp}_{<v,k-1>}[j]$ of $\text{PStemp}[<v, k\text{-}1>]$)
      if ($z \in \text{PStemp}_{<v,k-1>}[j]$)    //handle all the duplicated vertices of a singleton set $\{z\}$
        then   {  delete $z$ from $\text{PStemp}_{<v,k-1>}[j]$;
             "consecutive" vertex deleting-replenishing operations on the left action field;
             "consecutive" vertex deleting-replenishing operations on the right action field;
             if ($\emptyset \in \text{PStemp}[<v, k\text{-}1>]$)   then   { $\text{PStemp}[<v, k\text{-}1>]=\{\emptyset\}$, stop;}
         }
call **CHECK** to obtain the final $\text{PStemp}[<v, k\text{-}1>]$;
if ($\text{PStemp}[<v, k\text{-}1>]=\{\emptyset\}$)
   then   {   $\text{PStemp}[<u, k>]=\{\{u\}\}$, stop;}
   else   {   $\text{PStemp}[<u, k>]=\text{PStemp}[<v, k\text{-}1>]\otimes\{\{u\}\}$, stop;}
END

---

**Pseudocode of Operator: CM**    //Consistency Maintain (V3.0)
Input: $PS[\ ]$, $<v, k\text{-}1>$, $<u, k>$
Output: $\text{PStemp}[<u, k>]$

BEGIN
1.  flag1=true, flag2=false, flag3=false, $\text{PStemp}[<v, k\text{-}1>]=PS[<v, k\text{-}1>]$;
2.  for ($j=k\text{-}2$ to 1 step -1) {
3.    if ($\text{PStemp}_{<v,k-1>}[j] \in \text{PStemp}[<v, k\text{-}1>]$ contains the vertex $u$) {
4.       flag1=false, flag2=true;
5.       delete $u$ from $\text{PStemp}_{<v,k-1>}[j]$;
6.       $\text{PStemp}[<v, k\text{-}1>]=$**LAFDR**$(\text{PStemp}[<v, k\text{-}1>], <u, j>)$;
7.       $\text{PStemp}[<v, k\text{-}1>]=$**RAFDR**$(\text{PStemp}[<v, k\text{-}1>], <u, j>)$;
8.       if ($\emptyset \in \text{PStemp}[<v, k\text{-}1>]$) {
9.          $\text{PStemp}[<u, k>]=\{\{u\}\}$, return($\text{PStemp}[<u, k>]$), stop;
10.      }  //end if
11.    }  //end if
12.  }  //end for $j$
13.  if (flag1) {
14.    $\text{PStemp}[<u, k>]=\text{PStemp}[<v, k\text{-}1>]\otimes\{\{u\}\}$, return($\text{PStemp}[<u, k>]$), stop;
15.  }  //end if (flag1)
16.  While (flag2) {
17.    flag3=false;
18.    for ($j=k\text{-}2$ to 1 step -1) {
19.       if ($\text{PStemp}_{<v,k-1>}[j]$ has only one element such as $z$) {
20.          for ($i=k\text{-}2$ & $i \neq j$ to 1 step -1) {

21.            if (PStemp$_{<v,k-1>}$[$i$]∈ PStemp[<$v$, $k$-1>] contains the vertex $z$) {
22.                flag3=true;
23.                delete $z$ from  PStemp$_{<v,k-1>}$[$i$];
24.                PStemp[<$v$, $k$-1>]=**LAFDR**(PStemp[<$v$, $k$-1>], <$z$, $i$>);
25.                PStemp[<$v$, $k$-1>]=**RAFDR**(PStemp[<$v$, $k$-1>], <$z$, $i$>);
26.                if (∅∈ PStemp[<$v$, $k$-1>]) {
27.                    PStemp[<$u$, $k$>]={{$u$}}, return(PStemp[<$u$, $k$>]), stop;
28.                }  //end if
29.            }
30.        }   //end for $i$
31.        }
32.        if (flag3) break;
33.    }   //end for $j$
34.    if (flag3) flag2=true;
35.    else flag2=false;
36. }   //end while (flag2)
37. PStemp[<$v$, $k$-1>]=**CHECK**(*PS*[ ], PStemp[<$v$, $k$-1>]);
38. if (PStemp[<$v$, $k$-1>]={∅}) {
39.    PStemp[<$u$, $k$>]={{$u$}}, return(PStemp[<$u$, $k$>]), stop;
40. }
41. else {
42.    PStemp[<$u$, $k$>]=PStemp[<$v$, $k$-1>]⊗{{$u$}}, return(PStemp[<$u$, $k$>]), stop;
43. }   //end if
END

**Operator: LAFDR**
BEGIN
calculate the left action field $S$1 of the deleted vertex <$w$, $j$>;
calculate the set $S$2 of the replenished vertices;
while ($S$1-$S$2≠{})     //i.e., there exists some deleted vertices, but not replenished
   {  delete the vertices, which belong to $S$1-$S$2, from the segment set;
      calculate the left action field $S$1 of these deleted vertices belonging to $S$1-$S$2;
      calculate the set $S$2 of the replenished vertices; }
END

**Pseudocode of Operator: LAFDR**    //Left Action Field Deleting-Replenishing (V1.0)
Input: PStemp[<$v$, $k$-1>], <$w$, $j$>
Output: PStemp[<$v$, $k$-1>]
BEGIN
 1. if ($j$=0)
 2.    {  return PStemp[<$v$, $k$-1>], stop;   }
 3. else {
 4.        $x$=$w$, $i$=$j$, $S$1={}, $S$2={};
 5.        $S$1=$N^{-}$($x$)∩PStemp$_{<v,k-1>}$[$i$-1];    //$AF^{-}$(<$w$, $j$>)

```
6.         for (each vertex y∈ PStemp_{<v,k-1>}[i] & y≠x)
7.             S2=S2∪(N⁻(y)∩PStemp_{<v,k-1>}[i-1]);    //AF⁻(<y, i>)
8.         while (i>0) {
9.             if (S1-S2={})
10.                {  return PStemp[<v, k-1>], stop;   }   //S1⊆S2, all vertices are replenished
11.            else {
12.                S1temp=S1-S2, S1={}, S2={}, i=i-1;
13.                for (each vertex q∈ S1temp) {
14.                    delete q from  PStemp_{<v,k-1>}[i];
15.                    S1=S1∪(N⁻(q)∩PStemp_{<v,k-1>}[i-1]);
16.                }
17.                for (each vertex y∈ PStemp_{<v,k-1>}[i] & y∉ S1temp)
18.                    S2=S2∪(N⁻(y)∩PStemp_{<v,k-1>}[i-1]);   //AF⁻(<y, i>)
19.            }
20.        }  //end while
21.        return PStemp[<v, k-1>];
22.    }
END
```

**Operator: RAFDR**
BEGIN
calculate the right action field S1 of the deleted vertex <w, j>;
calculate the set S2 of the replenished vertices;
while (S1-S2≠{})   //i.e., there exists some deleted vertices, but not replenished
   {  delete the vertices, which belong to S1-S2, from the segment set;
      calculate the right action field S1 of these deleted vertices belonging to S1-S2;
      calculate the set S2 of the replenished vertices; }
END

---

**Pseudocode of Operator: RAFDR**   //Right Action Field Deleting-Replenishing (V1.0)
Input: PStemp[<v, k-1>], <w, j>
Output: PStemp[<v, k-1>]

```
BEGIN
 1.  if (j=k-1)
 2.      {  return PStemp[<v, k-1>], stop;  }
 3.  else {
 4.      x=w, i=j, S1={}, S2={};
 5.      S1=N⁺(x)∩PStemp_{<v,k-1>}[i+1];   //AF⁻(<w, j>)
 6.      for (each vertex y∈ PStemp_{<v,k-1>}[i] & y≠x)
 7.          S2=S2∪(N⁺(y)∩PStemp_{<v,k-1>}[i+1]);   //AF⁻(<y, i>)
 8.      while (i<k-1) {
 9.          if (S1-S2={})
10.             {  return PStemp[<v, k-1>], stop;  }   //S1⊆S2, all vertices are replenished
11.         else {
```

```
12.              S1temp=S1-S2, S1={}, S2={}, i=i+1;
13.              for (each vertex q∈S1temp) {
14.                  delete q from PStemp_{<v,k-1>}[i];
15.                  S1=S1∪(N⁺(q)∩PStemp_{<v,k-1>}[i+1]);
16.              }
17.              for (each vertex y∈PStemp_{<v,k-1>}[i] & y∉S1temp)
18.                  S2=S2∪(N⁺(y)∩PStemp_{<v,k-1>}[i+1]);   //AF⁺(<y, i>)
19.          }
20.      }   //end while
21.      return PStemp[<v, k-1>];
22.  }
END
```

**Operator: CHECK**

```
BEGIN
for (each parent vertex <w, k-2> of <v, k-1> in PStemp[<v, k-1>])
    { PStemp[<w, k-2>]=PS[<w, k-2>]∩_{min}PStemp[<v, k-1>];
      for (each parent vertex <x, k-3> of <w, k-2> in PStemp[<w, k-2>])
          { PStemp[<x, k-3>]=PS[<x, k-3>]∩_{min}PStemp[<w, k-2>];
            call CHECK1 to calculate the final PStemp[<x, k-3>];
            PStemp[<w, k-2>]=[∪_{max}(PStemp[<x, k-3>]⊗{{w}})]∩_{min}PStemp[<w, k-2>];
          }
      PStemp[<v, k-1>]=[∪_{max}(PStemp[<w, k-2>]⊗{{v}})]∩_{min}PStemp[<v, k-1>];
    }
END
```

---

**Pseudocode of Operator: CHECK**   //Check the case of PStemp[<v, k-1>]={∅} (V3.0)
Input: *PS*[ ], PStemp[<v, k-1>]
Output: PStemp[<v, k-1>]

```
BEGIN
 1.  PStemp1={∅};
 2.  for (each parent vertex <w, k-2> of <v, k-1> in PStemp[<v, k-1>]) {
 3.      PStemp[<w, k-2>]=PS[<w, k-2>]∩_{min}PStemp[<v, k-1>];
 4.      PStemp2={∅};
 5.      for (each parent vertex <x, k-3> of <w, k-2> in PStemp[<w, k-2>]) {
 6.          PStemp[<x, k-3>]=PS[<x, k-3>]∩_{min}PStemp[<w, k-2>];
 7.          PStemp[<x, k-3>]=CHECK1(PStemp[<x, k-3>]);
 8.          if (PStemp[<x, k-3>]≠{∅})
 9.              PStemp2=PStemp2∪_{max}PStemp[<x, k-3>];
10.      }   //for each <x, k-3>
11.      if (PStemp2={∅})
12.          PStemp[<w, k-2>]={∅};
13.      else PStemp[<w, k-2>]=(PStemp2⊗{{w}})∩_{min}PStemp[<w, k-2>];
14.      if (PStemp[<w, k-2>]≠{∅})
```

15.                PStemp1=PStemp1$\cup_{max}$PStemp[<w, k-2>];
16.      }     //for each <w, k-2>
17.      if (PStemp1={∅})
18.          PStemp[<v, k-1>]={∅};
19.      else PStemp[<v, k-1>]=(PStemp1⊗{{v}})$\cap_{min}$PStemp[<v, k-1>];
20.      return (PStemp[<v, k-1>]), stop;
END

**Operator: CHECK1**
BEGIN
for (scan the PStemp[<x, k-3>] to find the empty segment set)
    if (there exists an empty segment set)    then { PStemp[<x, k-3>]={∅}, stop; }
for (each segment level *i*)
    for (each element p in $PStemp_{<x, k-3>}[i]$)
        {   PStemp[<p, i>]=PS[<p, i>]$\cap_{min}$PStemp[<x, k-3>];
            handle all the duplicated vertices of a singleton set in PStemp[<p, i>];
            if (∅∈ PStemp[<p, i>])    then    PStemp[<p, i>]={∅};
            PStemp[<p, i>]=[$\cup_{max}$(PStemp[<q, i-1>]⊗{{p}})]$\cap_{min}$PStemp[<p, i>];   //<q, i-1> is the parent vertex of [<p, i>] in PStemp[<p, i>]
        }
END

---

**Pseudocode of Operator: CHECK1**    //Check the case of PStemp[<x, k-3>]={∅} (V1.0)
Input: PS[ ], PStemp[<x, k-3>]
Output: PStemp[<x, k-3>]

---
BEGIN
 1.  PStemp = PStemp[<x, k-3>], PStemp[<1, 0>]={{1}};
 2.  for (*i*=k-3 to 1 step -1)
 3.      if (PStemp[*i*]=∅) {
 4.          PStemp[<x, k-3>]={∅}, return(PStemp[<x, k-3>]), stop;
 5.      }
 6.  for (*i*=1 to k-3 step +1)
 7.      for (each element p in PStemp[*i*]) {
 8.          PStemp[<p, i>]=PS[<p, i>]$\cap_{min}$PStemp;
 9.          flag2=true, flag3=false;
10.          While (flag2) {
11.              flag3=false;
12.              for (*j*=k-3 to 1 step -1) {
13.                  if ($PStemp_{<p,i>}[j]$ has only one element such as *z*) {
14.                      for (*l*=k-3 & *l*≠*j* to 1 step -1) {
15.                          if ($PStemp_{<p,i>}[l]$∈ PStemp[<p, i>] contains the vertex *z*) {
16.                              flag3=true;
17.                              delete *z* from $PStemp_{<p,i>}[l]$;
18.                              PStemp[<p, i>]=**LAFDR**(PStemp[<p, i>], <z, l>);

```
19.                         PStemp[<p, i>]=RAFDR(PStemp[<p, i>], <z, l>);
20.                         if (∅∈ PStemp[<p, i>])   PStemp[<p, i>]={∅};
21.                     }   //end if
22.                 }       //end for l
23.             }       //end if
24.             if (flag3) break;
25.         }   //end for j
26.         if (flag3) flag2=true;
27.         else flag2=false;
28.     }   //end while (flag2)
29.     if (∅∈ PStemp[<p, i>])   PStemp[<p, i>]={∅};
30.     else {
31.         PStemp1={∅};
32.         for (each parent vertex <q, i-1> of <p, i> & q∈ PStemp[i-1])
33.             if (PStemp[<q, i-1>]≠{∅})
34.                 PStemp1=PStemp1∪_{max}PStemp[<q, i-1>];
35.             if (PStemp1={∅})
36.                 PStemp[<p, i>]={∅};
37.             else PStemp[<p, i>]=(PStemp1⊗{{p}})∩_{min}PStemp[<p, i>];
38.     }
39.   } //end for each element p
40.   return (PStemp[<x, k-3>]), stop;
END
```

Consider the **CM** operator. In step 2, the segment sets of the path set can be handled in time $O(n-1)$ steps. In step 6 and 7, the **LAFDR** operator and **RAFDR** operator can be performed in time $O(n^2)$ steps with disjoint set technique. Thus the total time complexity of step 2 is $O(n^3)$. So is the total time complexity of step 20. The analysis of step 18 is a bit complicated. Case 1, if the "If" statement of step 19 is false, then step 20 does not execute. So the total time complexity of step 18 is $O(n)$. Case 2, if the "If" statement of step 19 is true, but the "If" statement of step 21 is false, then step 20 runs in $O(n)$ time, resulting in that the total time complexity of step 18 is $O(n^2)$. Case 3, if both the "If" statement of step 19 and the "If" statement of step 20 are true, then step 20 runs in $O(n^3)$ time, resulting in that the total time complexity of step 18 is $O(n)+O(n^2)+O(n^3)=O(n^3)$.

Theoretically, the iterations of the step 16 is $O(1)$ in the best case, $O(n)$ on average, and $O(n^2)$ in the worst case, respectively. So the total time complexity of step 16 is theoretically $O(n^3)$ in the best case, $O(n^4)$ on average, and $O(n^5)$ in the worst case, respectively.

For the same reason, in step 37, the **CHECK** operator can be performed in time $O(n^6 \times d^2)$ on average and $O(n^7 \times d^2)$ in the worst case. However, with the parallel computing of $n \times n$ computers, the total time complexity of step 37 can be improved to $O(n^4)$ on average and $O(n^5)$ in the worst case. Thus the time complexity of the **CM** operator is $O(n^6 \times d^2)$ on average and $O(n^7 \times d^2)$ in the worst case. With the parallel computing of $n \times n$ computers, the total time complexity of the **CM** operator can be improved to $O(n^4)$ on average and $O(n^5)$ in the worst case.

The path set of a vertex needs $O(n^2)$ space and the path hologram has $2+(n-1)^2$ vertices. Thus the space complexity of the **CM** operator is $O(n^4)$. With the parallel computing of $n$ computers, the space complexity of the **CM** operator can be improved to $O(n^3)$.

### 3.3 LPM Operator

After judging the consistency condition about the vertex $<u, k>$ joining the path set $PS[<v, k-1>]$ of its each parent vertex $<v, k-1>$ respectively, we can obtain all the basic paths from the initial vertex ***S*** or a vertex on segment level $j$, where $1\leq j\leq k-1$, to the vertex $<u, k>$ through the vertex $<v, k-1>$. By comparing these paths with greedy strategy, we consider the longest path as the final path from the initial vertex ***S*** or a vertex on segment level $j$, where $1\leq j\leq k-1$, to the vertex $<u, k>$. If there exist many such longest paths of the same length, the final $PS[<u, k>]$ is generated by calculating the union operation of these paths, i.e., the segment sets on the same segment level in the longest paths execute the merging operation. In fact, the **LPM** operator completes the calculation of the **Formula (1)**.

We know that the **CM** operator returns the PStemp[$<u, k>$] which contains the longest basic paths from the initial vertex ***S*** or a vertex on segment level $j$, where $1\leq j\leq k-1$, to the vertex $<u, k>$ through its parent vertex $<v, k-1>$. On the other hand, $PS[<u, k>]$ has stored the known longest basic paths from the initial vertex ***S*** or a vertex on segment level $j$, where $1\leq j\leq k-1$, to the vertex $<u, k>$ up to now. So that, if $PS[<u, k>]$ and PStemp[$<u, k>$] have the same length, then the union operation on $PS[<u, k>]$ and PStemp[$<u, k>$] would be calculated, i.e., $PS_{<u, k>}[i]$=PStemp$_{<u, k>}[i]\cup PS_{<u, k>}[i]$ for $0\leq i\leq k$. If PStemp[$<u, k>$] has a longer length, then $PS[<u, k>]$=PStemp[$<u, k>$]; otherwise, the $PS[<u, k>]$ remains.

**Operator: LPM**
BEGIN
$PS[<u, k>]=PS[<u, k>]\cup_{max}$PStemp[$<u, k>$];
END

---

**Pseudocode of Operator: LPM**   //Longest Path Merging (V1.0)
Input: $PS[<u, k>]$, PStemp[$<u, k>$]
Output: $PS[<u, k>]$

---

BEGIN
1.  length1 = length($PS[<u, k>]$), length2 = length(PStemp[$<u, k>$]);
2.  if (PStemp[$<u, k>$]={{u}} or length2<length1)
3.      return $PS[<u, k>]$;
4.  else {
5.      if (length2>length1) {
6.          $PS[<u, k>]$ = PStemp[$<u, k>$];
7.          return $PS[<u, k>]$
8.      }
9.      else {
10.         for ($i$=length1 to 1 step -1)
11.             $PS_{<u,k>}[i] = PS_{<u,k>}[i] \cup$ PStemp$_{<u,k>}[i]$;
12.         return $PS[<u, k>]$
13.     }  //end if
14. }  //end if
END

In step 10, the segment sets of the path set can be handled in time $O(n\text{-}1)$ steps. In step 11, the union operation on two sets can be handled in time $O(n\text{-}1)$ steps. The path set of a vertex needs $O(n^2)$ space. Thus the time complexity and space complexity of the **LPM** operator are $O(n^2)$ and $O(n^2)$, respectively.

### 3.4  FHC Operator

The **FHC** operator finds the basic path from the initial vertex ***S*** to the final vertex ***D*** in the path hologram ***H*** by backward searching for a non-redundant valid parent vertex based on the intersection of its path set and the path set of its parent vertex recursively at each step.

The Hamiltonian cycle was backward searched from the final vertex ***D*** to the initial vertex ***S***. At each step, we select such a non-redundant valid parent vertex <*v*, *k*-1> of <*u*, *k*> based on the temporary path set PStemp that $\varnothing \notin$ **CHECK**($PS[<v, k\text{-}1>] \cap_{min} \text{PStemp}$). Next, the PStemp will be updated by $\text{PStemp} \cap_{min} PS[<v, k\text{-}1>]$, where PStemp=$PS[<1, n>]$ initially.

**Operator: FHC**
BEGIN
<*u*, *k*>=<1, *n*>, PStemp=$PS[<1, n>]$, print(<1, *n*>);
for (scan each segment level *i* from level *n*-1 to level 0)
    { select such a parent vertex <*v*, *i*> of <*u*, *k*> that **CHECK**($PS[<v, i>] \cap_{min} \text{PStemp}$)≠{∅};
      print(<*v*, *i* >), <*u*, *k*>=<*v*, *i*>, PStemp=$PS[<v, i>] \cap_{min} \text{PStemp}$;
    }
END

---

**Pseudocode of Operator: FHC**   //Find Hamilton Cycle (V3.0)
Input: the path set array *PS*[ ], D=<1, *n*>   //array *PS*[ ] saves the path sets of all vertices
Output: a Hamiltonian cycle

---

BEGIN
1. Init(SS);   // Initialize the stack SS which saves the Hamiltonian cycle
2. PUSH(SS, 1);
3. <*u*, *k*>=<1, *n*>;
4. PStemp=$PS[<1, n>]$;
5. for (*i*=*n*-1 to 1 step -1) {
6.    select such a non-redundant valid parent vertex <*v*, *i*> of <*u*, *k*>, where *v*∈ PStemp[*i*], that $\varnothing \notin$ **CHECK**($PS[<v, i>] \cap_{min} \text{PStemp}$);
7.    PUSH(SS, *v*);
8.    <*u*, *k*>=<*v*, *i*>;
9.    for (*j*=*i* to 0 step -1)
10.       PStemp[*j*]=PStemp[*j*]∩PS$_{<v,i>}$[*j*];
11. }   // end for *i*
12. PUSH(SS, 1);
13. return SS;
END

In step 5, the segments of the path hologram can be handled in time $O(n\text{-}1)$ steps. In step 6, the time complexity of the **CHECK** operator is $O(n^6 \times d^2)$ on average and $O(n^7 \times d^2)$ in the worst case. With the parallel computing of $n \times n$ computers, the total time complexity of the **CHECK** operator can be improved to $O(n^4)$ on average and $O(n^5)$ in the worst case. Thus the time complexity of the **FHC** operator is $O(n^7 \times d^2)$ on average and $O(n^8 \times d^2)$ in the worst case. With the parallel computing of $n \times n$ computers, the total time complexity of the **FHC** operator can be improved to $O(n^5)$ on average and $O(n^6)$ in the worst case.

The path set of a vertex needs $O(n^2)$ space and the path hologram has $2+(n-1)^2$ vertices. So the space complexity of the **FHC** operator is $O(n^4)$. With the parallel computing of $n$ computers, the space complexity of the **FHC** operator can be improved to $O(n^3)$.

### 3.5 GPHG Algorithm

We describe other auxiliary algorithm **GPHG**, which transforms the original graph $G=(V, E)$ into the path hologram $H=<V_H, E_H, S, D, L>$. First, we select any vertex $v$ of $V$ in the original graph $G=(V, E)$ to be as a starting vertex and generate the initial vertex $S=<v, 0>$ and the final vertex $D=<v, L>$ of $V_H$ in the path hologram $H=<V_H, E_H, S, D, L>$ respectively, where $S$ is located on segment level 0 and $D$ located on segment level $L=|V|$. Second, according to each edge $vu \in E(G)$ incident with $v$ in $G$, we generate two directed edges $(<v, 0>, <u, 1>)$ and $(<u, L\text{-}1>, <v, L>)$ in $H$ respectively. Third, according to each edge $wu \in E(G)$ which neither $w$ nor $u$ is $v$ in $G$, we generate $2(L\text{-}2)$ directed edges $(<u, k>, <w, k+1>)$ and $(<w, k>, <u, k+1>)$ in $H$ respectively, where $1 \leq k \leq L\text{-}2$ and $L=|V(G)|$.

**Algorithm: GPHG**
BEGIN
select any vertex $v$ as the initial vertex $S=<v, 0>$ and the final vertex $D=<v, n>$;
for (all adjacency vertex $u$ of vertex $v$)
   { generate the directed edge $(<v, 0>, <u, 1>)$;
      calculate the in-neighbourhood $N^-(<u, 1>)$ and the out-neighbourhood $N^+(<v, 0>)$;
      generate the directed edge $(<u, n\text{-}1>, <v, n>)$;
      calculate the in-neighbourhood $N^-(<v, n>)$ and the out-neighbourhood $N^+(<u, n\text{-}1>)$;
   }
for (each edge $wx \in E$ && $w \neq v$ && $x \neq v$)
   for ($i=1$ to $n\text{-}2$ step $+1$)    //considering each segment
      { generate the directed edge $(<w, i>, <x, i+1>)$;
        calculate the in-neighbourhood $N^-(<x, i+1>)$ and the out-neighbourhood $N^+(<w, i>)$;
        generate the directed edge $(<x, i>, <w, i+1>)$;
        calculate the in-neighbourhood $N^-(<w, i+1>)$ and the out-neighbourhood $N^+(<x, i>)$;
      }
END

---

**Pseudocode of Algorithm: GPHG**   //Generate Path HoloGram (V1.0)
Input: the original graph $G=(V, E)$
Output: the path hologram $H=<V_H, E_H, S, D, L>$
BEGIN
  1. Label all vertices of $V(G)$ in ascending order as $1, 2, \cdots, n=|V(G)|$;

2. Let $V_H=\{\}$, $E_H=\{\}$, $L=n$;
3. No loss of generality, select vertex 1 as the start point, let $S=<1, 0>$ and $D=<1, n>$, $V_H=V_H\cup\{S\}$, $V_H=V_H\cup\{D\}$, $N^-(S)=N^+(S)=\{\}$, $N^-(D)=N^+(D)=\{\}$;
4. for ($i=1$ to $n$-1 step +1)   //considering each segment
5.     for ($j=2$ to $n$ step +1) {   //considering each vertex
6.         $V_H=V_H\cup\{<j, i>\}$;
7.         $N^-(<j, i>)=N^+(<j, i>)=\{\}$;
8.     }
9. for (all adjacency vertex $u$ of vertex 1 in graph $G$) {
10.     $E_H=E_H\cup\{(<1, 0>, <u, 1>)\}$;
11.     $N^-(<u, 1>)=N^-(<u, 1>)\cup\{1\}$;
12.     $N^+(<1, 0>)=N^+(<1, 0>)\cup\{u\}$;
13.     $E_H=E_H\cup\{(<u, n-1>, <1, n>)\}$;
14.     $N^-(<1, n>)=N^-(<1, n>)\cup\{u\}$;
15.     $N^+(<u, n-1>)=N^+(<u, n-1>)\cup\{1\}$;
16. } //end for all
17. for (any edge $e=uv\in E$ && $u\neq 1$ && $v\neq 1$)   //generate all other directed edges
18.     for ($i=1$ to $n$-2 step +1) {   //considering each segment
19.         $E_H=E_H\cup\{(<u, i>, <v, i+1>)\}$;
20.         $N^-(<v, i+1>)=N^-(<v, i+1>)\cup\{u\}$;
21.         $N^+(<u, i>)=N^+(<u, i>)\cup\{v\}$;
22.         $E_H=E_H\cup\{(<v, i>, <u, i+1>)\}$;
23.         $N^-(<u, i+1>)=N^-(<u, i+1>)\cup\{v\}$;
24.         $N^+(<v, i>)=N^+(<v, i>)\cup\{u\}$;
25.     }//end for i
26. return $H=<V_H, E_H, S, D, L>$
END

The total time complexity of step 4 is $O(n^2)$. In step 9, adjacent vertex can be handled in time $O(d)$ steps, where $d$ is the maximum degree of vertex. The total time complexity of step 17 is $O(n\times e)$ steps, where $e$ is the size of the original graph. The path hologram has $2d+2(n-2)(e-d)$ edges and $(n-1)^2+2$ vertices. Thus the time complexity and space complexity of the **GPHG** algorithm are $O(n\times e)$ respectively.

## 4 Instances

In this section, we discuss some instances to illustrate the principle of our method.

**Example 1.** A finite connected undirected original graph $G=(V, E)$ and its path hologram $H=<V_H, E_H, S, D, L>$ are shown in **Figure 2(a)** and **2(b)**, respectively. In view of the size of the drawing, the arrow is removed from the line.

Select any vertex $v$ of $V(G)$, without loss of generality assume that $v=1$, to be as a starting vertex. Thus, $S=<1, 0>$ and $D=<1, 8>$. Initially, $PS[S]=PS[<1, 0>]=\{\{1\}\}$, $PS[<2, 1>]=\{\{2\}\}$, $PS[<3, 1>]=\{\{3\}\}$, $PS[<4, 1>]=\{\{4\}\}$, $PS[<5, 1>]=\{\{5\}\}$, $PS[<6, 1>]=\{\{6\}\}$, $PS[<7, 1>]=\{\{7\}\}$, $PS[<8, 1>]=\{\{8\}\}$, $PS[<2, 2>]=\{\{2\}\}$, and so on.

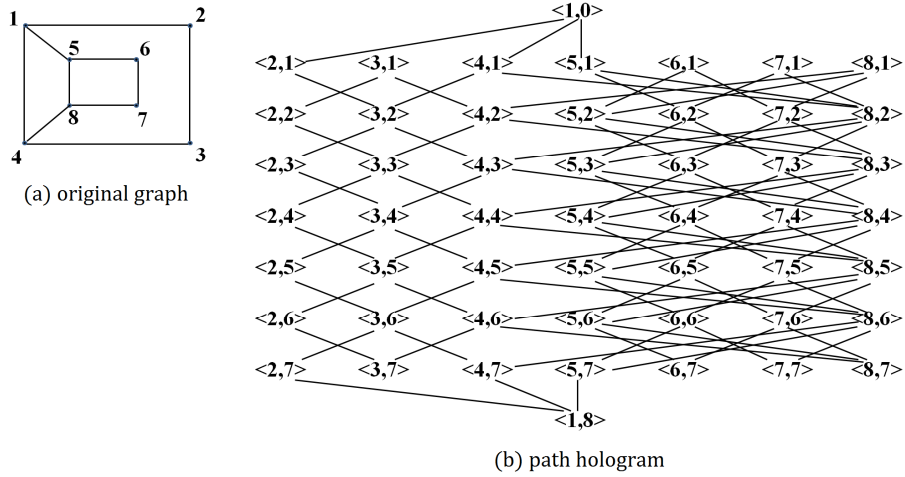

(a) original graph

(b) path hologram

**Figure 2** A finite connected undirected original graph *G* and its path hologram *H* of **Example 1**

Iteration 1: Processing segment level *k*=1.

The vertices <3, 1>, <6, 1>, <7, 1> and <8, 1> have no parent vertices, so they are not handled. In other words, they keep the initial value. The vertex <2, 1> has a parent vertex <1, 0>, thus the vertex 2 will be considered whether to cause a conflict or not if it joins the path set *PS*[<1, 0>]. That is, we call the **CM**(*PS*[<1, 0>], <2, 1>) operator to judge the conflict. Because the vertex 2 does not occur in *PS*[<1, 0>]={{1}}, a basic path from the vertex <1, 0> to the vertex <2, 1> can be obtained. So that PStemp[<2, 1>]=**CM**(*PS*[<1, 0>], <2, 1>) ={{1}, {2}} is constructed by joining the vertex 2 in *PS*[<1, 0>]. Then we call the **LPM**(*PS*[<2, 1>], PStemp[<2, 1>]) operator to compute the path set union. That is, *PS*[<2, 1>]=**LPM**(*PS*[<2, 1>], PStemp[<2, 1>])=∪$_{max}$ {*PS*[<2, 1>], PStemp[<2, 1>]}=∪$_{max}$ {{{2}}, {{1}, {2}}}={{1}, {2}}, where *PS*[<2, 1>]={{2}} is shorter than PStemp[<2, 1>] ={{1}, {2}}. Analogously, *PS*[<4, 1>]={{1}, {4}}, *PS*[<5, 1>]={{1}, {5}}.

Iteration 2: Processing segment level *k*=2.

1. The vertex <2, 2> has one parent vertex <3, 1>. Since the vertex 2 does not occur in *PS*[<3, 1>]= {{3}}, the vertex 2 can join *PS*[<3, 1>] directly. We obtain *PS*[<2, 2>]={{3}, {2}}.

2. The vertex <3, 2> has two parent vertices <2, 1> and <4, 1> and cannot cause a conflict when it joins *PS*[<2, 1>] or *PS*[<4, 1>] respectively. So that, we obtain PStemp[<3, 2>]=**CM**(*PS*[<2, 1>], <3, 2>)={{1}, {2}, {3}} and *PS*[<3, 2>]=**LPM**(*PS*[<3, 2>], PStemp[<3, 2>])=∪$_{max}${{{3}}, {{1}, {2}, {3}}}={{1}, {2}, {3}} after the vertex 3 joining *PS*[<2, 1>]. Further, we obtain PStemp[<3, 2>]= **CM**(*PS*[<4, 1>], <3, 2>)={{1}, {4}, {3}} and *PS*[<3, 2>]=**LPM**(*PS*[<3, 2>], PStemp[<3, 2>])= ∪$_{max}${{{1}, {2}, {3}}, {{1}, {4}, {3}}}={{1}, {2, 4}, {3}} after the vertex 3 joining *PS*[<4, 1>]. Here both *PS*[<3, 2>]={{1}, {2}, {3}} and PStemp[<3, 2>]={{1}, {4}, {3}} have the same longest length.

3. Analogously, *PS*[<4, 2>]={{3, 8}, {4}}, *PS*[<5, 2>]={{6, 8}, {5}}, *PS*[<6, 2>]={{1}, {5}, {6}}, *PS*[<7, 2>]={{6, 8}, {7}}, *PS*[<8, 2>]={{1}, {4, 5}, {8}}.

Iteration 3: Processing segment level *k*=3.

1. The vertex <2, 3> has one parent vertex <3, 2>. Since the vertex 2 appears in *PS*[<3, 2>]= {{1}, {2, 4}, {3}}, the vertex 2 will cause a conflict when it joins *PS*[<3, 2>], this means that there exists a path 1-2-3-2, which contains the vertex 2 appeared in front, from the initial vertex *S* to the vertex <2, 3> through the vertex <3, 2>. After deleting the vertex 2 from PS$_{<3,2>}$[1], the vertex 1∈ $AF^-$(<2, 1>) and the vertex 3∈ $AF^+$(<2, 1>) are replenished due to the reason of 1∈ $AF^-$(<4, 1>) and 3∈ $AF^+$(<4, 1>), respectively. So that the process of the "consecutive" deleting-replenishing operations recursively on the left/right action field of the duplicative vertex of 2 in $PS[<3, 2>]$ has been accomplished. At this time, PStemp[<3, 2>]={{1}, {4}, {3}}. Then the While(flag2) loop will be executed only once for handling the duplicates of singletons. Since there do not exist the duplicates of singletons, resulting in that PStemp[<3, 2>]={{1}, {4}, {3}}.

Then we call the **CHECK** operator to check the validity of PStemp[<3, 2>]={{1}, {4}, {3}}. PStemp[<4, 1>]=$PS[<4, 1>]\cap_{min}$PStemp[<3, 2>]={{1}, {4}}$\cap_{min}${{1}, {4}, {3}}={{1}, {4}} and does not contain the empty set $\varnothing$. PStemp[<1, 0>]=$PS[<1, 0>]\cap_{min}$PStemp[<3, 2>]={{1}}$\cap_{min}${{1}, {4}, {3}}={{1}} and does not contain the empty set $\varnothing$. Therefore, PStemp[<3, 2>]={{1}, {4}, {3}}, resulting in that PStemp[<2, 3>]=PStemp[<3, 2>]⊗{{2}}={{1}, {4}, {3}, {2}}. Noticed that <2, 1>∉ PStemp[<3, 2>], so we need not check PStemp[<2, 1>]. In fact, PStemp[<2, 1>]=$PS[<2, 1>]\cap_{min}$ PStemp[<3, 2>]={{1}, {2}}$\cap_{min}${{1}, {4}, {3}}={{1}, $\varnothing$}, containing the empty set $\varnothing$.

Next, we have $PS[<2, 3>]$=**LPM**($PS[<2, 3>]$, PStemp[<2, 3>])=$\cup_{max}${{{2}}, {{1}, {4}, {3}, {2}}} ={{1}, {4}, {3}, {2}} after the vertex 2 joining the remainder of $PS[<3, 2>]$.

2. The vertex <3, 3> has two parent vertices <2, 2> and <4, 2>. (1) Since the vertex 3 appears in $PS[<2, 2>]$={{3}, {2}}, the vertex 3 will cause a conflict when it joins $PS[<2, 2>]$. After deleting the duplicative vertex of 3 from PS$_{<2,2>}$[1], an empty segment set $\varnothing$ in PStemp[<2, 2>]={{}, {2}} occurs, resulting in that the <2, 2> is abandoned. (2) Since the vertex 3 appears in $PS[<4, 2>]$= {{3, 8}, {4}}, the vertex 3 will cause a conflict when it joins $PS[<4, 2>]$. After deleting the duplicates of 3 from PS$_{<4,2>}$[1], PStemp[<4, 2>]={{8}, {4}}. The While(flag2) loop will be executed only once. Then we call the **CHECK** operator to check the validity of PStemp[<4, 2>]={{8}, {4}}. PStemp[<8, 1>]=PS[<8, 1>]$\cap_{min}$PStemp[<4, 2>]={{8}}$\cap_{min}${{8}, {4}}={{8}} and does not contain the empty set $\varnothing$. Thus, PStemp[<4, 2>]={{8}, {4}}, resulting in that PStemp[<3, 3>]=PStemp[<4, 2>]⊗{{3}}= {{8}, {4}, {3}}. Noticed that <1, 0>∉ PStemp[<4, 2>], so we need not check PStemp[<1, 0>].

Then we have $PS[<3, 3>]$=**LPM**($PS[<3, 3>]$, PStemp[<3, 3>])=$\cup_{max}${{{3}}, {{8}, {4}, {3}}}= {{8}, {4}, {3}} after the vertex 3 joining the remainder of $PS[<4, 2>]$.

3. Analogously, $PS[<4, 3>]$={{1}, {2, 5}, {3, 8}, {4}}, $PS[<5, 3>]$={{1}, {4}, {8}, {5}}, $PS[<6, 3>]$={{8}, {5, 7}, {6}}, $PS[<7, 3>]$={{1}, {4, 5}, {6, 8}, {7}}, $PS[<8, 3>]$={{3, 6}, {4, 5, 7}, {8}}.

Iteration 4: Processing segment level $k$=4.

1. The vertex <2, 4> has one parent vertex <3, 3> and the vertex 2 does not occur in $PS[<3, 3>]$={{8}, {4}, {3}}, so PStemp[<2, 4>]=**CM**($PS[<3, 3>]$, <2, 4>)={{8}, {4}, {3}, {2}} and $PS[<2, 4>]$=**LPM**($PS[<2, 4>]$, PStemp[<2, 4>])=$\cup_{max}${{{2}}, {{8}, {4}, {3}, {2}}}={{8}, {4}, {3}, {2}}. In this case, the **CHECK** operator is not executed.

2. The vertex <3, 4> has two parent vertices <2, 3> and <4, 3>. (1) Since the vertex 3 appears in $PS[<2, 3>]$={{1}, {4}, {3}, {2}}, the vertex 3 will cause a conflict when it joins $PS[<2, 3>]$. After deleting the duplicative vertex of 3 from PS$_{<2,3>}$[2], an empty segment set $\varnothing$ in PStemp[<2, 3>] ={{1}, {4}, {}, {2}} occurs, resulting in that the <2, 3> is abandoned. (2) Since the vertex 3 appears in $PS[<4, 3>]$={{1}, {2, 5}, {3, 8}, {4}}, the vertex 3 will cause a conflict when it joins $PS[<4, 3>]$. After deleting the duplicative vertex of 3 from PS$_{<4,3>}$[2] and the invalid vertex 2∈$AF^-$(<3, 2>), we have PStemp[<4, 3>]={{1}, {5}, {8}, {4}}. The While(flag2) loop will be executed only once. Then we call the **CHECK** operator to check the validity of PStemp[<4, 3>]={{1}, {5}, {8}, {4}} and we finally determined that PStemp[<4, 3>]={{1}, {5}, {8}, {4}}.

Next we have $PS[<3, 4>]$=**LPM**($PS[<3, 4>]$, PStemp[<3, 4>])=$\cup_{max}${{{3}}, {{1}, {5}, {8}, {4}, {3}}}={{1}, {5}, {8}, {4}, {3}} after the vertex 3 joining the remainder of $PS[<4, 3>]$.

3. Analogously, *PS*[<4, 4>]={{6}, {5, 7}, {8}, {4}}, *PS*[<5, 4>]={{3, 6, 8}, {4, 7}, {6, 8}, {5}}, *PS*[<6, 4>]={{1}, {4, 5}, {8}, {5, 7}, {6}}, *PS*[<7, 4>]={{3, 6, 8}, {4, 5}, {6, 8}, {7}}, *PS*[<8, 4>]={{1}, {2, 5}, {3, 6}, {4, 7}, {8}}.

Iteration 5: Processing segment level *k*=5.

1. As the analysis above, we have *PS*[<2, 5>]={{1}, {5}, {8}, {4}, {3}, {2}}, *PS*[<3, 5>]={{6}, {5, 7}, {8}, {4}, {3}}, *PS*[<4, 5>]={{1}, {5}, {6}, {7}, {8}, {4}}, *PS*[<5, 5>]={{1}, {2, 4}, {3, 8}, {4, 7}, {6, 8}, {5}}, *PS*[<6, 5>]={{3}, {4}, {8}, {5, 7}, {6}}, *PS*[<7, 5>]={{1}, {2, 4}, {3, 8}, {4, 5}, {6, 8}, {7}}.

2. The vertex <8, 5> has three parent vertices <4, 4>, <5, 4> and <7, 4>. (1) Since the vertex 8 appears in *PS*[<4, 4>]={{6}, {5, 7}, {8}, {4}}, the vertex 8 will cause a conflict when it joins *PS*[<4, 4>]. After deleting the duplicates of 8 from $PS_{<4,4>}[3]$, an empty segment set ∅ in PStemp[<4, 4>]={{6}, {5, 7}, {}, {4}} occurs, resulting in that the <4, 4> is abandoned. (2) Since the vertex 8 appears in *PS*[<5, 4>]={{3, 6, 8}, {4, 7}, {6, 8}, {5}}, the vertex 8 will cause a conflict when it joins *PS*[<5, 4>]. After deleting the duplicative vertex of 8 from $PS_{<5,4>}[3]$ and $PS_{<5,4>}[1]$, as well as the invalid vertices in left/right action fields, we have PStemp[<5, 4>]={{6}, {7}, {6}, {5}}. So that the process of the "consecutive" deleting-replenishing operations recursively on the left/right action field of the duplicative vertex of 8 in *PS*[<5, 4>] has been accomplished. The While(flag2) loop will be executed only once. We find that the singleton {6}=$PStemp_{<5,4>}[3]$ has a duplicative vertex <6, 1>, so that PStemp[<5, 4>]={∅, {7}, {6}, {5}} after deleting the duplicative vertex of the singletons, resulting in that the <5, 4> is abandoned. (3) Since the vertex 8 appears in *PS*[<7, 4>]={{3, 6, 8}, {4, 5}, {6, 8}, {7}}, the vertex 8 will cause a conflict when it joins *PS*[<7, 4>]. After deleting the duplicative vertex of 8 from $PS_{<7,4>}[3]$ and $PS_{<7,4>}[1]$, as well as the invalid vertices in left/right action fields, we have PStemp[<7, 4>]={{6}, {5}, {6}, {7}}. Similar to the case (2), we have PStemp[<7, 4>]={∅} and abandon the parent vertex <7, 4>.

So far, all the parent vertices of <8, 5> are invalid, this means that there does not exist a basic path from the initial vertex **S** to the vertex <8, 5> through any one of its parent vertices <4, 4>, <5, 4>, or <7, 4>. Therefore, *PS*[<8, 5>]={{8}}, which is the initial value. In these cases, the **CHECK** operator is not executed.

Iteration 6: Processing segment level *k*=6.

As the analysis above, we have *PS*[<2, 6>]={{6}, {5, 7}, {8}, {4}, {3}, {2}}, *PS*[<3, 6>]={{1}, {5}, {6}, {7}, {8}, {4}, {3}}, *PS*[<4, 6>]={{8}, {4}}, *PS*[<5, 6>]={{3}, {4}, {8}, {7}, {6}, {5}}, *PS*[<6, 6>]={{1}, {2}, {3}, {4}, {8}, {5, 7}, {6}}, *PS*[<7, 6>]={{3}, {4}, {8}, {5}, {6}, {7}}, *PS*[<8, 6>]={{8}}.

Iteration 7: Processing segment level *k*=7.

As the analysis above, we have *PS*[<2, 7>]={{1}, {5}, {6}, {7}, {8}, {4}, {3}, {2}}, *PS*[<3, 7>]= {{8}, {4}, {3}}, *PS*[<4, 7>]={{8}, {4}}, *PS*[<5, 7>]={{1}, {2}, {3}, {4}, {8}, {7}, {6}, {5}}, *PS*[<6, 7>]={{6}}, *PS*[<7, 7>]={{1}, {2}, {3}, {4}, {8}, {5}, {6}, {7}}, *PS*[<8, 7>]= {{8}}.

Iteration 8: Processing segment level *k*=8.

Since the vertex <1, 8> has three parent vertices <2, 7>, <4, 7> and <5, 7>, the vertex 1 will be considered whether to cause a conflict or not if it joins the path set *PS*[<2, 7>], *PS*[<4, 7>], or *PS*[<5, 7>], respectively. Because the vertex <1, 8> is the final vertex **D**, the vertex 1 can naturally join *PS*[<2, 7>], *PS*[<4, 7>] and *PS*[<5, 7>] respectively. Thus we obtain PStemp[<1, 8>]=**CM**(*PS*[<2, 7>], <1, 8>)={{1}, {5}, {6}, {7}, {8}, {4}, {3}, {2}, {1}} and *PS*[<1, 8>]=**LPM**(*PS*[<1, 8>], PStemp[<1, 8>])={{1}}∪$_{max}${{1}, {5}, {6}, {7}, {8}, {4}, {3}, {2}, {1}}={{1}, {5}, {6}, {7}, {8}, {4}, {3},

{2}, {1}} after the vertex 1 joining $PS[<2, 7>]$. In the same way, we obtain PStemp[<1, 8>]= **CM**($PS[<4, 7>]$, <1, 8>)={{8}, {4}, {1}} and $PS[<1, 8>]$=**LPM**($PS[<1, 8>]$, PStemp[<1, 8>]) ={{1}, {5}, {6}, {7}, {8}, {4}, {3}, {2}, {1}}$\cup_{max}${{8}, {4}, {1}}={{1}, {5}, {6}, {7}, {8}, {4}, {3}, {2}, {1}} after the vertex 1 joining $PS[<4, 7>]$. Further, PStemp[<1, 8>]=**CM**($PS[<5, 7>]$, <1, 8>)={{1}, {2}, {3}, {4}, {8}, {7}, {6}, {5}, {1}} and $PS[<1, 8>]$=**LPM**($PS[<1, 8>]$, PStemp[<1, 8>]) ={{1}, {5}, {6}, {7}, {8}, {4}, {3}, {2}, {1}}$\cup_{max}${{1}, {2}, {3}, {4}, {8}, {7}, {6}, {5}, {1}}= {{1}, {2, 5}, {3, 6}, {4, 7}, {8}, {4, 7}, {3, 6}, {2, 5}, {1}} after the vertex 1 joining $PS[<5, 7>]$.

Since the length of $PS[<1, 8>]$={{1}, {2, 5}, {3, 6}, {4, 7}, {8}, {4, 7}, {3, 6}, {2, 5}, {1}} is 9=$L$+1, where $L$=8= $|V(G)|$, there must exist a basic path from the initial vertex $S$ to the final vertex $D$ in the path hologram $H$, inferring that the original graph $G$ is Hamiltonian. Besides, $PS[<7, 7>]$={{1}, {2}, {3}, {4}, {8}, {5}, {6}, {7}}, there must exist a Hamiltonian path in the original graph $G$.

Next, we call the **FHC**($PS[\ ]$, $D$) operator to find a Hamiltonian cycle.

Initially, we let PStemp=$PS[<1, 8>]$={{1}, {2, 5}, {3, 6}, {4, 7}, {8}, {4, 7}, {3, 6}, {2, 5}, {1}} and the path 1. The vertex <1, 8> has two parent vertices <2, 7> and <5, 7> due to PStemp. Then we select any a non-redundant valid parent vertex of <1, 8>, without loss of generality assume that <2, 7>, based on the PStemp[7]={2, 5} for backward searching. Since $\varnothing \notin$ **CHECK**($PS[<2, 7>]\cap_{min}$PStemp) ={{1}, {5}, {6}, {7}, {8}, {4}, {3}, {2}}, we can finally decide to choose 2 and obtain the path 2-1 as well as PStemp=PStemp$\cap_{min}PS[<2, 7>]$={{1}$\cap${1}, {2, 5}$\cap${5}, {3, 6}$\cap${6}, {4, 7}$\cap${7}, {8}$\cap${8}, {4, 7} $\cap${4}, {3, 6}$\cap${3}, {2, 5}$\cap${2}}={{1}, {5}, {6}, {7}, {8}, {4}, {3}, {2}}. Next we select any a non-redundant valid parent vertex <3, 6> of <2, 7> from PStemp[6]={3} due to $\varnothing \notin$ **CHECK**($PS[<3, 6>]\cap_{min}$PStemp) and obtain the path 3-2-1 as well as PStemp=PStemp$\cap_{min}PS[<3, 6>]$={{1}$\cap${1}, {5}$\cap${5}, {6}$\cap${6}, {7}$\cap${7}, {8}$\cap$ {8}, {4}$\cap${4}, {3}$\cap${3}}={{1}, {5}, {6}, {7}, {8}, {4}, {3}}. Iteratively, we obtain the path 4-3-2-1 as well as PStemp=PStemp$\cap_{min}PS[<4, 5>]$= {{1}, {5}, {6}, {7}, {8}, {4}}. Finally, we obtain the path 1-5-6-7-8-4-3-2-1. So that we can find a Hamiltonian cycle 1-5-6-7-8-4-3-2-1 in the original graph $G$.

On the other hand, if we select the parent vertex <5, 7> of <1, 8> instead of the <2, 7> of <1, 8>, then we can find the Hamiltonian cycle 1-2-3-4-8-7-6-5-1 as follows. Initially, we let PStemp=$PS[<1, 8>]$={{1}, {2, 5}, {3, 6}, {4, 7}, {8}, {4, 7}, {3, 6}, {2, 5}, {1}} and the path 1. The vertex <1, 8> has two parent vertices <2, 7> and <5, 7> and we select the <5, 7> instead of <2, 7> for backward searching. Since $\varnothing \notin$ **CHECK**($PS[<5, 7>]\cap_{min}$PStemp)={{1}, {2}, {3}, {4}, {8}, {7}, {6}, {5}}, we finally decide to choose 5 and obtain the path 5-1 as well as PStemp=PStemp$\cap_{min}PS[<5, 7>]$= {{1}$\cap${1}, {2, 5}$\cap${2}, {3, 6}$\cap${3}, {4, 7}$\cap${4}, {8}$\cap${8}, {4, 7}$\cap${7}, {3, 6}$\cap${6}, {2, 5}$\cap${5}}={{1}, {2}, {3}, {4}, {8}, {7}, {6}, {5}}. Next, we select <6, 6> for backward searching, since $\varnothing \notin$ **CHECK**($PS[<6, 6>]\cap_{min}$PStemp)={{1}, {2}, {3}, {4}, {8}, {7}, {6}}. We obtain the path 6-5-1 as well as PStemp=PStemp$\cap_{min}PS[<6, 6>]$={{1}, {2}, {3}, {4}, {8}, {7}, {6}}. Similarly, we obtain the path 7-6-5-1 as well as PStemp=PStemp$\cap_{min}PS[<7, 5>]$={{1}, {2}, {3}, {4}, {8}, {7}}. Further, we obtain the path 8-7-6-5-1 as well as PStemp=PStemp$\cap_{min}PS[<8, 4>]$={{1}, {2}, {3}, {4}, {8}}. Finally, we obtain the path 1-2-3-4-8-7-6-5-1 and the Hamiltonian cycle 1-2-3-4-8-7-6-5-1.

As a result, we can find any a Hamiltonian cycle by selecting a different parent vertex $v_p$ for backward searching at each step $t$ such that $\varnothing \notin$ **CHECK**(PStemp$\cap_{min}PS[<v_p, t>]$), and this computation of finding a Hamiltonian cycle can be accomplished in deterministic polynomial time.

In fact, we can also call **FHC**($PS[\ ]$, <7, 7>) to find the Hamiltonian path 1-2-3-4-8-5-6-7 due to the length of <7, 7> being 8.

**Example 2.** An undirected complete graph of order 5, $K_5=(V, E)$, and its path hologram $H=<V_H, E_H, S, D, L>$ are shown in **Figure 3(a)** and **3(b)**, respectively. As one known, a complete graph has the maximum number of edges.

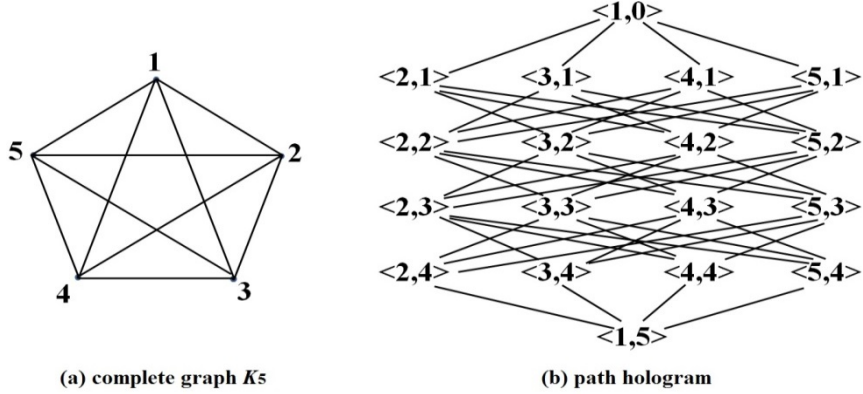

**Figure 3** An undirected complete graph $K_5$ and its path hologram $H$ of **Example 2**

Select any vertex $v$ of $V(K_5)$, without loss of generality assume that $v=1$, to be as a starting vertex. The path hologram $H=<V_H, E_H, S, D, L>$ transformed from the graph $K_5$ is shown in **Figure 3(b)**. In view of the size of the drawing, the arrow is removed from the line. Here, $S$=<1, 0> and $D$=<1, 5>.

We call the **PHG-BP** algorithm to judge whether a basic path from the initial vertex $S$ to the final vertex $D$ exists or not in the path hologram $H$. If this path exists, it is saved in the path set $PS$[<1, 5>]. Then we call the **FHC** operator to find this basic path from the initial vertex $S$ to the final vertex $D$ with backward searches.

Initially, we set $PS$[<1, 0>]={{1}}, $PS$[<2, 1>]={{2}}, $PS$[<3, 1>]={{3}}, $PS$[<4, 1>]= {{4}}, $PS$[<5, 1>]={{5}}, $PS$[<2, 2>]={{2}}, $PS$[<3, 2>]={{3}}, $PS$[<4, 2>]={{4}}, $PS$[<5, 2>]={{5}}, and so on.

Iteration 1: Processing segment level $k$=1.

The vertex <2, 1> has only one parent vertex <1, 0>, we will call the **CM**($PS$[<1, 0>], <2, 1>) operator to judge whether the conflict will occur or not if the vertex <2, 1> joins the $PS$[<1, 0>]. Because the vertex 2 does not appear in $PS$[<1, 0>]={{1}}, a basic path from the initial vertex <1, 0> to the vertex <2, 1> can be obtained. So that PStemp[<2, 1>]=**CM**($PS$[<1, 0>], <2, 1>)={{1}, {2}} is constructed. After calling the **LPM**($PS$[<2, 1>], PStemp[<2, 1>]) operator, we obtain $PS$[<2, 1>]= {{1}, {2}}. In this case, the ″While (flag2)″ loop is not executed.

Analogically, $PS$[<3, 1>]={{1}, {3}}, $PS$[<4, 1>]={{1}, {4}}, $PS$[<5, 1>]={{1}, {5}}.

Iteration 2: Processing segment level $k$=2.

The vertex <2, 2> has three parent vertices <3, 1>, <4, 1>, and <5, 1>, respectively.

1. We call the **CM**($PS$[<3, 1>], <2, 2>) operator to judge whether the conflict will occur or not if the vertex <2, 2> joins the $PS$[<3, 1>]. Because the vertex 2 does not appear in $PS$[<3, 1>]={{1}, {3}}, a basic path from the initial vertex <1, 0> through the vertex <3, 1> to the vertex <2, 2> can be obtained. So that PStemp[<2, 2>]=**CM**($PS$[<3, 1>], <2, 2>)={{1}, {3}, {2}} is constructed. After calling the **LPM**($PS$[<2, 2>], PStemp[<2, 2>]) operator, we obtain $PS$[<2, 2>]={{1}, {3}, {2}}. In this case, the ″While (flag2)″ loop is not executed.

2. We call the **CM**($PS$[<4, 1>], <2, 2>) operator to judge whether the conflict will occur or not if the vertex <2, 2> joins the $PS$[<4, 1>]. Fortunately, the vertex 2 does not appear in $PS$[<4, 1>]={{1}, {4}}. Thus we obtain $PS$[<2, 2>]={{1}, {3, 4}, {2}}. In this case, the ″While (flag2)″ loop is not executed.

3. We call the **CM**(*PS*[<5, 1>], <2, 2>). The analysis is similar to case 2. Finally, we obtain *PS*[<2, 2>]={{1}, {3, 4, 5}, {2}}. In this case, the "While (flag2)" loop is not executed.

Analogously, *PS*[<3, 2>]={{1}, {2, 4, 5}, {3}}, *PS*[<4, 2>]={{1}, {2, 3, 5}, {4}}, *PS*[<5, 2>]= {{1}, {2, 3, 4}, {5}}.

Iteration 3: Processing segment level *k*=3.

The vertex <2, 3> has three parent vertices <3, 2>, <4, 2>, and <5, 2>, respectively.

1. We call the **CM**(*PS*[<3, 2>], <2, 3>) operator to judge whether the conflict will occur or not if the vertex <2, 3> joins the *PS*[<3, 2>]. We find that the vertex 2 appears in *PS*[<3, 2>]={{1}, {2, 4, 5}, {3}}, i.e., $2 \in PS_{<3,2>}[1]=\{2, 4, 5\}$. After deleting the duplicate of vertex 2 from $PS_{<3,2>}[1]$, the vertex $1 \in AF^-(<2, 1>)$ and the vertex $3 \in AF^+(<2, 1>)$ are replenished due to the reason of $1 \in AF^-(<4, 1>)$ and $3 \in AF^+(<4, 1>)$, respectively. Therefore, the "consecutive" deleting-replenishing operations on the left/right action field respectively will halt. At this time, we obtain PStemp[<3, 2>]={{1}, {4, 5}, {3}}. Since there does not exist any a singleton except for $PStemp_{<3,2>}[0]$, i.e., the leftmost end, and $PStemp_{<3,2>}[2]$, i.e., the rightmost end, the "While (flag2)" loop runs only once. Further, the step 18 in the **CM** operator runs in *n*-2 steps, while the step 20 in the **CM** operator does not execute. Then we call the **CHECK** operator. PStemp[<3, 2>]={{1}, {4, 5}, {3}} does not contain the empty set ∅. PStemp[<4, 1>]=PS[<4, 1>]$\cap_{min}$PStemp[<3, 2>]={{1}, {4}}, PStemp[<5, 1>]=PS[<5, 1>] $\cap_{min}$PStemp[<3, 2>]={{1}, {5}}, PStemp[<1, 0>]=PS[<1, 0>]$\cap_{min}$PStemp[<3, 2>]={{1}}. They all do not contain the empty set ∅. Finally, we obtain PStemp[<2, 3>]=PStemp[<3, 2>]⊗{{2}}={{1}, {4, 5}, {3}, {2}}. Then we call the **LPM**(*PS*[<2, 3>], PStemp[<2, 3>]) operator to obtain *PS*[<2, 3>]= {{1}, {4, 5}, {3}, {2}}. In this case, the "While (flag2)" loop runs only once when we consider the parent vertex <3, 2> of <2, 3>.

2. We call the **CM**(*PS*[<4, 2>], <2, 3>) operator to judge whether the conflict will occur or not if the vertex <2, 3> joins the *PS*[<4, 2>]. The process is similar to case 1. We have *PS*[<2, 3>]={{1}, {3, 4, 5}, {3, 4}, {2}}, while the "While (flag2)" loop runs only once.

3. We continue to handle the third parent vertex <5, 2> of <2, 3>. Similarly, we obtain *PS*[<2, 3>]= {{1}, {3, 4, 5}, {3, 4, 5}, {2}} and the "While (flag2)" loop runs only once.

Thus far, we have handled all parent vertices of <2, 3> and we find that the "While (flag2)" loop runs only in 3, i.e., *n*-2=5-2=3, times in total. In other words, the "While (flag2)" loop runs only once when any a parent vertex of <2, 3> is considered.

Analogously, *PS*[<3, 3>]={{1}, {2, 4, 5}, {2, 4, 5}, {3}}, *PS*[<4, 3>]={{1}, {2, 3, 5}, {2, 3, 5}, {4}}, *PS*[<5, 3>]={{1}, {2, 3, 4}, {2, 3, 4}, {5}}.

Iteration 4: Processing segment level *k*=4.

The vertex <2, 4> has three parent vertices <3, 3>, <4, 3>, and <5, 3>, respectively.

1. We call the **CM**(*PS*[<3, 3>], <2, 4>) operator to judge whether the conflict will occur or not if the vertex <2, 4> joins the *PS*[<3, 3>]. We find that the vertex 2 appears in *PS*[<3, 3>]={{1}, {2, 4, 5}, {2, 4, 5}, {3}} twice, i.e., $2 \in PS_{<3,3>}[2]=\{2, 4, 5\}$ and $2 \in PS_{<3,3>}[1]=\{2, 4, 5\}$, respectively. After deleting the duplicate of vertex 2, all the vertices in left/right action field have been replenished. That is, the vertex $4 \in AF^-(<2, 2>)$ and $5 \in AF^-(<2, 2>)$ are replenished due to the reason of $5 \in AF^-(<4, 2>)$ and $4 \in AF^-(<5, 2>)$, respectively. Similarly, the vertex $3 \in AF^+(<2, 2>)$ is replenished due to the reason of $3 \in AF^+(<4, 2>)$ or $3 \in AF^+(<5, 2>)$. The same is true for $2 \in PS_{<3,3>}[1]$. Thus, the "consecutive" deleting-replenishing operations on the left/right action field respectively will halt. At this time, we obtain PStemp[<3, 3>]={{1}, {4, 5}, {4, 5}, {3}}. Since there does not exist any a singleton except for PStemp$_{<3,3>}$[0], i.e., the leftmost end, and PStemp$_{<3,3>}$[3], i.e., the rightmost end, the ″While (flag2)″ loop runs only once. Then we call the **CHECK** operator. PStemp[<3, 3>]= {{1}, {4, 5}, {4, 5}, {3}} does not contain the empty set ∅. PStemp[<4, 2>]=*PS*[<4, 2>] ∩$_{min}$PStemp[<3, 3>]={{1}, {5}, {4}}, PStemp[<5, 2>]=*PS*[<5, 2>]∩$_{min}$PStemp[<3, 3>]={{1}, {4}, {5}}, PStemp[<4, 1>]=*PS*[<4, 1>]∩$_{min}$PStemp[<3, 3>]={{1}, {4}}, PStemp[<5, 1>]=*PS*[<5, 1>] ∩$_{min}$PStemp[<3, 3>]={{1}, {5}}, PStemp[<1, 0>]=*PS*[<1, 0>]∩$_{min}$PStemp[<3, 3>]={{1}}. They all do not contain the empty set ∅. Further, the step 18 in the **CM** operator runs in *n*-2 steps, while the step 20 in the **CM** operator does not execute. Finally, we obtain PStemp[<2, 4>]=PStemp[<3, 3>] ⊗{{2}}={{1}, {4, 5}, {4, 5}, {3}, {2}}. Then we call the **LPM**(*PS*[<2, 4>], PStemp[<2, 4>]) operator to obtain *PS*[<2, 4>]={{1}, {4, 5}, {4, 5}, {3}, {2}}. In this case, the ″While (flag2)″ loop runs only once when we consider the parent vertex <3, 3> of <2, 4>.

2. We call the **CM**(*PS*[<4, 3>], <2, 4>) operator to judge whether the conflict will occur or not if the vertex <2, 4> joins the *PS*[<4, 3>]. The process is similar to case 1. Finally, we obtain *PS*[<2, 4>] ={{1}, {3, 4, 5}, {3, 4, 5}, {3, 4}, {2}} and the ″While (flag2)″ loop runs only once.

3. We continue to handle the third parent vertex <5, 3> of <2, 4>. Similarly, we obtain *PS*[<2, 4>]= {{1}, {3, 4, 5}, {3, 4, 5}, {3, 4, 5}, {2}} and the ″While (flag2)″ loop runs only once.

Thus far, we have handled all parent vertices of <2, 4> and we find that the ″While (flag2)″ loop runs only in 3, i.e., *n*-2=5-2=3, times in total. In other words, the ″While (flag2)″ loop runs only once when any a parent vertex of <2, 4> is considered.

Analogously, *PS*[<3, 4>]={{1}, {2, 4, 5}, {2, 4, 5}, {2, 4, 5}, {3}}, *PS*[<4, 4>]={{1}, {2, 3, 5}, {2, 3, 5}, {2, 3, 5}, {4}}, *PS*[<5, 4>]={{1}, {2, 3, 4}, {2, 3, 4}, {2, 3, 4}, {5}}.

Iteration 5: Processing segment level *k*=5.

Since the vertex <1, 5> is the final vertex **D**, it can join the path set of its all parent vertices naturally. Thus we obtain *PS*[<1, 5>]={{1}, {2, 3, 4, 5}, {2, 3, 4, 5}, {2, 3, 4, 5}, {2, 3, 4, 5}, {1}}, while the ″While (flag2)″ loop does not execute.

Since the length of *PS*[<1, 5>]={{1}, {2, 3, 4, 5}, {2, 3, 4, 5}, {2, 3, 4, 5}, {2, 3, 4, 5}, {1}} is 6=*L*+1, where *L*=5=|$V(K_5)$|, there must exist a basic path from the initial vertex **S** to the final vertex **D** in the path hologram **H**, inferring that the original graph $K_5$ is Hamiltonian.

Next, we call the **FHC**(*PS*[ ], **D**) operator to find a Hamiltonian cycle.

Initially, we have PStemp=*PS*[<1, 5>]={{1}, {2, 3, 4, 5}, {2, 3, 4, 5}, {2, 3, 4, 5}, {2, 3, 4, 5}, {1}} and the path 1. The vertex <1, 5> has four parent vertices <2, 4>, <3, 4>, <4, 4> and <5, 4>. Then we select a non-redundant valid parent vertex <2, 4> of <1, 5> due to ∅∉**CHECK**(PStemp∩$_{min}$*PS*[<2, 4>]), without loss of generality, for backward searching based on the PStemp[4]={2, 3, 4, 5} and obtain the path 2-1 as well as PStemp=PStemp∩$_{min}$*PS*[<2, 4>] ={{1}∩{1}, {2, 3, 4, 5}∩{3, 4, 5}, {2, 3, 4, 5}∩{3, 4, 5}, {2, 3, 4, 5}∩{3, 4, 5}, {2, 3, 4, 5}∩{2}}={{1}, {3, 4, 5}, {3, 4, 5}, {3, 4, 5}, {2}}. Next we select a non-redundant valid parent vertex <3, 3> of <2, 4> from PStemp[3]={3, 4, 5} due to ∅∉**CHECK**(PStemp∩$_{min}$*PS*[<3, 3>]) and obtain the path 3-2-1 as well as PStemp=PStemp∩$_{min}$ *PS*[<3, 3>]={{1}∩{1}, {3, 4, 5}∩{2, 4, 5}, {3, 4, 5}∩{2, 4, 5}, {3, 4, 5}∩{3}}={{1}, {4, 5}, {4, 5}, {3}}. Iteratively, we obtain the path 4-3-2-1 as well as PStemp=PStemp∩$_{min}$*PS*[<4, 2>]={{1}, {5}, {4}}, and the path 5-4-3-2-1 as well as PStemp=PStemp∩$_{min}$*PS*[<5, 1>]={{1}, {5}}. Finally, we consider a parent vertex of <5, 1> and obtain the path 1-5-4-3-2-1. So that we find a Hamiltonian cycle 1-5-4-3-2-1 in the original graph $K_5$. Of course, there is another Hamiltonian cycle 1-5-4-2-3-1 in the original graph $K_5$. If we select the parent vertex <3, 4> of <1, 5> instead of the <2, 4> of <1, 5>, we can find the Hamiltonian cycle 1-5-4-2-3-1.

As a result, we can find any a Hamiltonian cycle by selecting a different parent vertex $v_p$ for backward searching at each step $t$ such that $\varnothing \notin \mathbf{CHECK}(PStemp \cap_{min} PS[<v_p, t>])$, and this computation of finding a Hamiltonian cycle can be accomplished in deterministic polynomial time. Further, $(n-1)!$ Hamiltonian cycles have been stored in the $PS[D]=PS[<1, n>]$ with $O(n^2)$ space.

**Example 3.** A finite connected undirected original graph $G=(V, E)$ and its (part) path hologram $H=<V_H, E_H, S, D, L>$ are shown in Figure 4(a) and 4(b), respectively.

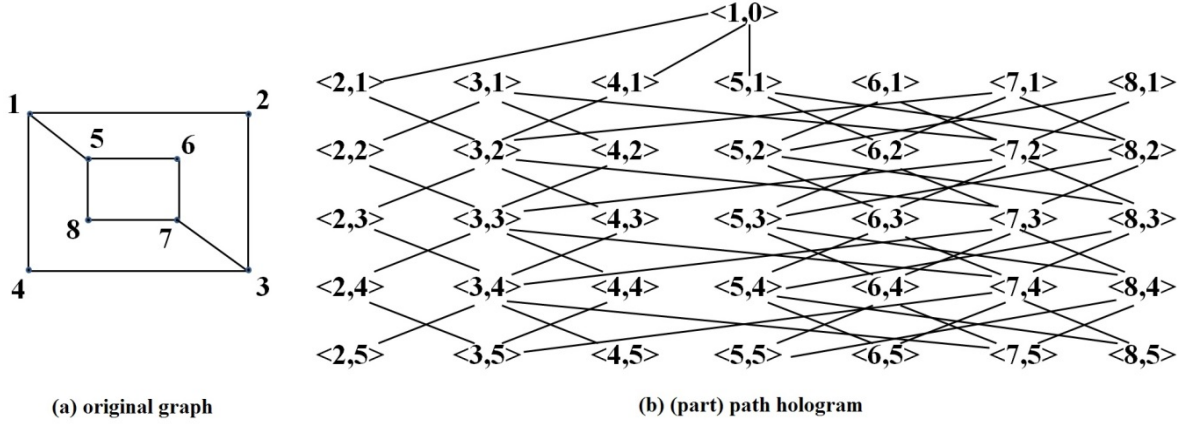

(a) original graph      (b) (part) path hologram

**Figure 4**    A finite connected undirected original graph $G$ and its (part) path hologram $H$ of **Example 3**

Select any vertex $v$ of $V(G)$, without loss of generality assume that $v=1$, to be as a starting vertex. The (part) path hologram $H=<V_H, E_H, S, D, L>$ constructed from the graph $G$ is shown in **Figure 4(b)**. In view of the size of the drawing, the vertices on the last three segment levels are omitted and the arrow is removed from the line. Here, $S=<1, 0>$ and $D=<1, 8>$. Initially, $PS[S]=PS[<1, 0>]=\{\{1\}\}$, $PS[<2, 1>]=\{\{2\}\}$, $PS[<3, 1>]= \{\{3\}\}$, $PS[<4, 1>]=\{\{4\}\}$, $PS[<5, 1>]=\{\{5\}\}$, $PS[<6, 1>]=\{\{6\}\}$, $PS[<7, 1>]=\{\{7\}\}$, $PS[<8, 1>]=\{\{8\}\}$, $PS[<2, 2>]=\{\{2\}\}$, and so on.

Iteration 1: Processing segment level $k=1$.

$PS[<2, 1>]=\{\{1\}, \{2\}\}$, $PS[<3, 1>]=\{\{3\}\}$, $PS[<4, 1>]=\{\{1\}, \{4\}\}$, $PS[<5, 1>]=\{\{1\}, \{5\}\}$, $PS[<6, 1>]=\{\{6\}\}$, $PS[<7, 1>]=\{\{7\}\}$, $PS[<8, 1>]=\{\{8\}\}$.

Iteration 2: Processing segment level $k=2$.

$PS[<2, 2>]=\{\{3\}, \{2\}\}$, $PS[<3, 2>]=\{\{1, 2, 4\}, \{3\}\}$, $PS[<4, 2>]=\{\{3\}, \{4\}\}$, $PS[<5, 2>]=\{\{6, 8\}, \{5\}\}$, $PS[<6, 2>]=\{\{1, 5\}, \{6\}\}$, $PS[<7, 2>]=\{\{3, 6, 8\}, \{7\}\}$, $PS[<8, 2>]=\{\{1, 5\}, \{8\}\}$.

Iteration 3: Processing segment level $k=3$.

$PS[<2, 3>]=\{\{1, 4\}, \{3\}, \{2\}\}$, $PS[<3, 3>]=\{\{6, 8\}, \{7\}, \{3\}\}$, $PS[<4, 3>]=\{\{1, 2\}, \{3\}, \{4\}\}$, $PS[<5, 3>]=\{\{5\}\}$, $PS[<6, 3>]=\{\{3, 8\}, \{5, 7\}, \{6\}\}$, $PS[<7, 3>]=\{\{1, 2, 4, 5\}, \{3, 6, 8\}, \{7\}\}$, $PS[<8, 3>]=\{\{3, 6\}, \{5, 7\}, \{8\}\}$.

Iteration 4: Processing segment level $k=4$.

$PS[<2, 4>]=\{\{6, 8\}, \{7\}, \{3\}, \{2\}\}$, $PS[<3, 4>]=\{\{1, 5\}, \{6, 8\}, \{7\}, \{3\}\}$, $PS[<4, 4>]=\{\{6, 8\}, \{7\}, \{3\}, \{4\}\}$, $PS[<5, 4>]=\{\{3, 6, 8\}, \{7\}, \{6, 8\}, \{5\}\}$, $PS[<6, 4>]=\{\{1, 2, 4, 5\}, \{3, 8\}, \{7\}, \{6\}\}$, $PS[<7, 4>]=\{\{6, 8\}, \{5\}, \{6, 8\}, \{7\}\}$, $PS[<8, 4>]=\{\{1, 2, 4, 5\}, \{3, 6\}, \{7\}, \{8\}\}$.

Iteration 5: Processing segment level $k=5$.

$PS[<2, 5>]=\{\{1, 5\}, \{6, 8\}, \{7\}, \{3\}, \{2\}\}$, $PS[<3, 5>]=\{\{6, 8\}, \{5\}, \{6, 8\}, \{7\}, \{3\}\}$, $PS[<4, 5>]=\{\{1, 5\}, \{6, 8\}, \{7\}, \{3\}, \{4\}\}$, $PS[<5, 5>]=\{\{1, 2, 4\}, \{3\}, \{7\}, \{6, 8\}, \{5\}\}$, $PS[<6, 5>]=\{\{3\}, \{7\}, \{8\}, \{5\}, \{6\}\}$, $PS[<7, 5>]=\{\{7\}\}$, $PS[<8, 5>]=\{\{3\}, \{7\}, \{6\}, \{5\}, \{8\}\}$.

Iteration 6: Processing segment level *k*=6.

*PS*[<2, 6>]={{6, 8}, {5}, {6, 8}, {7}, {3}, {2}}, *PS*[<3, 6>]={{7}, {3}}, *PS*[<4, 6>]={{6, 8}, {5}, {6, 8}, {7}, {3}, {4}}, *PS*[<5, 6>]={{5}}, *PS*[<6, 6>]={{1}, {2, 4}, {3}, {7}, {8}, {5}, {6}}, *PS*[<7, 6>]={{7}}, *PS*[<8, 6>]={{1}, {2, 4}, {3}, {7}, {6}, {5}, {8}}.

Iteration 7: Processing segment level *k*=7.

*PS*[<2, 7>]={{7}, {3}, {2}}, *PS*[<3, 7>]={{7}, {3}}, *PS*[<4, 7>]={{7}, {3}, {4}}, *PS*[<5, 7>]={{5}}, *PS*[<6, 7>]={{5, 7}, {6}}, *PS*[<7, 7>]={{7}}, *PS*[<8, 7>]={{5, 7}, {8}}.

Iteration 8: Processing segment level *k*=8.

The vertex <1, 8> has three parent vertices <2, 7>, <4, 7> and <5, 7>. Since the <1, 8> is the final vertex ***D***, we have *PS*[<1, 8>]={{7}, {3}, {2, 4}, {1}}. Since the length of *PS*[<1, 8>]≠9, the original graph ***G*** is not Hamiltonian.

**Example 4.** A finite connected undirected original graph ***G***=(***V***, ***E***) and its (part) path hologram ***H***=<$V_H$, $E_H$, ***S***, ***D***, ***L***> are shown in **Figure 5** and **6**, respectively. The ***G*** is a maximal nonHamiltonian graph with |***V***|=11 & |***E***|=40. In view of the size of the drawing, the arrow is removed from the line.

In the clique $K_5$, the 5 vertices of degree *n*-1 are adjacent to all others. In the independent set $\overline{K_5}$, the 5 vertices of degree 5 are adjacent to all vertices of $K_5$. In the clique $K_{n-2\times 5}=K_1$, the *n*-10=1 vertex of degree 5 is adjacent to all vertices of $K_5$. The ***G*** is ($\overline{K_5}+K_{n-2\times 5}$)∨$K_5$.

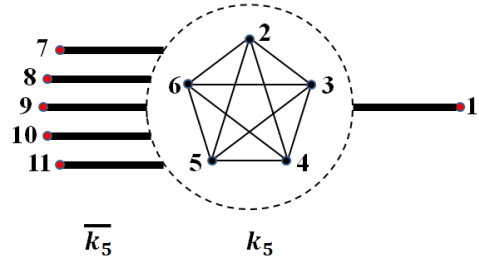

**Figure 5**  The maximal nonHamiltonian graph ***G*** of order *n*=11 and size *e*=40

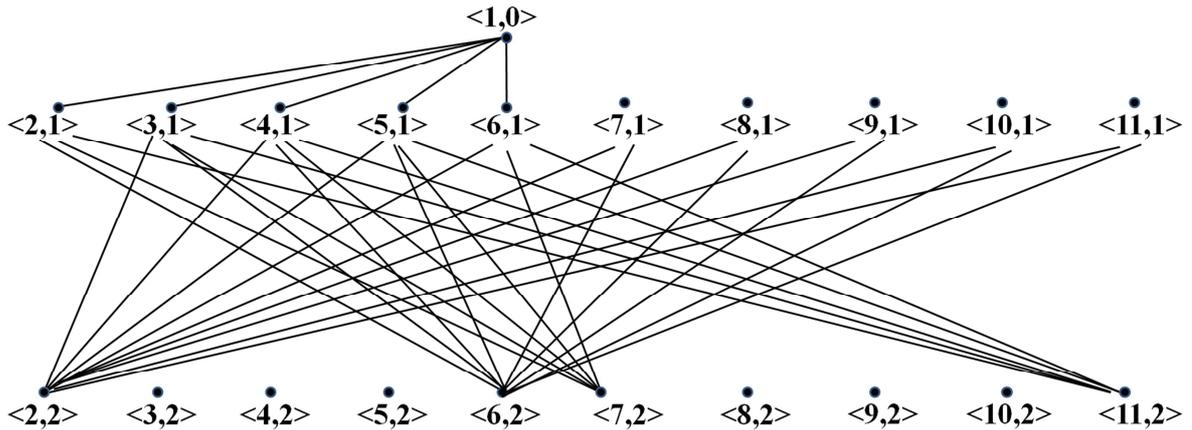

**Figure 6**  Its (part) path hologram of the maximal nonHamiltonian graph ***G*** of order *n*=11 and size *e*=40

Select any vertex *v* of *V*(***G***), without loss of generality assume that *v*=1, to be as a starting vertex. Thus, ***S***=<1, 0> and ***D***=<1, 11>. Initially, *PS*[***S***]=*PS*[<1, 0>]={{1}}, *PS*[<2, 1>]={{2}}, *PS*[<3, 1>]={{3}}, *PS*[<4, 1>]={{4}}, *PS*[<5, 1>]={{5}}, *PS*[<6, 1>]={{6}}, *PS*[<7, 1>]={{7}}, *PS*[<8, 1>]={{8}}, *PS*[<9, 1>]={{9}}, *PS*[<10, 1>]={{10}}, *PS*[<11, 1>]={{11}}, *PS*[<2, 2>]={{2}}, and so on.

Iteration 1: Processing segment level *k*=1.

*PS*[<2, 1>]={{1}, {2}}, *PS*[<3, 1>]={{1}, {3}}, *PS*[<4, 1>]={{1}, {4}}, *PS*[<5, 1>]={{1}, {5}}, *PS*[<6, 1>]={{1}, {6}}, *PS*[<7, 1>]={{7}}, *PS*[<8, 1>]={{8}}, *PS*[<9, 1>]={{9}}, *PS*[<10, 1>]={{10}}, *PS*[<11, 1>]={{11}}.

Iteration 2: Processing segment level *k*=2.

*PS*[<2, 2>]={{1}, {3,4,5,6}, {2}}, *PS*[<3, 2>]={{1}, {2,4,5,6}, {3}}, *PS*[<4, 2>]={{1}, {2,3,5,6}, {4}}, *PS*[<5, 2>]={{1}, {2,3,4,6}, {5}}, *PS*[<6, 2>]={{1}, {2,3,4,5}, {6}}, *PS*[<7, 2>]={{1}, {2,3,4,5,6}, {7}}, *PS*[<8, 2>]={{1}, {2,3,4,5,6}, {8}}, *PS*[<9, 2>]={{1}, {2,3,4,5,6}, {9}}, *PS*[<10, 2>]={{1}, {2,3,4,5,6}, {10}}, *PS*[<11, 2>]={{1}, {2,3,4,5,6}, {11}}.

Iteration 3: Processing segment level *k*=3.

*PS*[<2, 3>]={{1}, {3,4,5,6}, {3,4,5,6,7,8,9,10,11}, {2}}, *PS*[<3, 3>]={{1}, {2,4,5,6}, {2,4,5,6,7,8,9,10,11}, {3}}, *PS*[<4, 3>]={{1}, {2,3,5,6}, {2,3,5,6,7,8,9,10,11}, {4}}, *PS*[<5, 3>]={{1}, {2,3,4,6}, {2,3,4,6,7,8,9,10,11}, {5}}, *PS*[<6, 3>]= {{1}, {2,3,4,5}, {2,3,4,5,7,8,9,10,11}, {6}}. *PS*[<7, 3>]= {{1}, {2,3,4,5,6}, {2,3,4,5,6}, {7}}, *PS*[<8, 3>]={{1}, {2,3,4,5,6}, {2,3,4,5,6}, {8}}, *PS*[<9, 3>]={{1}, {2,3,4,5,6}, {2,3,4,5,6}, {9}}, *PS*[<10, 3>]={{1}, {2,3,4,5,6}, {2,3,4,5,6}, {10}}, *PS*[<11, 3>]={{1}, {2,3,4,5,6}, {2,3,4,5,6}, {11}}.

Iteration 4: Processing segment level *k*=4.

*PS*[<2, 4>]={{1}, {3,4,5,6}, {3,4,5,6,7,8,9,10,11}, {3,4,5,6,7,8,9,10,11}, {2}}, *PS*[<3, 4>]={{1}, {2,4,5,6}, {2,4,5,6,7,8,9,10,11}, {2,4,5,6,7,8,9,10,11}, {3}}, *PS*[<4, 4>]={{1}, {2,3,5,6}, {2,3,5,6,7,8,9,10,11}, {2,3,5,6,7,8,9,10,11}, {4}}, *PS*[<5, 4>]={{1}, {2,3,4,6}, {2,3,4,6,7,8,9,10,11}, {2,3,4,6,7,8,9,10,11}, {5}}, *PS*[<6, 4>]={{1}, {2,3,4,5}, {2,3,4,5,7,8,9,10,11}, {2,3,4,5,7,8,9,10,11}, {6}}, *PS*[<7, 4>]={{1}, {2,3,4,5,6}, {2,3,4,5,6,8,9,10,11}, {2,3,4,5,6}, {7}}, *PS*[<8, 4>]={{1}, {2,3,4,5,6}, {2,3,4,5,6,7,9,10,11}, {2,3,4,5,6}, {8}}, *PS*[<9, 4>]= {{1}, {2,3,4,5,6}, {2,3,4,5,6,7,8,10,11}, {2,3,4,5,6}, {9}}, *PS*[<10, 4>]={{1}, {2,3,4,5,6}, {2,3,4,5,6,7,8,9,11}, {2,3,4,5,6}, {10}}, *PS*[<11, 4>]={{1}, {2,3,4,5,6}, {2,3,4,5,6,7,8,9,10}, {2,3,4,5,6}, {11}}.

Iteration 5: Processing segment level *k*=5.

*PS*[<2, 5>]={{1}, {3,4,5,6}, {3,4,5,6,7,8,9,10,11}, {3,4,5,6,7,8,9,10,11}, {3,4,5,6,7,8,9,10,11}, {2}}, *PS*[<3, 5>]={{1}, {2,4,5,6}, {2,4,5,6,7,8,9,10,11}, {2,4,5,6,7,8,9,10,11}, {2,4,5,6,7, 8,9,10,11}, {3}}, *PS*[<4, 5>]={{1}, {2,3,5,6}, {2,3,5,6,7,8,9,10,11}, {2,3,5,6,7,8,9,10,11}, {2,3,5,6,7, 8,9,10,11}, {4}}, *PS*[<5, 5>]={{1}, {2,3,4,6}, {2,3,4,6,7,8,9,10,11}, {2,3,4,6,7,8,9,10,11}, {2,3,4,6,7,8,9,10,11}, {5}}, *PS*[<6, 5>]={{1}, {2,3,4,5}, {2,3,4,5,7,8,9,10,11}, {2,3,4,5,7,8,9,10,11}, {2,3,4,5,7,8,9,10,11}, {6}}, *PS*[<7, 5>]={{1}, {2,3,4,5,6}, {2,3,4,5,6,8,9,10,11}, {2,3,4,5,6,8,9,10,11}, {2,3,4,5,6}, {7}}, *PS*[<8, 5>]={{1}, {2,3,4,5,6}, {2,3,4,5,6,7,9,10,11}, {2,3,4,5,6,7,9,10,11}, {2,3,4,5, 6}, {8}}, *PS*[<9, 5>]={{1}, {2,3,4,5,6}, {2,3,4,5,6,7,8,10,11}, {2,3,4,5,6,7,8,10,11}, {2,3,4,5,6}, {9}}, *PS*[<10, 5>]= {{1}, {2,3,4,5,6}, {2,3,4,5,6,7,8,9,11}, {2,3,4,5,6,7,8,9,11}, {2,3,4,5,6}, {10}}, *PS*[<11, 5>]={{1}, {2,3,4,5,6}, {2,3,4,5,6,7,8,9,10}, {2,3,4,5,6,7,8,9,10}, {2,3,4,5,6}, {11}}.

Iteration 6: Processing segment level *k*=6.

*PS*[<2, 6>]={{1}, {3,4,5,6}, {3,4,5,6,7,8,9,10,11}, {3,4,5,6,7,8,9,10,11}, {3,4,5,6,7,8,9,10,11}, {3, 4,5,6,7,8,9,10,11}, {2}}, *PS*[<3, 6>]={{1}, {2,4,5,6}, {2,4,5,6,7,8,9,10,11}, {2,4,5,6,7,8,9,10,11}, {2, 4,5,6,7,8,9,10,11}, {2,4,5,6,7,8,9,10,11}, {3}}, *PS*[<4, 6>]={{1}, {2,3,5,6}, {2,3,5,6,7,8,9,10,11}, {2, 3,5,6,7,8, 9,10,11}, {2,3,5,6,7,8,9,10,11}, {2,3,5,6,7,8,9,10,11}, {4}}, *PS*[<5, 6>]={{1}, {2,3,4,6}, {2, 3,4,6,7,8, 9,10,11}, {2,3,4,6,7,8,9,10,11}, {2,3,4,6,7,8,9, 10,11}, {2,3,4,6,7,8,9,10,11}, {5}}, *PS*[<6, 6>]={{1}, {2,3,4,5}, {2,3,4,5,7,8,9,10,11}, {2,3,4,5,7,8,9,10,11}, {2,3,4,5,7,8,9,10,11}, {2,3,4,5,7,8, 9,10,11}, {6}}, *PS*[<7, 6>]={{1}, {2,3,4,5,6}, {2,3,4,5,6,8,9,10,11}, {2,3,4,5,6,8,9,10,11}, {2,3,4,5,6,

8,9,10,11},{2,3,4,5,6}, {7}}, *PS*[<8, 6>]={{1}, {2,3,4,5,6}, {2,3,4,5,6,7,9,10,11}, {2,3,4,5,6,7,9,10, 11}, {2,3,4,5,6,7,9,10,11}, {2,3,4,5,6}, {8}}, *PS*[<9, 6>]={{1}, {2,3,4,5,6}, {2,3,4,5,6,7,8,10,11}, {2, 3,4,5,6,7,8,10, 11}, {2,3,4,5,6,7,8,10,11},{2,3,4,5,6}, {9}}, *PS*[<10, 6>]={{1}, {2,3,4,5,6}, {2,3,4,5,6, 7,8,9,11}, {2,3,4,5,6,7,8,9,11}, {2,3,4,5,6,7,8,9,11},{2,3,4,5,6}, {10}}, *PS*[<11, 6>]={{1}, {2,3,4,5, 6}, {2,3,4,5,6,7,8,9,10}, {2,3,4,5,6,7,8,9,10}, {2,3,4,5,6,7,8,9,10},{2,3,4,5,6}, {11}}.

Iteration 7: Processing segment level $k=7$.

*PS*[<2, 7>]={{1}, {3,4,5,6}, {3,4,5,6,7,8,9,10,11}, {3,4,5,6,7,8,9,10,11}, {3,4,5,6,7,8,9,10,11}, {3, 4,5,6,7,8,9,10,11}, {3,4,5,6,7,8,9,10,11}, {2}}, *PS*[<3, 7>]={{1}, {2,4,5,6}, {2,4,5,6,7,8,9,10,11}, {2, 4,5,6,7,8,9,10,11}, {2,4,5,6,7,8,9,10,11}, {2,4,5,6,7,8,9,10,11}, {2,4,5,6,7,8,9,10,11}, {3}}, *PS*[<4, 7>]={{1}, {2,3,5,6}, {2,3,5,6,7,8,9,10,11}, {2,3,5,6,7,8,9,10,11}, {2,3,5,6,7,8,9,10,11}, {2,3,5,6,7,8, 9,10,11}, {2,3,5,6,7,8,9,10,11}, {4}}, *PS*[<5, 7>]={{1}, {2,3,4,6}, {2,3,4,6,7,8,9,10,11}, {2,3,4,6,7,8, 9,10,11}, {2,3,4,6,7,8,9,10,11}, {2,3,4,6,7,8,9,10,11}, {2,3,4,6,7,8,9,10,11}, {5}}, *PS*[<6, 7>]={{1}, {2,3,4,5}, {2,3,4,5,7,8,9,10,11}, {2,3,4,5,7,8,9,10,11}, {2,3,4,5,7,8,9,10,11}, {2,3,4,5,7,8,9,10,11}, {2, 3,4,5,7,8,9,10,11}, {6}}, *PS*[<7, 7>]={{1}, {2,3,4,5,6}, {2,3,4,5,6,8,9,10,11}, {2,3,4,5,6,8,9,10,11}, {2,3,4,5,6,8,9,10,11}, {2,3,4,5,6,8,9,10,11}, {2,3,4,5,6}, {7}}, *PS*[<8, 7>]={{1}, {2,3,4,5,6}, {2,3,4, 5,6,7,9,10,11}, {2,3,4,5,6,7,9,10,11}, {2,3,4,5,6, 7,9,10,11}, {2,3,4,5,6,7,9,10,11}, {2,3,4,5,6}, {8}}, *PS*[<9, 7>]={{1}, {2,3,4,5,6}, {2,3,4,5,6,7,8,10,11}, {2,3,4,5,6,7,8,10,11}, {2,3,4,5,6,7,8,10,11}, {2, 3,4,5,6,7,8,10,11}, {2,3,4,5,6}, {9}}, *PS*[<10, 7>]={{1}, {2,3,4,5,6}, {2,3,4,5,6,7,8,9,11}, {2,3,4,5,6, 7,8,9,11}, {2,3,4,5,6,7,8,9,11}, {2,3,4,5,6,7,8,9,11}, {2,3,4,5,6}, {10}}, *PS*[<11, 7>]={{1}, {2,3,4,5, 6}, {2,3,4,5,6,7,8,9,10}, {2,3,4,5,6,7,8,9,10}, {2,3,4,5,6,7,8,9,10}, {2,3,4,5,6,7,8,9,10}, {2,3,4,5,6}, {11}}.

Iteration 8: Processing segment level $k=8$.

*PS*[<2, 8>]={{1}, {3,4,5,6}, {3,4,5,6,7,8,9,10,11}, {3,4,5,6,7,8,9,10,11}, {3,4,5,6,7,8,9,10,11}, {3, 4,5,6,7,8,9,10,11}, {3,4,5,6,7,8,9,10,11}, {3,4,5,6,7,8,9,10,11}, {2}}, *PS*[<3, 8>]={{1}, {2,4,5,6}, {2, 4,5,6,7,8,9,10,11}, {2,4,5,6,7,8,9,10,11}, {2,4,5,6,7,8,9,10,11}, {2,4,5,6,7,8,9,10,11}, {2,4,5,6,7,8,9, 10,11}, {2,4,5,6,7,8,9,10,11}, {3}}, *PS*[<4, 8>]={{1}, {2,3,5,6}, {2,3,5,6,7,8,9,10,11}, {2,3,5,6,7,8,9, 10,11}, {2,3,5,6,7,8,9,10,11}, {2,3,5,6,7,8,9,10,11}, {2,3,5,6,7,8,9,10,11}, {2,3,5,6,7,8,9,10,11}, {4}}, *PS*[<5, 8>]={{1}, {2,3,4,6}, {2,3,4,6,7,8,9,10,11}, {2,3,4,6,7,8,9,10,11}, {2,3,4,6,7,8,9,10,11}, {2,3, 4,6,7,8,9,10,11}, {2,3,4,6,7,8,9,10,11}, {2,3,4,6,7,8,9,10,11}, {5}}, *PS*[<6, 8>]={{1}, {2,3,4,5}, {2,3, 4,5,7,8,9,10,11}, {2,3,4,5,7,8,9,10,11}, {2,3,4,5,7,8,9,10,11}, {2,3,4,5,7,8,9,10,11}, {2,3,4,5,7,8,9,10, 11}, {2,3,4,5,7,8,9,10,11}, {6}}, *PS*[<7, 8>]={{1}, {2,3,4,5,6}, {2,3,4,5,6,8,9,10,11}, {2,3,4,5,6,8,9, 10,11}, {2,3,4,5,6,8,9,10,11}, {2,3,4,5,6,8,9,10,11}, {2,3,4,5,6,8,9,10,11}, {2,3,4,5,6}, {7}}, *PS*[<8, 8>]={{1}, {2,3,4,5,6}, {2,3,4,5,6,7,9,10,11}, {2,3,4,5,6,7,9,10,11}, {2,3,4,5,6,7,9,10,11}, {2,3,4,5,6, 7,9,10,11}, {2,3,4,5,6,7,9,10,11}, {2,3,4,5,6}, {8}}, *PS*[<9, 8>]={{1}, {2,3,4, 5,6}, {2,3,4,5,6,7,8,10, 11}, {2,3,4,5,6,7,8,10,11}, {2,3,4,5,6,7,8,10,11}, {2,3,4,5,6,7,8,10,11}, {2,3,4,5,6,7,8,10,11}, {2,3,4, 5,6}, {9}}, *PS*[<10, 8>]={{1}, {2,3,4,5,6}, {2,3,4,5,6,7,8,9,11}, {2,3,4,5,6,7,8,9,11}, {2,3,4,5,6,7,8, 9,11}, {2,3,4,5,6,7,8, 9,11}, {2,3,4,5,6,7,8, 9,11}, {2,3,4,5,6}, {10}}, *PS*[<11, 8>]={{1}, {2,3,4,5,6}, {2,3,4,5,6,7,8,9,10}, {2,3,4,5,6,7,8,9,10}, {2,3,4,5,6,7,8,9,10}, {2,3,4,5,6,7,8,9,10}, {2,3,4,5,6,7,8,9, 10}, {2,3,4,5,6}, {11}}.

Iteration 9: Processing segment level $k=9$.

*PS*[<2, 9>]={{1}, {3,4,5,6}, {3,4,5,6,7,8,9,10,11}, {3,4,5,6,7,8,9,10,11}, {3,4,5,6,7,8,9,10,11}, {3, 4,5,6,7,8,9,10,11}, {3,4,5,6,7,8,9,10,11}, {3,4,5,6,7,8,9,10,11}, {3,4,5,6,7,8,9,10,11}, {2}}={{1}, $K_5 \setminus$ {2}, $(K_5 \cup \overline{K_5}) \setminus$ {2}, $(K_5 \cup \overline{K_5}) \setminus$ {2}, $(K_5 \cup \overline{K_5}) \setminus$ {2}, $(K_5 \cup \overline{K_5}) \setminus$ {2}, $(K_5 \cup \overline{K_5}) \setminus$ {2}, $(K_5 \cup \overline{K_5}) \setminus$ {2}, $(K_5 \cup \overline{K_5}) \setminus$ {2}, {2}}, *PS*[<3, 9>]={{1}, {2,4,5,6}, {2,4,5,6,7,8,9,10,11}, {2,4,5,6,7,8,9,10,11}, {2,4,

5,6,7,8,9,10,11}, {2,4,5,6,7,8,9,10,11}, {2,4,5,6,7,8,9,10,11}, {2,4,5,6,7,8,9,10,11}, {2,4,5,6,7,8,9,10,11}, {3}}, *PS*[<4, 9>]={{1}, {2,3,5,6}, {2,3,5,6,7,8,9,10,11}, {2,3,5,6,7,8,9,10,11}, {2,3,5,6,7,8,9,10,11}, {2,3,5,6,7,8,9,10,11}, {2,3,5,6,7,8,9,10,11}, {2,3,5,6,7,8,9,10,11}, {2,3,5,6,7,8,9,10,11}, {4}}, *PS*[<5, 9>]={{1}, {2,3,4,6}, {2,3,4,6,7,8,9,10,11}, {2,3,4,6,7,8,9,10,11}, {2,3,4,6,7,8,9,10,11}, {2,3,4,6,7,8,9,10,11}, {2,3,4,6,7,8,9,10,11}, {2,3,4,6,7,8,9,10,11}, {5}}, *PS*[<6, 9>]={{1}, {2,3,4,5}, {2,3,4,5,7,8,9,10,11}, {2,3,4,5,7,8,9,10,11}, {2,3,4,5,7,8,9,10,11}, {2,3,4,5,7,8,9,10,11}, {2,3,4,5,7,8,9,10,11}, {2,3,4,5,7,8,9,10,11}, {6}}, *PS*[<7, 9>]={{1}, {2,3,4,5,6}, {2,3,4,5,6,8,9,10,11}, {2,3,4,5,6,8,9,10,11}, {2,3,4,5,6,8,9,10,11}, {2,3,4,5,6,8,9,10,11}, {2,3,4,5,6,8,9,10,11}, {2,3,4,5,6}, {7}}, *PS*[<8, 9>]={{1}, {2,3,4,5,6}, {2,3,4,5,6,7,9,10,11}, {2,3,4,5,6,7,9,10,11}, {2,3,4,5,6,7,9,10,11}, {2,3,4,5,6,7,9,10,11}, {2,3,4,5,6,7,9,10,11}, {2,3,4,5,6}, {8}}, *PS*[<9, 9>]={{1}, {2,3,4,5,6}, {2,3,4,5,6,7,8,10,11}, {2,3,4,5,6,7,8,10,11}, {2,3,4,5,6,7,8,10,11}, {2,3,4,5,6,7,8,10,11}, {2,3,4,5,6,7,8,10,11}, {2,3,4,5,6}, {9}}, *PS*[<10, 9>]={{1}, {2,3,4,5,6}, {2,3,4,5,6,7,8,9,11}, {2,3,4,5,6,7,8,9,11}, {2,3,4,5,6,7,8,9,11}, {2,3,4,5,6,7,8,9,11}, {2,3,4,5,6,7,8,9,11}, {2,3,4,5,6}, {10}}, *PS*[<11, 9>]={{1}, {2,3,4,5,6}, {2,3,4,5,6,7,8,9,10}, {2,3,4,5,6,7,8,9,10}, {2,3,4,5,6,7,8,9,10}, {2,3,4,5,6,7,8,9,10}, {2,3,4,5,6,7,8,9,10}, {2,3,4,5,6}, {11}}.

Iteration 10: Processing segment level *k*=10.

The vertex <2, 10> has nine parent vertices <3, 9>, <4, 9>, <5, 9>, <6, 9>, <7, 9>, <8, 9>, <9, 9>, <10, 9> and <11, 9>, respectively.

When 2 joins *PS*[<3, 9>], we obtain PStemp[<3, 9>]={{1}, {4,5,6}, {4,5,6,7,8,9,10,11}, {4,5,6,7,8,9,10,11}, {4,5,6,7,8,9,10,11}, {4,5,6,7,8,9,10,11}, {4,5,6,7,8,9,10,11}, {4,5,6,7,8,9,10,11}, {3}}. Without doubt, no matter which vertex of $PStemp_{<3,9>}[1]$={4,5,6} we choose, we cannot get the basic path from <1, 0> to <3, 9>, although *PS*[<3, 9>] contains basic paths from <1, 0> to <3, 9>. This is because PStemp[<3, 9>] is constrained by the choice of 2. Now we prove that we can obtain PStemp[<3, 9>]={∅} by calling the **CHECK** operator.

PStemp[<4, 8>]=*PS*[<4, 8>]$\cap_{min}$PStemp[<3, 9>]={{1}, {5,6}, {5,6,7,8,9,10,11}, {5,6,7,8,9,10,11}, {5,6,7,8,9,10,11}, {5,6,7,8,9,10,11}, {5,6,7,8,9,10,11}, {5,6,7,8,9,10,11}, {4}}, PStemp[<5, 7>] =*PS*[<5, 7>]$\cap_{min}$PStemp[<4, 8>]={{1}, {6}, {6,7,8,9,10,11}, {6,7,8,9,10,11}, {6,7,8,9,10,11}, {6,7,8,9,10,11}, {6,7,8,9,10,11}, {5}}. After deleting the duplicates of singleton {6}, we get PStemp[<5, 7>]={∅}. So is PStemp[<6, 7>].

PStemp[<7, 7>]=*PS*[<7, 7>]$\cap_{min}$PStemp[<4, 8>]={{1}, {5,6}, {5,6,8,9,10,11}, {5,6,8,9,10,11}, {5,6,8,9,10,11}, {5,6,8,9,10,11}, {5,6}, {7}}. Then we call **CHECK1** to detect the PStemp[<7, 7>]. PStemp[<1, 0>]=*PS*[<1, 0>] $\cap_{min}$ PStemp[<4, 7>]={{1}}. PStemp[<5, 1>]=*PS*[<5, 1>] $\cap_{min}$ PStemp[<4, 7>]={{1}, {5}}. PStemp[<6, 1>]=*PS*[<6, 1>] $\cap_{min}$ PStemp[<4, 7>]={{1}, {6}}. PStemp[<5, 2>]=*PS*[<5, 2>]$\cap_{min}$PStemp[<4, 7>]={{1}, {6}, {5}}. PStemp[<6, 2>]=*PS*[<6, 2>] $\cap_{min}$PStemp[<4, 7>]={{1}, {5}, {6}}. PStemp[<8, 2>]=*PS*[<8, 2>]$\cap_{min}$PStemp[<4, 7>]={{1}, {5,6}, {8}}. PStemp[<9, 2>]=*PS*[<9, 2>]$\cap_{min}$PStemp[<4, 7>]={{1}, {5, 6}, {9}}. PStemp[<10, 2>]= *PS*[<10, 2>]$\cap_{min}$ PStemp[<4, 7>]={{1}, {5, 6}, {10}}. PStemp[<11, 2>]=*PS*[<11, 2>]$\cap_{min}$ PStemp[<4, 7>]={{1}, {5, 6}, {11}}. PStemp[<5, 3>]=*PS*[<5, 3>]$\cap_{min}$PStemp[<4, 7>]={{1}, {6}, {8,9,10,11}, {5}}. PStemp[<6, 3>]=*PS*[<6, 3>]$\cap_{min}$PStemp[<4, 7>]={{1}, {5}, {8,9,10,11}, {6}}. PStemp[<8, 3>]=*PS*[<8, 3>]$\cap_{min}$PStemp[<4, 7>]={{1}, {5,6}, {5,6}, {8}}. PStemp[<9, 3>]= *PS*[<9, 3>]$\cap_{min}$PStemp[<4, 7>]={{1}, {5,6}, {5,6}, {9}}. PStemp[<10, 3>]=*PS*[<10, 3>]$\cap_{min}$ PStemp[<4, 7>]={{1}, {5,6}, {5,6}, {10}}. PStemp[<11, 3>]=*PS*[<11, 3>]$\cap_{min}$PStemp[<4, 7>]= {{1}, {5,6}, {5,6}, {11}}. PStemp[<5, 4>]=*PS*[<5, 4>]$\cap_{min}$PStemp[<4, 7>]={{1}, {6}, {6,8,9,10, 11}, {6,8,9,10,11}, {5}}, resulting in PStemp[<5, 4>]={∅}. So is PStemp[<6, 4>]. PStemp[<8, 4>] =[(PStemp[<5, 3>]∪$_{max}$PStemp[<6, 3>])⊗{{8}}]∩$_{min}$PStemp[<8, 4>]={1}, {5,6}, {9,10,11}, {5, 6}, {8}}. PStemp[<9, 4>]={1}, {5,6}, {8,10,11}, {5,6}, {9}}. PStemp[<10, 4>]={1}, {5,6}, {8,9, 11}, {5,6}, {10}}. PStemp[<11, 4>]={1}, {5,6}, {8,9,10}, {5,6}, {11}}. PStemp[<5, 5>]=$PS$[<5, 5>] ∩$_{min}$PStemp[<4, 7>]={{1}, {6}, {6,8,9,10,11}, {6,8,9,10,11}, {6,8,9,10,11}, {5}}, resulting in PStemp[<5, 5>]={∅}. So is PStemp[<6, 5>]. PStemp[<8, 5>]=(PStemp[<5, 4>]∪$_{max}$PStemp[<6, 4>])⊗{{8}}={∅}. So are PStemp[<9, 5>], PStemp[<10, 5>], PStemp[<11, 5>]. Since PStemp[<p, 5>]={∅} where 5≤p≤11, we can infer PStemp[<7, 7>]={∅}. So are PStemp[<8, 7>], PStemp[<9, 7>], PStemp[<10, 7>], PStemp[<11, 7>].

Since PStemp[<4, 8>]={{1}, {5,6}, {5,6,7,8,9,10,11}, {5,6,7,8,9,10, 11}, {5,6,7,8,9,10,11}, {5,6,7, 8,9,10,11}, {5,6,7,8,9,10,11}, {5,6,7,8,9,10,11}, {4}}, now PStemp[<p, 7>]={∅} where 5≤p≤11, we infer PStemp[<4, 8>]={∅}. So are PStemp[<5, 8>], PStemp[<6, 8>], PStemp[<7, 8>], PStemp[<8, 8>], PStemp[<9, 8>], PStemp[<10, 8>], PStemp[<11, 8>].

Since PStemp[<3, 9>]={{1}, {4,5,6}, {4,5,6,7,8,9,10,11}, {4,5,6, 7,8,9,10,11}, {4,5,6,7,8,9,10,11}, {4,5,6,7,8,9,10,11}, {4,5,6,7,8,9,10,11}, {4,5,6,7,8,9,10,11}, {4,5,6,7,8,9,10,11}, {3}}, now PStemp[<p, 8>]={∅} where 4≤p≤11, we infer PStemp[<3, 9>]={∅}. So are PStemp[<4, 9>], PStemp[<5, 9>], PStemp[<6, 9>], PStemp[<7, 9>], PStemp[<8, 9>], PStemp[<9, 9>], PStemp[<10, 9>], PStemp[<11, 9>]. Thus we obtain $PS$[<2, 10>]={{2}}.

Analogously, $PS$[<3, 10>]={{3}}, $PS$[<4, 10>]={{4}}, $PS$[<5, 10>]={{5}}, $PS$[<6, 10>]={{6}}, $PS$[<7, 10>]={{1}, {2,3,4,5,6}, {2,3,4,5,6,8,9,10,11}, {2,3,4,5,6,8,9,10,11}, {2,3,4,5,6,8,9,10,11}, {2, 3,4,5,6,8,9,10,11}, {2,3,4,5,6,8,9,10,11}, {2,3,4,5,6,8,9,10,11}, {2,3,4,5,6,8,9,10,11}, {2,3,4,5,6}, {7}}, $PS$[<8, 10>]={{1}, $K_5$, ($K_5 ∪ \overline{K_5}$)\{8}, ($K_5 ∪ \overline{K_5}$)\{8}, ($K_5 ∪ \overline{K_5}$)\{8}, ($K_5 ∪ \overline{K_5}$)\{8}, ($K_5 ∪ \overline{K_5}$)\{8}, ($K_5 ∪ \overline{K_5}$)\{8}, ($K_5 ∪ \overline{K_5}$)\{8}, $K_5$, {8}}, $PS$[<9, 10>]={{1}, $K_5$, ($K_5 ∪ \overline{K_5}$)\{9}, ($K_5 ∪ \overline{K_5}$)\{9}, ($K_5 ∪ \overline{K_5}$)\{9}, ($K_5 ∪ \overline{K_5}$)\{9}, ($K_5 ∪ \overline{K_5}$)\{9}, ($K_5 ∪ \overline{K_5}$)\{9}, ($K_5 ∪ \overline{K_5}$)\{9}, $K_5$, {9}}, $PS$[<10, 10>]={{1}, $K_5$, ($K_5 ∪ \overline{K_5}$)\{10}, ($K_5 ∪ \overline{K_5}$)\{10}, ($K_5 ∪ \overline{K_5}$)\{10}, ($K_5 ∪ \overline{K_5}$)\{10}, ($K_5 ∪ \overline{K_5}$)\{10}, ($K_5 ∪ \overline{K_5}$)\{10}, ($K_5 ∪ \overline{K_5}$)\{10}, $K_5$, {10}}, $PS$[<11, 10>]={{1}, $K_5$, ($K_5 ∪ \overline{K_5}$)\{11}, ($K_5 ∪ \overline{K_5}$)\{11}, ($K_5 ∪ \overline{K_5}$)\{11}, ($K_5 ∪ \overline{K_5}$)\{11}, ($K_5 ∪ \overline{K_5}$)\{11}, ($K_5 ∪ \overline{K_5}$)\{11}, ($K_5 ∪ \overline{K_5}$)\{11}, $K_5$, {11}}.

Iteration 11: Processing segment level $k$=11.

The vertex <1, 11> has five parent vertices <2, 10>, <3, 10>, <4, 10>, <5, 10> and <6, 10>, respectively, we obtain $PS$[<1, 11>]={{2,3,4,5,6}, {1}}. Since the length of $PS$[<1, 11>]≠12, the original graph $G$=($\overline{K_5}+K_{n-2×5}$)∨$K_5$ is not Hamiltonian.

On the other hand, we can obtain a Hamiltonion path based on $PS$[<7, 10>], $PS$[<8, 10>], $PS$[<9, 10>], $PS$[<10, 10>] or $PS$[<11, 10>], respectively.

Without doubt, for any maximal nonHamiltonian graph ($\overline{K_m}+K_{n-2m}$)∨$K_m$, we can reach the same conclusion above.

## 5  Conclusion

We conclude the first deterministic polynomial time algorithm for Hamiltonian cycle problem and then discuss potential directions for future work.

1. During the construction of the path hologram $H$, if the original graph $G$ is a digraph, then $n$-2 directed edges (<$u$, $k$>, <$v$, $k$+1>) will be generated for each directed edge ($u$, $v$)∈$E$($G$), where $n$=|$V$($G$)| and 1≤$k$≤$n$-2. If the original graph $G$ is a mixed graph, then $n$-2 directed edges (<$u$, $k$>, <$v$, $k$+1>) will be generated for each directed edge ($u$, $v$)∈$E$($G$) and $n$-2 directed edges (<$u$, $k$>, <$v$, $k$+1>) as well as $n$-2 directed edges (<$v$, $k$>, <$u$, $k$+1>) will be generated for each undirected edge $uv$∈$E$($G$), where $n=|V(G)|$ and $1 \leq k \leq n$-2. Thus, the **PHG-BP** algorithm can handle the Hamiltonian cycle problem of undirected graph, digraph, or missed graph.

In fact, the path hologram proposed in this paper is only to help readers understand the idea of our method. Thus, the **PHG-BP** algorithm can be directly applied to the original graph *G* instead of the path hologram *H*.

2. If the original graph *G* has a Hamiltonian cycle, then any a vertex on this cycle can be used as the starting vertex. Thus, we can select any vertex *u* in *G* to generate the initial vertex *S*=<*u*, 0> and the final vertex *D*=<*u*, *n*> in path hologram *H*. On the contrary, the Hamiltonian path in a traceable graph uses a particular vertex as its starting vertex. Thus, the **PHG-BP** algorithm should been run on each vertex as the initial vertex respectively.

In fact, the **PHG-BP** algorithm can find not only the Hamiltonian cycle in a Hamiltonian graph or a Hamiltonian path in a traceable graph, but also the longest cycle or the longest path in any an original graph *G* in deterministic polynomial time.

3. The greatest contribution of our method is the concept of a path set in the path hologram where the (*n*-1)! paths can be stored in $O(n^2)$ space. If the original graph *G* is Hamiltonian, then the path set *PS*[*D*]=*PS*[<*u*, *n*>] of the final vertex *D*=<*u*, *n*> in the path hologram *H* stores all the Hamiltonian cycles from the initial vertex *S*=<*u*, 0> to the final vertex *D*=<*u*, *n*>. To ensure that each valid path stored in the path set can be found in backward searching while each invalid broken path fragment stored in the path set cannot be visited in backward searching, the key strategy of our method is the "consecutive" deleting-replenishing operations recursively on the left/right action field of an "invalid" vertex, respectively. During the backward searching, only need one parent vertex of the current vertex be selected based on **CHECK**(PStemp$\cap_{min}$PS[<parent vertex>]). Therefore, the backward search is deterministic and only one vertex is selected at each step.

4. The **PHG-BP** algorithm is a greedy algorithm, which meets the optimality principle and can be realized by using ordinary loop structure. The path set of vertices is only related to the path set of ancestor vertices and is independent of the path set of sibling vertices on the same segment level. Therefore, parallel algorithms can be designed very easily, so as to improve the time complexity and space complexity of the **PHG-BP** algorithm.

5. As the theoretically analysis above, the iterations of the "While (flag2)" loop has an upper bound $O(n^2)$ time in the worst case. We know that the actual iterations of the "While (flag2)" loop is determined by the number of singletons. Without loss of generality, we suppose that the number of singletons is *ns*, where $1 \leq ns < k$-$1 \leq n$. So the number of non-singleton is *k*-*ns*. If the *ns* is large, then in the path set, the number of the segment sets of a big size will be sparse and an empty segment set will be generated quickly or the execution of the duplicates of the singletons will be finished quickly. If the *ns* is small, then the number of the singletons will be sparse and the process of the duplicates of the singletons will be finished quickly, too. Thus, we guess the actual time complexity of the "While (flag2)" loop is $O(1)$ on average and in the worst case. The **Examples 1**~4 support such conjecture. Whether it is true or not for any graphs of order *n* is our future work.

6. The time complexity of **PHG-BP** algorithm depends on the time complexity of **CM** operator, while the latter depends on the While(flag2) loop and the **CHECK** operator. As discussed above, the While(flag2) loop might run at $O(1)$ steps, or $O(n^2)$ steps in the worst case. In the **CHECK** operator, if all PStemp[<*x*, *i*-1>]={∅} for each parent vertex <*x*, *i*-1> of <*w*, *i*>, then we can let PStemp[<*w*, *i*>] ={∅}, although *PS*[<*w*, *i*>]$\cap_{min}$PStemp might not be {∅}. Further, if all PStemp[<*w*, *i*>]={∅} for some segment level *i* (1≤*i*≤*k*-2), then we can infer PStemp[<*v*, *k*-1>]={∅}. How to improve the efficiency of the **CHECK** operator is our future work.

    7. If we only find the Hamiltonian cycle/path, we can consider only such parent vertex <*v*, *k*-1> of <*u*, *k*> that length(*PS*[<*v*, *k*-1>])=*k*, i.e., we consider only the basic path from the initial vertex *S* through its parent vertex <*v*, *k*-1> to <*u*, *k*>.

    8. We can optimize the four operations $\cup_{max}$, $\cap_{min}$, **LAFRD** and **RAFRD** with disjoint set. Using programming language to realize the **PHG-BP** algorithm is our future work.

## A   Appendix

In this Appendix, we will prove the correctness of our algorithm based on the proof of **Theorem 1~8** stated in Section 1.2.

**Theorem 1.** A finite connected undirected graph *G*=(*V*, *E*) has a Hamiltonian cycle if and only if the corresponding path hologram *H*=<$V_H$, $E_H$, *S*, *D*, *L*> has a basic path from the initial vertex *S* to the final vertex *D*.

Proof. For any a Hamiltonian cycle $u_1$-$u_2$-…-$u_n$-$u_1$ in *G*, where |*V*(*G*)|=*n*, we can let the initial vertex *S*=<$u_1$, 0> and the final vertex *D*=<$u_1$, *n*> during the process of constructing the corresponding path hologram *H*, respectively. This means that there exist a basic path <$u_1$, 0>-<$u_2$, 1>-…-<$u_n$, *n*-1>-<$u_1$, *n*> from *S* to *D* in *H*. Vice versa.   □

**Theorem 2.** A finite connected undirected graph *G* is a Hamiltonian graph if and only if the **PHG-BP** algorithm must save the basic path from the initial vertex *S* to the final vertex *D* in the path set of the vertex *D* in the corresponding path hologram *H*.

Proof. First of all, we know that any vertex can be selected as the starting vertex and the end vertex on any Hamiltonian cycle in graph *G*. Thus, we can select any vertex in *G* to generate the initial vertex *S* and the final vertex *D* in path hologram *H*.

    As the proof of the **Theorem 1**, there is a Hamiltonian cycle $u_1$-$u_2$-…-$u_n$-$u_1$ in *G*, where |*V*(*G*)| =*n*, if and only if, there exist a basic path <$u_1$, 0>-<$u_2$, 1>-…-<$u_n$, *n*-1>-<$u_1$, *n*> from *S* to *D* in *H*. This means that there exist these directed edges (<$u_1$, 0>, <$u_2$, 1>), (<$u_2$, 1>, <$u_3$, 2>),…,(<$u_n$, *n*-1>, <$u_1$, n>) in $E_H$.

    The **PHG-BP** algorithm begins with the initial vertex *S*=<$u_1$, 0> to find the descendant vertices on the next segment level, such that these descendant vertices do not reappear on the generated basic path $u_1$. Since (<$u_1$, 0>, <$u_2$, 1>)∈$E_H$, the **PHG-BP** algorithm can find such a descendant vertex $u_2$ in *O*(deg($u_1$)) time steps and obtain *PS*[<$u_2$, 1>]={{$u_1$}, {$u_2$}} when the vertex <$u_2$, 1> joining the tail of *PS*[<$u_1$, 0>]. Next, the **PHG-BP** algorithm begins with the vertex <$u_2$, 1> to find the descendant vertices on the next segment level, such that these descendant vertices do not reappear on the known basic path $u_1$-$u_2$. Since (<$u_2$, 1>, <$u_3$, 2>)∈$E_H$, the **PHG-BP** algorithm can find such a descendant vertex $u_3$ in *O*(deg($u_2$)) time steps and obtain *PS*[<$u_3$, 2>] ={{$u_1$}, {$u_2$,…}, {$u_3$}} when the vertex <$u_3$, 2> joining the tail of *PS*[<$u_2$, 1>]. This process continues until the **PHG-BP** algorithm obtains finally *PS*[*D*]=*PS*[<$u_1$, *n*>]={{$u_1$}, {$u_2$,…}, {$u_3$,…},…, {$u_n$,…}, {$u_1$}}. So *PS*[*D*] saves the basic path <$u_1$, 0>-<$u_2$, 1>-<$u_3$, 2>-…-<$u_n$, *n*-1>-<$u_1$, *n*> in *H*. Vice versa.   □

    Before proving the **Theorem 3**, we analyze the following cases first. As one knows, when the "consecutive" deleting-replenishing operations on the duplicative vertex of *u* have halted, the original "invalid" path on which the duplicative vertex of *u* is located may be broken. That is, some fragments of the original "invalid" path were not actually removed from PStemp[<*v*, *k*-1>]. In other words, some "broken" invalid path fragments may remain in *PS*[<*u*, *k*>]. The analysis of the following cases will

show that these "broken" invalid path fragments cannot be visited in Backward Searches even if they are saved in $PS[<u, k>]$.

Consider the situations shown in **Figure 7**.

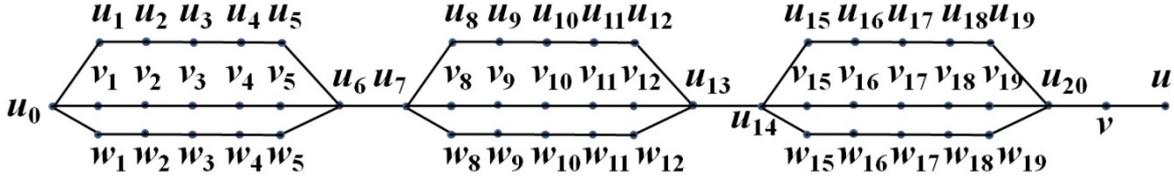

**Figure 7**  Duplicative vertex of $<u, 22>$ on the path set $PS[<v, 21>]$

Case 1. There are exactly 3 unique basic paths in $PS[<u_{20}, 20>]$, i.e., $P_1=u_0$-$u_1$-$u_2$-$u_3$-$u_4$-$u_5$-$u_6$-$u_7$-$u_8$-$u_9$-$u_{10}$-$u_{11}$-$u_{12}$-$u_{13}$-$u_{14}$-$u_{15}$-$u_{16}$-$u_{17}$-$u_{18}$-$u_{19}$-$u_{20}$, $P_2=u_0$-$v_1$-$v_2$-$v_3$-$v_4$-$v_5$-$u_6$-$u_7$-$v_8$-$v_9$-$v_{10}$-$v_{11}$-$v_{12}$-$u_{13}$-$u_{14}$-$v_{15}$-$v_{16}$-$v_{17}$-$v_{18}$-$v_{19}$-$u_{20}$, $P_3=u_0$-$w_1$-$w_2$-$w_3$-$w_4$-$w_5$-$u_6$-$u_7$-$w_8$-$w_9$-$w_{10}$-$w_{11}$-$w_{12}$-$u_{13}$-$u_{14}$-$w_{15}$-$w_{16}$-$w_{17}$-$w_{18}$-$w_{19}$-$u_{20}$.

In this case, we must have $PS[<u_{19}, 19>]=\{\{u_0\},\{u_1\},\{u_2\},\{u_3\},\{u_4\},\{u_5\},\{u_6\},\{u_7\},\{u_8\},\{u_9\},\{u_{10}\},\{u_{11}\},\{u_{12}\},\{u_{13}\},\{u_{14}\},\{u_{15}\},\{u_{16}\},\{u_{17}\},\{u_{18}\},\{u_{19}\}\}$. $PS[<v_{19}, 19>]=\{\{u_0\},\{v_1\},\{v_2\},\{v_3\},\{v_4\},\{v_5\},\{u_6\},\{u_7\},\{v_8\},\{v_9\},\{v_{10}\},\{v_{11}\},\{v_{12}\},\{u_{13}\},\{u_{14}\},\{v_{15}\},\{v_{16}\},\{v_{17}\},\{v_{18}\},\{v_{19}\}\}$. $PS[<w_{19}, 19>]=\{\{u_0\},\{w_1\},\{w_2\},\{w_3\},\{w_4\},\{w_5\},\{u_6\},\{u_7\},\{w_8\},\{w_9\},\{w_{10}\},\{w_{11}\},\{w_{12}\},\{u_{13}\},\{u_{14}\},\{w_{15}\},\{w_{16}\},\{w_{17}\},\{w_{18}\},\{w_{19}\}\}$.

So $PS[<u_{20}, 20>]=\{\{u_0\},\{u_1, v_1, w_1\},\{u_2, v_2, w_2\},\{u_3, v_3, w_3\},\{u_4, v_4, w_4\},\{u_5, v_5, w_5\},\{u_6\},\{u_7\},\{u_8, v_8, w_8\},\{u_9, v_9, w_9\},\{u_{10}, v_{10}, w_{10}\},\{u_{11}, v_{11}, w_{11}\},\{u_{12}, v_{12}, w_{12}\},\{u_{13}\},\{u_{14}\},\{u_{15}, v_{15}, w_{15}\},\{u_{16}, v_{16}, w_{16}\},\{u_{17}, v_{17}, w_{17}\},\{u_{18}, v_{18}, w_{18}\},\{u_{19}, v_{19}, w_{19}\},\{u_{20}\}\}$. $PS[<v, 21>]=PS[<u_{20}, 20>]\otimes\{\{v\}\}$.

Case 1.1. If $u=v_3$, the deleted path $P_{del}=v_1$-$v_2$-$v_3$-$v_4$-$v_5$ will be removed from PStemp$[<v, 21>]$, resulting in that two "broken" invalid path fragments $v_8$-$v_9$-$v_{10}$-$v_{11}$-$v_{12}$ and $v_{15}$-$v_{16}$-$v_{17}$-$v_{18}$-$v_{19}$ remained in PStemp$[<v, 21>]$. That is, PStemp$[<v, 21>]=\{\{u_0\},\{u_1, w_1\},\{u_2, w_2\},\{u_3, w_3\},\{u_4, w_4\},\{u_5, w_5\},\{u_6\},\{u_7\},\{u_8, v_8, w_8\},\{u_9, v_9, w_9\}, \{u_{10}, v_{10}, w_{10}\}, \{u_{11}, v_{11}, w_{11}\},\{u_{12}, v_{12}, w_{12}\}, \{u_{13}\}, \{u_{14}\},\{u_{15}, v_{15}, w_{15}\},\{u_{16}, v_{16}, w_{16}\},\{u_{17}, v_{17}, w_{17}\},\{u_{18}, v_{18}, w_{18}\},\{u_{19}, v_{19}, w_{19}\},\{u_{20}\},\{v\}\}$. So $PS[<u, 22>]=\{\{u_0\},\{u_1, w_1\},\{u_2, w_2\},\{u_3, w_3\},\{u_4, w_4\},\{u_5, w_5\},\{u_6\},\{u_7\},\{u_8, v_8, w_8\},\{u_9, v_9, w_9\}, \{u_{10}, v_{10}, w_{10}\}, \{u_{11}, v_{11}, w_{11}\},\{u_{12}, v_{12}, w_{12}\}, \{u_{13}\}, \{u_{14}\},\{u_{15}, v_{15}, w_{15}\},\{u_{16}, v_{16}, w_{16}\},\{u_{17}, v_{17}, w_{17}\},\{u_{18}, v_{18}, w_{18}\},\{u_{19}, v_{19}, w_{19}\},\{u_{20}\},\{v\},\{u\}\}$ contains the above two "broken" invalid path fragments.

When we backward search from $<u, 22>$ to $<u_0, 0>$, we have that PStemp$=PS[<u, 22>]$ at iteration 1, PStemp$=$PStemp$\cap_{min}PS[<v, 21>]$ at iteration 2, and PStemp$=$PStemp$\cap_{min}PS[<u_{20}, 20>]$ at iteration 3. Now, PStemp$=\{\{u_0\},\{u_1,w_1\},\{u_2,w_2\},\{u_3,w_3\},\{u_4,w_4\},\{u_5,w_5\},\{u_6\},\{u_7\},\{u_8,v_8,w_8\},\{u_9,v_9,w_9\},\{u_{10},v_{10},w_{10}\},\{u_{11},v_{11},w_{11}\},\{u_{12},v_{12},w_{12}\},\{u_{13}\},\{u_{14}\},\{u_{15},v_{15},w_{15}\},\{u_{16},v_{16},w_{16}\},\{u_{17},v_{17},w_{17}\},\{u_{18},v_{18},w_{18}\},\{u_{19},v_{19},w_{19}\},\{u_{20}\}\}$.

$<u_{20}, 20>$ is the meeting point of the rightmost "parallel path cluster". Since PStemp[19]$=\{u_{19}, v_{19}, w_{19}\}$, the $<u_{20}, 20>$ has three parent vertices $<u_{19}, 19>$, $<v_{19}, 19>$, and $<w_{19}, 19>$. If we choose the $<u_{19}, 19>$ for backward searching, we have PStemp$=$PStemp$\cap_{min}PS[<u_{19}, 19>]=\{\{u_0\}, \{u_1\},\{u_2\},\{u_3\},\{u_4\},\{u_5\},\{u_6\},\{u_7\},\{u_8\},\{u_9\},\{u_{10}\},\{u_{11}\},\{u_{12}\},\{u_{13}\},\{u_{14}\},\{u_{15}\},\{u_{16}\},\{u_{17}\},\{u_{18}\},\{u_{19}\}\}$. Therefore, we can backward search from $<u_{19}, 19>$ to $<u_0, 0>$ along this valid path $P_1=u_0$-$u_1$-$u_2$-$u_3$-$u_4$-$u_5$-$u_6$-$u_7$-$u_8$-$u_9$-$u_{10}$-$u_{11}$-$u_{12}$-$u_{13}$-$u_{14}$-$u_{15}$-$u_{16}$-$u_{17}$-$u_{18}$-$u_{19}$. Similarly, we can choose the $<w_{19}, 19>$ for backward searching from $<w_{19}, 19>$ to $<u_0, 0>$ along another valid path $P_3$.

On the other hand, $<v_{19}, 19>$ is the rightmost vertex of the "broken" invalid path fragment $v_{15}$-$v_{16}$-$v_{17}$-$v_{18}$-$v_{19}$ located on the rightmost "parallel path cluster", and $<v_{12}, 12>$ is the rightmost vertex of the "broken" invalid path fragment $v_8$-$v_9$-$v_{10}$-$v_{11}$-$v_{12}$ located on the inner "parallel path cluster".

Since PStemp at $<u_{20}, 20>$ does not contain the deleted path $P_{del}=v_1$-$v_2$-$v_3$-$v_4$-$v_5$, while $PS[<v_{19}, 19>]$ contains only the deleted path $P_{del}=v_1$-$v_2$-$v_3$-$v_4$-$v_5$, $\varnothing$ must occur in PStemp$\cap_{min}PS[<v_{19}, 19>]$, i.e., $\varnothing \in$ PStemp$\cap_{min}PS[<v_{19}, 19>]$ and PStemp[3]$\cap PS_{<v19,19>}[3]=\varnothing$. In other words, we have PStemp=PStemp$\cap_{min}PS[<v_{19}, 19>]$={$\{u_0\},\{\ \},\{\ \},\{\ \},\{\ \},\{\ \}, \{u_6\}, \{u_7\}, \{v_8\},\{v_9\},\{v_{10}\},\{v_{11}\}$, $\{v_{12}\},\{u_{13}\},\{u_{14}\},\{v_{15}\},\{v_{16}\},\{v_{17}\},\{v_{18}\},\{v_{19}\}\}$. So we will not choose the parent vertex $<v_{19}, 19>$ of the $<u_{20}, 20>$ for backward searching due to $\varnothing \in$ PStemp$\cap_{min}PS[<v_{19}, 19>]$.

Further, since the $<v_{19}, 19>$ is not chosen for backward searching, this will result in that PStemp= PStemp$\cap_{min}PS[<u_{19}, 19>]$ or PStemp=PStemp$\cap_{min}PS[<w_{19}, 19>]$, respectively. In either case, $v_{12}$ is not contained in the PStemp=PStemp$\cap_{min}PS[<u_{13}, 13>]$ and $v_5$ is not contained in the PStemp= PStemp$\cap_{min}PS[<u_6, 6>]$. So both $<v_{12}, 12>$ and $<v_5, 5>$ cannot be visited for backward searching. Therefore, the two "broken" invalid path fragments $v_{15}$-$v_{16}$-$v_{17}$-$v_{18}$-$v_{19}$ and $v_8$-$v_9$-$v_{10}$-$v_{11}$-$v_{12}$ cannot be visited for backward searching.

Case 1.2. If $u=v_{10}$, the deleted path $P_{del}=v_8$-$v_9$-$v_{10}$-$v_{11}$-$v_{12}$ will be removed from PStemp[$<v, 21>$], resulting in that two "broken" invalid path fragments $v_1$-$v_2$-$v_3$-$v_4$-$v_5$ and $v_{15}$-$v_{16}$-$v_{17}$-$v_{18}$-$v_{19}$ remained in PStemp[$<v, 21>$]. So $PS[<u, 22>]$={$\{u_0\},\{u_1, v_1, w_1\},\{u_2, v_2, w_2\},\{u_3, v_3, w_3\}, \{u_4, v_4, w_4\}, \{u_5, v_5, w_5\},\{u_6\},\{u_7\},\{u_8,w_8\},\{u_9, w_9\},\{u_{10},w_{10}\}, \{u_{11}, w_{11}\},\{u_{12},w_{12}\}, \{u_{13}\}, \{u_{14}\}, \{u_{15}, v_{15}, w_{15}\},\{u_{16}, v_{16}, w_{16}\},\{u_{17}, v_{17}, w_{17}\}, \{u_{18}, v_{18}, w_{18}\},\{u_{19}, v_{19}, w_{19}\},\{u_{20}\},\{v\},\{u\}\}$.

Similar to case 1.1, $\varnothing \in$ PStemp$\cap_{min}PS[<v_{19}, 19>]$, i.e., PStemp[10]$\cap PS_{<v19,19>}[10]=\varnothing$. So we will not choose the parent vertex $<v_{19}, 19>$ of the $<u_{20}, 20>$ for backward searching. Further, $v_{12}$ is not contained in the PStemp=PStemp$\cap_{min}PS[<u_{13}, 13>]$ and $v_5$ is not contained in the PStemp= PStemp$\cap_{min}PS[<u_6, 6>]$. So both $<v_{12}, 12>$ and $<v_5, 5>$ cannot be visited for backward searching. Therefore, the two "broken" invalid path fragments $v_{15}$-$v_{16}$-$v_{17}$-$v_{18}$-$v_{19}$ and $v_1$-$v_2$-$v_3$-$v_4$-$v_5$ cannot be visited for backward searching.

Case 1.3. If $u=v_{17}$, the deleted path $P_{del}=v_{15}$-$v_{16}$-$v_{17}$-$v_{18}$-$v_{19}$ will be removed from PStemp[$<v, 21>$], resulting in that two "broken" invalid path fragments $v_1$-$v_2$-$v_3$-$v_4$-$v_5$ and $v_8$-$v_9$-$v_{10}$-$v_{11}$-$v_{12}$ remained in PStemp[$<v, 21>$]. So $PS[<u, 22>]$={$\{u_0\},\{u_1, v_1, w_1\},\{u_2, v_2, w_2\},\{u_3, v_3, w_3\}, \{u_4, v_4, w_4\},\{u_5, v_5, w_5\},\{u_6\},\{u_7\},\{u_8, v_8, w_8\},\{u_9, v_9, w_9\}, \{u_{10}, v_{10}, w_{10}\}, \{u_{11}, v_{11}, w_{11}\}, \{u_{12}, v_{12}, w_{12}\},\{u_{13}\},\{u_{14}\},\{u_{15},w_{15}\},\{u_{16},w_{16}\}, \{u_{17},w_{17}\}, \{u_{18},w_{18}\}, \{u_{19}, w_{19}\},\{u_{20}\},\{v\},\{u\}\}$.

Similar to case 1.1, $v_{19}$ is not contained in PStemp$\cap_{min}PS[<u_{20}, 20>]$, $v_{12}$ is not contained in PStemp=PStemp$\cap_{min}PS[<u_{13}, 13>]$ and $v_5$ is not contained in PStemp=PStemp$\cap_{min}PS[<u_6, 6>]$. So that $<v_{19}, 19>$, $<v_{12}, 12>$ and $<v_5, 5>$ cannot be visited for backward searching. Therefore, the two "broken" invalid path fragments $v_1$-$v_2$-$v_3$-$v_4$-$v_5$ and $v_8$-$v_9$-$v_{10}$-$v_{11}$-$v_{12}$ cannot be visited for backward searching.

Case 1.4. If $u=u_3$, $u=v_{10}$, and $u=w_{17}$, the deleted path $P_{del1}=u_1$-$u_2$-$u_3$-$u_4$-$u_5$, $P_{del2}=v_8$-$v_9$-$v_{10}$-$v_{11}$-$v_{12}$, and $P_{del3}=w_{15}$-$w_{16}$-$w_{17}$-$w_{18}$-$w_{19}$ will be removed from PStemp[$<v, 21>$], resulting in that six "broken" invalid path fragments $u_8$-$u_9$-$u_{10}$-$u_{11}$-$u_{12}$, $u_{15}$-$u_{16}$-$u_{17}$-$u_{18}$-$u_{19}$, $v_1$-$v_2$-$v_3$-$v_4$-$v_5$, $v_{15}$-$v_{16}$-$v_{17}$-$v_{18}$-$v_{19}$, $w_1$-$w_2$-$w_3$-$w_4$-$w_5$, and $w_8$-$w_9$-$w_{10}$-$w_{11}$-$w_{12}$ remained in PStemp[$<v, 21>$]. So $PS[<u, 22>]$={$\{u_0\},\{v_1, w_1\},\{v_2, w_2\},\{v_3, w_3\},\{v_4, w_4\},\{v_5, w_5\},\{u_6\},\{u_7\},\{u_8, w_8\},\{u_9, w_9\},\{u_{10},w_{10}\},\{u_{11}, w_{11}\},\{u_{12}, w_{12}\},\{u_{13}\},\{u_{14}\},\{u_{15},v_{15}\},\{u_{16},v_{16}\}, \{u_{17},v_{17}\}, \{u_{18},v_{18}\}$, $\{u_{19}, v_{19}\},\{u_{20}\},\{v\},\{u\}\}$ contains the above six "broken" invalid path fragments.

Similar to case 1.1, 1.2, and 1.3, we have $\varnothing \in$ PStemp$\cap_{min}PS[<u_{19}, 19>]$, $\varnothing \in$ PStemp$\cap_{min}PS[<v_{19}, 19>]$ and $w_{19} \notin$ PStemp[19]. So we will not choose the parent vertex $<u_{19}, 19>$ or $<v_{19}, 19>$ of the $<u_{20}, 20>$ for backward searching due to $\varnothing \in$ PStemp$\cap_{min}PS[<u_{19}, 19>]$ or $\varnothing \in$ PStemp$\cap_{min}PS[<v_{19}, 19>]$, and we cannot choose $<w_{19}, 19>$ for backward searching due to $w_{19} \notin$ PStemp[19]. This means that there does not exist a basic path from $<u_0, 0>$ through $<v, 21>$ to $<u, 22>$, although PStemp[$<v, 21>$] does not contain the empty set $\varnothing$. So we let PStemp[$<v, 21>$]={$\varnothing$}.

On the other hand, PStemp=PStemp[$<v, 21>$]={$\{u_0\},\{v_1, w_1\},\{v_2, w_2\},\{v_3, w_3\},\{v_4, w_4\},\{v_5, w_5\},\{u_6\},\{u_7\},\{u_8, w_8\},\{u_9, w_9\},\{u_{10}, w_{10}\},\{u_{11}, w_{11}\},\{u_{12}, w_{12}\},\{u_{13}\},\{u_{14}\},\{u_{15}, v_{15}\},\{u_{16}, v_{16}\},\{u_{17}, v_{17}\},\{u_{18}, v_{18}\},\{u_{19}, v_{19}\},\{u_{20}\},\{v\}$}. PStemp[$<u_{19}, 19>$]=$PS[<u_{19}, 19>] \cap_{min}$ PStemp ={$\varnothing$}, PStemp[$<v_{19}, 19>$]=$PS[<v_{19}, 19>] \cap_{min}$ PStemp={$\varnothing$}, PStemp[$<u_{20}, 20>$]=$PS[<u_{20}, 20>] \cap_{min}$PStemp$\neq$ {$\varnothing$}. However, since PStemp[$<u_{19}, 19>$]={$\varnothing$} and PStemp[$<v_{19}, 19>$]={$\varnothing$}, and $w_{19} \notin$ PStemp[19], we can infer that $<u_{20}, 20>$ cannot be reached for the reason of $<u, 22>$. So we should let PStemp[$<u_{20}, 20>$]={$\varnothing$}. This is the responsibility of the **CHECK** operator.

Case 2. There are exactly 6 unique basic paths in $PS[<u_{20}, 20>]$, i.e., $P_1=u_0$-$u_1$-$u_2$-$u_3$-$u_4$-$u_5$-$u_6$-$u_7$-$u_8$-$u_9$-$u_{10}$-$u_{11}$-$u_{12}$-$u_{13}$-$u_{14}$-$u_{15}$-$u_{16}$-$u_{17}$-$u_{18}$-$u_{19}$-$u_{20}$, $P_2=u_0$-$v_1$-$v_2$-$v_3$-$v_4$-$v_5$-$u_6$-$u_7$-$v_8$-$v_9$-$v_{10}$-$v_{11}$-$v_{12}$-$u_{13}$-$u_{14}$-$v_{15}$-$v_{16}$-$v_{17}$-$v_{18}$-$v_{19}$-$u_{20}$, $P_3=u_0$-$w_1$-$w_2$-$w_3$-$w_4$-$w_5$-$u_6$-$u_7$-$w_8$-$w_9$-$w_{10}$-$w_{11}$-$w_{12}$-$u_{13}$-$u_{14}$-$w_{15}$-$w_{16}$-$w_{17}$-$w_{18}$-$w_{19}$-$u_{20}$, $P_4=u_0$-$u_1$-$u_2$-$u_3$-$u_4$-$u_5$-$u_6$-$u_7$-$v_8$-$v_9$-$v_{10}$-$v_{11}$-$v_{12}$-$u_{13}$-$u_{14}$-$w_{15}$-$w_{16}$-$w_{17}$-$w_{18}$-$w_{19}$-$u_{20}$, $P_5=u_0$-$v_1$-$v_2$-$v_3$-$v_4$-$v_5$-$u_6$-$u_7$-$w_8$-$w_9$-$w_{10}$-$w_{11}$-$w_{12}$-$u_{13}$-$u_{14}$-$u_{15}$-$u_{16}$-$u_{17}$-$u_{18}$-$u_{19}$-$u_{20}$, $P_6=u_0$-$w_1$-$w_2$-$w_3$-$w_4$-$w_5$-$u_6$-$u_7$-$u_8$-$u_9$-$u_{10}$-$u_{11}$-$u_{12}$-$u_{13}$-$u_{14}$-$v_{15}$-$v_{16}$-$v_{17}$-$v_{18}$-$v_{19}$-$u_{20}$.

In this case, we must have $PS[<u_{19}, 19>]$={$\{u_0\},\{u_1,v_1\},\{u_2,v_2\},\{u_3,v_3\},\{u_4,v_4\},\{u_5,v_5\},\{u_6\},\{u_7\},\{u_8,w_8\},\{u_9,w_9\},\{u_{10},w_{10}\},\{u_{11},w_{11}\},\{u_{12},w_{12}\},\{u_{13}\},\{u_{14}\},\{u_{15}\},\{u_{16}\},\{u_{17}\},\{u_{18}\},\{u_{19}\}$}. $PS[<v_{19},19>]$={$\{u_0\},\{v_1,w_1\},\{v_2,w_2\},\{v_3,w_3\},\{v_4,w_4\},\{v_5,w_5\},\{u_6\},\{u_7\},\{u_8,v_8\},\{u_9,v_9\},\{u_{10},v_{10}\},\{u_{11},v_{11}\},\{u_{12},v_{12}\},\{u_{13}\},\{u_{14}\},\{v_{15}\},\{v_{16}\},\{v_{17}\},\{v_{18}\},\{v_{19}\}$}. $PS[<w_{19},19>]$={$\{u_0\},\{u_1,w_1\},\{u_2,w_2\},\{u_3,w_3\},\{u_4,w_4\},\{u_5,w_5\},\{u_6\},\{u_7\},\{v_8,w_8\},\{v_9,w_9\},\{v_{10},w_{10}\},\{v_{11},w_{11}\},\{v_{12},w_{12}\},\{u_{13}\},\{u_{14}\},\{w_{15}\},\{w_{16}\},\{w_{17}\},\{w_{18}\},\{w_{19}\}$}. $PS[<u_{12},12>]$={$\{u_0\},\{u_1,w_1\},\{u_2,w_2\},\{u_3,w_3\},\{u_4,w_4\},\{u_5,w_5\},\{u_6\},\{u_7\},\{u_8\},\{u_9\},\{u_{10}\},\{u_{11}\},\{u_{12}\}$}, $PS[<v_{12},12>]$={$\{u_0\},\{u_1,v_1\},\{u_2,v_2\},\{u_3,v_3\},\{u_4,v_4\},\{u_5,v_5\},\{u_6\},\{u_7\},\{v_8\},\{v_9\},\{v_{10}\},\{v_{11}\},\{v_{12}\}$}, $PS[<w_{12}, 12>]$={$\{u_0\},\{v_1,w_1\},\{v_2,w_2\},\{v_3,w_3\},\{v_4,w_4\},\{v_5,w_5\},\{u_6\},\{u_7\},\{w_8\},\{w_9\},\{w_{10}\},\{w_{11}\},\{w_{12}\}$}.

So $PS[<u_{20}, 20>]$={$\{u_0\},\{u_1, v_1, w_1\},\{u_2, v_2, w_2\},\{u_3, v_3, w_3\},\{u_4, v_4, w_4\},\{u_5, v_5, w_5\},\{u_6\},\{u_7\},\{u_8, v_8, w_8\},\{u_9, v_9, w_9\},\{u_{10}, v_{10}, w_{10}\},\{u_{11}, v_{11}, w_{11}\},\{u_{12}, v_{12}, w_{12}\},\{u_{13}\},\{u_{14}\},\{u_{15}, v_{15}, w_{15}\},\{u_{16}, v_{16}, w_{16}\},\{u_{17}, v_{17}, w_{17}\},\{u_{18}, v_{18}, w_{18}\},\{u_{19}, v_{19}, w_{19}\},\{u_{20}\}$}. $PS[<v, 21>]$ =$PS[<u_{20}, 20>] \otimes \{\{v\}\}$.

Case 2.1. If $u=v_3$, the deleted path $P_{del}=v_1$-$v_2$-$v_3$-$v_4$-$v_5$ will be removed from PStemp[$<v, 21>$], resulting in that four "broken" invalid path fragments $v_8$-$v_9$-$v_{10}$-$v_{11}$-$v_{12}$, $w_8$-$w_9$-$w_{10}$-$w_{11}$-$w_{12}$, $u_{15}$-$u_{16}$-$u_{17}$-$u_{18}$-$u_{19}$, and $v_{15}$-$v_{16}$-$v_{17}$-$v_{18}$-$v_{19}$ remained in PStemp[$<v, 21>$]. So $PS[<u, 22>]$={$\{u_0\},\{u_1, w_1\},\{u_2, w_2\},\{u_3, w_3\},\{u_4,w_4\},\{u_5, w_5\},\{u_6\},\{u_7\},\{u_8, v_8, w_8\},\{u_9, v_9, w_9\},\{u_{10}, v_{10}, w_{10}\},\{u_{11}, v_{11}, w_{11}\},\{u_{12}, v_{12}, w_{12}\},\{u_{13}\},\{u_{14}\},\{u_{15}, v_{15}, w_{15}\},\{u_{16}, v_{16}, w_{16}\},\{u_{17}, v_{17}, w_{17}\},\{u_{18}, v_{18}, w_{18}\},\{u_{19}, v_{19}, w_{19}\},\{u_{20}\},\{v\},\{u\}$} contains the above four "broken" invalid path fragments.

When we backward search from $<u, 22>$ to $<u_0, 0>$, we have that PStemp=$PS[<u, 22>]$ at iteration 1, PStemp=PStemp$\cap_{min}PS[<v, 21>]$ at iteration 2, and PStemp=PStemp$\cap_{min}PS[<u_{20}, 20>]$ at iteration 3. Now, PStemp={$\{u_0\},\{u_1, w_1\},\{u_2, w_2\},\{u_3, w_3\},\{u_4,w_4\},\{u_5, w_5\},\{u_6\},\{u_7\},\{u_8, v_8, w_8\},\{u_9, v_9, w_9\},\{u_{10}, v_{10}, w_{10}\},\{u_{11}, v_{11}, w_{11}\},\{u_{12}, v_{12}, w_{12}\},\{u_{13}\},\{u_{14}\},\{u_{15},v_{15},w_{15}\},\{u_{16}, v_{16},w_{16}\},\{u_{17},v_{17},w_{17}\},\{u_{18},v_{18},w_{18}\},\{u_{19},v_{19},w_{19}\},\{u_{20}\}$}.

<$u_{20}$, 20> is the meeting point of the rightmost "parallel path cluster". Since PStemp[19]={$u_{19}, v_{19}, w_{19}$}, the <$u_{20}$, 20> has three parent vertices <$u_{19}$, 19>, <$v_{19}$, 19>, and <$w_{19}$, 19>.

(a)<$u_{19}$, 19> is the rightmost vertex of the "broken" invalid path fragment $u_{15}$-$u_{16}$-$u_{17}$-$u_{18}$-$u_{19}$ located on the rightmost "parallel path cluster". If we choose the parent vertex <$u_{19}$, 19> of the <$u_{20}$, 20> to backward search, we have PStemp=PStemp$\cap_{min}$PS[<$u_{19}$, 19>] at iteration 4. Here, PStemp= {$\{u_0\},\{u_1\},\{u_2\},\{u_3\},\{u_4\},\{u_5\},\{u_6\},\{u_7\},\{u_8,w_8\},\{u_9,w_9\},\{u_{10},w_{10}\},\{u_{11}, w_{11}\},\{u_{12}, w_{12}\},\{u_{13}\},\{u_{14}\},\{u_{15}\},\{u_{16}\},\{u_{17}\},\{u_{18}\},\{u_{19}\}$}. Undoubtedly, the "broken" invalid path fragment $u_{15}$-$u_{16}$-$u_{17}$-$u_{18}$-$u_{19}$ is located not only on an invalid path $P_5$ (i.e., $u_{15}$-$u_{16}$-$u_{17}$-$u_{18}$-$u_{19}$ is a real broken invalid path fragment) but also on a valid path $P_1$ (i.e., $u_{15}$-$u_{16}$-$u_{17}$-$u_{18}$-$u_{19}$ is a real broken valid path fragment). For the valid path fragment $u_{15}$-$u_{16}$-$u_{17}$-$u_{18}$-$u_{19}$ located on the valid path $P_1$, we can choose the $u_{19}$, $u_{12}$, and $u_5$ respectively to backward search at each meeting point of the "parallel path cluster". On the contrary, for the real broken invalid path fragment $u_{15}$-$u_{16}$-$u_{17}$-$u_{18}$-$u_{19}$ located on the invalid path $P_5$, we cannot choose the $w_{12}$ to backward search at the meeting point of the "parallel path cluster", because the PStemp at <$w_{12}$, 12> contains an ∅. That is, ∅∈PStemp$\cap_{min}$PS[<$w_{12}$, 12>]. Therefore, we will choose the parent vertex <$u_{19}$, 19> of the <$u_{20}$, 20> and the parent vertex <$u_{12}$, 12> of the <$u_{13}$, 13> to backward search, while we will not choose the parent vertex <$w_{12}$, 12> of the <$u_{13}$, 13> to backward search due to ∅∈PStemp $\cap_{min}$PS[<$w_{12}$, 12>].

(b)<$v_{19}$, 19> is the rightmost vertex of the "broken" invalid path fragment $v_{15}$-$v_{16}$-$v_{17}$-$v_{18}$-$v_{19}$ located on the rightmost "parallel path cluster". If we choose the parent vertex <$v_{19}$, 19> of the <$u_{20}$, 20> to backward search, we have PStemp=PStemp$\cap_{min}$PS[<$v_{19}$, 19>] at iteration 4. Here, PStemp= {$\{u_0\},\{w_1\},\{w_2\},\{w_3\},\{w_4\},\{w_5\},\{u_6\},\{u_7\},\{u_8,v_8\},\{u_9,v_9\},\{u_{10},v_{10}\},\{u_{11}, v_{11}\},\{u_{12}, v_{12}\},\{u_{13}\},\{u_{14}\},\{v_{15}\},\{v_{16}\},\{v_{17}\},\{v_{18}\},\{v_{19}\}$}. Undoubtedly, the "broken" invalid path fragment $v_{15}$-$v_{16}$-$v_{17}$-$v_{18}$-$v_{19}$ is located not only on an invalid path $P_2$ (i.e., $v_{15}$-$v_{16}$-$v_{17}$-$v_{18}$-$v_{19}$ is a real broken invalid path fragment) but also on a valid path $P_6$ (i.e., $v_{15}$-$v_{16}$-$v_{17}$-$v_{18}$-$v_{19}$ is a real broken valid path fragment). For the valid path fragment $v_{15}$-$v_{16}$-$v_{17}$-$v_{18}$-$v_{19}$ located on the valid path $P_6$, we can choose the $v_{19}$, $u_{12}$, and $w_5$ respectively to backward search at each meeting point of the "parallel path cluster". On the contrary, for the real invalid path fragment $v_{15}$-$v_{16}$-$v_{17}$-$v_{18}$-$v_{19}$ located on the invalid path $P_2$, we cannot choose the $v_{12}$ to backward search at the meeting point of the "parallel path cluster", because the PStemp at <$v_{12}$, 12> contains an ∅. That is, ∅∈PStemp $\cap_{min}$PS[<$v_{12}$, 12>]. Therefore, we will choose the parent vertex <$v_{19}$, 19> of the <$u_{20}$, 20> and the parent vertex <$u_{12}$, 12> of the <$u_{13}$, 13> to backward search, while we will not choose the parent vertex <$v_{12}$, 12> of the <$u_{13}$, 13> to backward search due to ∅∈PStemp$\cap_{min}$PS[<$v_{12}$, 12>].

(c)<$w_{19}$, 19> is the rightmost vertex of the valid path fragment $w_{15}$-$w_{16}$-$w_{17}$-$w_{18}$-$w_{19}$ located on the rightmost "parallel path cluster". This valid path fragment $w_{15}$-$w_{16}$-$w_{17}$-$w_{18}$-$w_{19}$ belongs to two valid paths $P_3$ and $P_4$. Therefore, we can backward search from <$w_{19}$, 19> to <$u_0$, 0> along $P_3$ or $P_4$, respectively. In other words, we can choose the parent vertex <$w_{19}$, 19> of <$u_{20}$, 20>, the parent vertex <$v_{12}$, 12> or <$w_{12}$, 12> of <$u_{13}$, 13>, and the parent vertex <$u_5$, 5> or <$w_5$, 5> of <$u_6$, 6> for backward searching.

As discussion for backward searching above, we find that the broken "invalid" path fragments $v_8$-$v_9$-$v_{10}$-$v_{11}$-$v_{12}$ can be visited along the valid path $P_4$, but cannot be visited along the invalid path $P_2$. Similarly, the broken "invalid" path fragments $w_8$-$w_9$-$w_{10}$-$w_{11}$-$w_{12}$ can be visited along the valid path $P_3$, but cannot be visited along the invalid path $P_5$. The broken "invalid" path fragments $u_{15}$-$u_{16}$-$u_{17}$-$u_{18}$-$u_{19}$ can be visited along the valid path $P_1$, but cannot be visited along the invalid path $P_5$. The broken "invalid" path fragments $v_{15}$-$v_{16}$-$v_{17}$-$v_{18}$-$v_{19}$ can be visited along the valid path $P_6$, but cannot be visited along the invalid path $P_2$.

On the other hand, the valid path fragment $u_8$-$u_9$-$u_{10}$-$u_{11}$-$u_{12}$ or $w_{15}$-$w_{16}$-$w_{17}$-$w_{18}$-$w_{19}$ can be visited along their original valid paths respectively. Further, the path fragments $u_1$-$u_2$-$u_3$-$u_4$-$u_5$ and $w_1$-$w_2$-$w_3$-$w_4$-$w_5$ are located on the "parallel path cluster" from which the $P_{del}=v_1$-$v_2$-$v_3$-$v_4$-$v_5$ is deleted. So they are valid and can be visited along their original valid paths respectively.

So we obtain the following conclusions. The valid path fragments can be visited along their original valid path. The "broken" invalid path fragments cannot be visited along the original "invalid" path on which the duplicative vertex of $u$ is deleted from the path set $PS[<v, k-1>]$. Contrarily, the "broken" invalid path fragments as "valid" path fragments can be visited along another original valid path if it exists.

Case 2.2. If $u=v_{10}$, the deleted path $P_{del}=v_8$-$v_9$-$v_{10}$-$v_{11}$-$v_{12}$ will be removed from PStemp[<$v$, 21>], resulting in that four "broken" invalid path fragments $u_1$-$u_2$-$u_3$-$u_4$-$u_5$, $v_1$-$v_2$-$v_3$-$v_4$-$v_5$, $v_{15}$-$v_{16}$-$v_{17}$-$v_{18}$-$v_{19}$, and $w_{15}$-$w_{16}$-$w_{17}$-$w_{18}$-$w_{19}$ remained in PStemp[<$v$, 21>]. So $PS[<u, 22>]=$ $\{\{u_0\},\{u_1,v_1,w_1\},\{u_2,v_2,w_2\},\{u_3,v_3,w_3\},\{u_4,v_4,w_4\},\{u_5,v_5,w_5\},\{u_6\},\{u_7\},\{u_8,w_8\},\{u_9,w_9\},$ $\{u_{10},w_{10}\},\{u_{11},w_{11}\},\{u_{12},w_{12}\},\{u_{13}\},\{u_{14}\},\{u_{15},v_{15},w_{15}\},\{u_{16},v_{16},w_{16}\},\{u_{17},v_{17},w_{17}\},\{u_{18},$ $v_{18},w_{18}\},\{u_{19},v_{19},w_{19}\},\{u_{20}\},\{v\},\{u\}\}$ contains the above four "broken" invalid path fragments.

When we backward search from <$u$, 22> to <$u_0$, 0>, we have that PStemp=$PS[<u, 22>]$ at iteration 1, PStemp=PStemp$\cap_{min}PS[<v, 21>]$ at iteration 2, and PStemp=PStemp$\cap_{min}PS[<u_{20}, 20>]$ at iteration 3. Now, PStemp=$\{\{u_0\},\{u_1, v_1, w_1\},\{u_2, v_2, w_2\},\{u_3, v_3, w_3\},\{u_4, v_4,w_4\},$ $\{u_5, v_5,w_5\},$ $\{u_6\},\{u_7\},\{u_8,w_8\},\{u_9,w_9\},\{u_{10},w_{10}\},\{u_{11},w_{11}\},\{u_{12},w_{12}\},\{u_{13}\},\{u_{14}\},\{u_{15},v_{15},w_{15}\},\{u_{16},v_{16},$ $w_{16}\},$ $\{u_{17},v_{17},w_{17}\},\{u_{18},v_{18},w_{18}\},\{u_{19},v_{19},w_{19}\},\{u_{20}\}\}$.

<$u_{20}$, 20> is the meeting point of the rightmost "parallel path cluster". Since PStemp[19]= $\{u_{19}, v_{19}, w_{19}\}$, the <$u_{20}$, 20> has three parent vertices <$u_{19}$, 19>, <$v_{19}$, 19>, and <$w_{19}$, 19>.

(a)<$u_{19}$, 19> is the rightmost vertex of the valid path fragment $u_{15}$-$u_{16}$-$u_{17}$-$u_{18}$-$u_{19}$ located on the rightmost "parallel path cluster". This valid path fragment $u_{15}$-$u_{16}$-$u_{17}$-$u_{18}$-$u_{19}$ belongs to two valid paths $P_1$ and $P_5$. Therefore, we can backward search from <$u_{19}$, 19> to <$u_0$, 0> along $P_1$ or $P_5$, respectively. In other words, we can choose the parent vertex <$u_{19}$, 19> of <$u_{20}$, 20>, the parent vertex <$u_{12}$, 12> or <$w_{12}$, 12> of <$u_{13}$, 13>, and the parent vertex <$u_5$, 5> or <$v_5$, 5> of <$u_6$, 6> for backward searching.

(b)<$v_{19}$, 19> is the rightmost vertex of the "broken" invalid path fragment $v_{15}$-$v_{16}$-$v_{17}$-$v_{18}$-$v_{19}$ located on the rightmost "parallel path cluster". If we choose the parent vertex <$v_{19}$, 19> of the <$u_{20}$, 20> to backward search, we have PStemp=PStemp$\cap_{min}PS[<v_{19}, 19>]$ at iteration 4. Here, PStemp $=\{\{u_0\},\{v_1,w_1\},\{v_2,w_2\},\{v_3,w_3\},\{v_4,w_4\},\{v_5,w_5\},\{u_6\},\{u_7\},\{u_8\},\{u_9\},\{u_{10}\},\{u_{11}\},$ $\{u_{12}\},\{u_{13}\},\{u_{14}\},\{v_{15}\},\{v_{16}\},\{v_{17}\},\{v_{18}\},\{v_{19}\}\}$. Undoubtedly, the "broken" invalid path fragment $v_{15}$-$v_{16}$-$v_{17}$-$v_{18}$-$v_{19}$ is located not only on an invalid path $P_2$ but also on a valid path $P_6$. For the valid path fragment $v_{15}$-$v_{16}$-$v_{17}$-$v_{18}$-$v_{19}$ located on the valid path $P_6$, we can choose the $v_{19}$, $u_{12}$, and $w_5$ respectively to backward search at each meeting point of the "parallel path cluster". On the contrary, for the invalid path fragment $v_{15}$-$v_{16}$-$v_{17}$-$v_{18}$-$v_{19}$ located on the invalid path $P_2$, we cannot choose the $v_{12}$ to backward search at the meeting point of the "parallel path cluster", because the PStemp at <$u_{13}$, 13> does not contain the $v_{12}$. That is, $v_{12}\notin$ PStemp[12]$\cap PS_{<u_{13},13>}[12]$ due to the $P_{del}=v_8\text{-}v_9\text{-}v_{10}\text{-}v_{11}\text{-}v_{12}$. Therefore, we will choose the parent vertex $<v_{19}, 19>$ of the $<u_{20}, 20>$ and the parent vertex $<u_{12}, 12>$ of the $<u_{13}, 13>$ to backward search, while we cannot choose the parent vertex $<v_{12}, 12>$ of the $<u_{13}, 13>$ to backward search due to $v_{12} \notin \text{PStemp}[12] \cap PS_{<u13,13>}[12]$.

(c)$<w_{19}, 19>$ is the rightmost vertex of the "broken" invalid path fragment $w_{15}\text{-}w_{16}\text{-}w_{17}\text{-}w_{18}\text{-}w_{19}$ located on the rightmost "parallel path cluster". If we choose the parent vertex $<w_{19}, 19>$ of the $<u_{20}, 20>$ to backward search, we have PStemp=PStemp$\cap_{min}PS[<w_{19}, 19>]$ at iteration 4. Here, PStemp $=\{\{u_0\},\{u_1, w_1\},\{u_2, w_2\},\{u_3, w_3\},\{u_4, w_4\},\{u_5, w_5\},\{u_6\},\{u_7\},\{w_8\},\{w_9\},\{w_{10}\}, \{w_{11}\},\{w_{12}\},\{u_{13}\},\{u_{14}\},\{w_{15}\},\{w_{16}\},\{w_{17}\},\{w_{18}\},\{w_{19}\}\}$. Undoubtedly, the "broken" invalid path fragment $w_{15}\text{-}w_{16}\text{-}w_{17}\text{-}w_{18}\text{-}w_{19}$ is located not only on an invalid path $P_4$ but also on a valid path $P_3$. For the valid path fragment $w_{15}\text{-}w_{16}\text{-}w_{17}\text{-}w_{18}\text{-}w_{19}$ located on the valid path $P_3$, we can choose the $w_{19}$, $w_{12}$, and $w_5$ respectively to backward search at each meeting point of the "parallel path cluster". On the contrary, for the invalid path fragment $w_{15}\text{-}w_{16}\text{-}w_{17}\text{-}w_{18}\text{-}w_{19}$ located on the invalid path $P_4$, we cannot choose the $v_{12}$ to backward search at the meeting point of the "parallel path cluster", because the PStemp at $<u_{13}, 13>$ does not contain the $v_{12}$. That is, $v_{12} \notin \text{PStemp}[12] \cap PS_{<u13,13>}[12]$ due to the $P_{del}=v_8\text{-}v_9\text{-}v_{10}\text{-}v_{11}\text{-}v_{12}$. Therefore, we will choose the parent vertex $<w_{19}, 19>$ of the $<u_{20}, 20>$ and the parent vertex $<w_{12}, 12>$ of the $<u_{13}, 13>$ to backward search, while we cannot choose the parent vertex $<v_{12}, 12>$ of the $<u_{13}, 13>$ to backward search due to $v_{12} \notin \text{PStemp}[12] \cap PS_{<u13,13>}[12]$.

As discussion for backward searching above, we find that the "broken" invalid path fragments $u_1\text{-}u_2\text{-}u_3\text{-}u_4\text{-}u_5$ can be visited along the valid path $P_1$, but cannot be visited along the invalid path $P_4$. Similarly, the "broken" invalid path fragments $v_1\text{-}v_2\text{-}v_3\text{-}v_4\text{-}v_5$ can be visited along the valid path $P_5$, but cannot be visited along the invalid path $P_2$. The "broken" invalid path fragments $v_{15}\text{-}v_{16}\text{-}v_{17}\text{-}v_{18}\text{-}v_{19}$ can be visited along the valid path $P_6$, but cannot be visited along the invalid path $P_2$. The "broken" invalid path fragments $w_{15}\text{-}w_{16}\text{-}w_{17}\text{-}w_{18}\text{-}w_{19}$ can be visited along the valid path $P_3$, but cannot be visited along the invalid path $P_4$.

On the other hand, the valid path fragment $w_1\text{-}w_2\text{-}w_3\text{-}w_4\text{-}w_5$ or $u_{15}\text{-}u_{16}\text{-}u_{17}\text{-}u_{18}\text{-}u_{19}$ can be visited along their original valid paths respectively. Further, the path fragments $u_8\text{-}u_9\text{-}u_{10}\text{-}u_{11}\text{-}u_{12}$ and $w_8\text{-}w_9\text{-}w_{11}\text{-}w_{12}\text{-}w_{13}$ are located on the "parallel path cluster" from which the $P_{del}=v_8\text{-}v_9\text{-}v_{10}\text{-}v_{11}\text{-}v_{12}$ is deleted. So they are valid and can be visited along their original valid paths respectively.

So we obtain the following conclusions again. The valid path fragments can be visited along their original valid path. The "broken" invalid path fragments cannot be visited along the original "invalid" path on which the duplicative vertex of $u$ is deleted from the path set PS[$<v, k\text{-}1>$]. Contrarily, the "broken" invalid path fragments as "valid" path fragments can be visited along another original valid path if it exists.

Case 2.3. If $u=v_{17}$, the deleted path $P_{del}=v_{15}\text{-}v_{16}\text{-}v_{17}\text{-}v_{18}\text{-}v_{19}$ will be removed from PStemp[$<v, 21>$], resulting in that four "broken" invalid path fragments $v_1\text{-}v_2\text{-}v_3\text{-}v_4\text{-}v_5$, $w_1\text{-}w_2\text{-}w_3\text{-}w_4\text{-}w_5$, $u_8\text{-}u_9\text{-}u_{10}\text{-}u_{11}\text{-}u_{12}$, and $v_8\text{-}v_9\text{-}v_{10}\text{-}v_{11}\text{-}v_{12}$ remained in PStemp[$<v, 21>$]. So $PS[<u, 22>]=\{\{u_0\}, \{u_1,v_1,w_1\},\{u_2,v_2,w_2\},\{u_3,v_3,w_3\},\{u_4,v_4,w_4\},\{u_5,v_5,w_5\},\{u_6\},\{u_7\},\{u_8,v_8,w_8\},\{u_9,v_9,w_9\},\{u_{10},v_{10},w_{10}\},\{u_{11},v_{11},w_{11}\},\{u_{12},v_{12},w_{12}\},\{u_{13}\},\{u_{14}\},\{u_{15},w_{15}\},\{u_{16},w_{16}\},\{u_{17},w_{17}\}, \{u_{18},w_{18}\},\{u_{19},w_{19}\},\{u_{20}\},\{v\},\{u\}\}$ contains the above four "broken" invalid path fragments.

When we backward search from $<u, 22>$ to $<u_0, 0>$, we have that PStemp=$PS[<u, 22>]$ at iteration 1, PStemp=PStemp$\cap_{min}PS[<v, 21>]$ at iteration 2, and PStemp=PStemp$\cap_{min}PS[<u_{20}, 20>]$ at iteration 3. Now, PStemp=$\{\{u_0\},\{u_1,v_1,w_1\},\{u_2,v_2,w_2\},\{u_3,v_3,w_3\},\{u_4,v_4,w_4\}, \{u_5,v_5,w_5\},$

$\{u_6\},\{u_7\},\{u_8,v_8,w_8\},\{u_9,v_9,w_9\},\{u_{10},v_{10},w_{10}\},\{u_{11},v_{11},w_{11}\},\{u_{12},v_{12},w_{12}\},\{u_{13}\},\{u_{14}\},$
$\{u_{15},w_{15}\},\{u_{16},w_{16}\},\{u_{17},w_{17}\},\{u_{18},w_{18}\},\{u_{19},w_{19}\},\{u_{20}\}\}$.

$<u_{20}, 20>$ is the meeting point of the rightmost "parallel path cluster". Since PStemp[19]= $\{u_{19},w_{19}\}$, the $<u_{20}, 20>$ has two parent vertices $<u_{19}, 19>$ and $<w_{19}, 19>$.

(a)$<u_{19}, 19>$ is the rightmost vertex of the valid path fragment $u_{15}$-$u_{16}$-$u_{17}$-$u_{18}$-$u_{19}$ located on the rightmost "parallel path cluster" from which the $P_{del}$=$v_{15}$-$v_{16}$-$v_{17}$-$v_{18}$-$v_{19}$ is deleted. This valid path fragment $u_{15}$-$u_{16}$-$u_{17}$-$u_{18}$-$u_{19}$ belongs to two valid paths $P_1$ and $P_5$. Therefore, we can backward search from $<u_{19}, 19>$ to $<u_0, 0>$ along $P_1$ or $P_5$, respectively. In other words, we can choose the parent vertex $<u_{19}, 19>$ of $<u_{20}, 20>$, the parent vertex $<u_{12}, 12>$ or $<w_{12}, 12>$ of $<u_{13}, 13>$, and the parent vertex $<u_5, 5>$ or $<v_5, 5>$ of $<u_6, 6>$ for backward searching.

That is, PStemp=PStemp∩$_{min}$PS[$<u_{20}, 20>$]={$\{u_0\},\{u_1,v_1,w_1\},\{u_2,v_2,w_2\},\{u_3,v_3,w_3\}, \{u_4,v_4,w_4\},\{u_5,v_5,w_5\},\{u_6\},\{u_7\},\{u_8,v_8,w_8\},\{u_9,v_9,w_9\},\{u_{10},v_{10},w_{10}\},\{u_{11},v_{11},w_{11}\}, \{u_{12},v_{12},w_{12}\},\{u_{13}\},\{u_{14}\},\{u_{15},w_{15}\},\{u_{16},w_{16}\},\{u_{17},w_{17}\},\{u_{18},w_{18}\},\{u_{19},w_{19}\},\{u_{20}\}\}$.

PStemp=PStemp∩$_{min}$PS[$<u_{19}, 19>$]={$\{u_0\},\{u_1,v_1\},\{u_2,v_2\},\{u_3,v_3\},\{u_4,v_4\},\{u_5,v_5\}, \{u_6\},\{u_7\},\{u_8,w_8\},\{u_9,w_9\},\{u_{10},w_{10}\},\{u_{11},w_{11}\},\{u_{12},w_{12}\},\{u_{13}\},\{u_{14}\},\{u_{15}\},\{u_{16}\},\{u_{17}\}, \{u_{18}\},\{u_{19}\}\}$.

PStemp=PStemp∩$_{min}$PS[$<u_{13}, 13>$]={$\{u_0\},\{u_1,v_1\},\{u_2,v_2\},\{u_3,v_3\},\{u_4,v_4\},\{u_5,v_5\},\{u_6\}, \{u_7\},\{u_8,w_8\},\{u_9,w_9\},\{u_{10},w_{10}\},\{u_{11},w_{11}\},\{u_{12},w_{12}\},\{u_{13}\}\}$.

PStemp=PStemp∩$_{min}$PS[$<u_{12}, 12>$]={$\{u_0\},\{u_1\},\{u_2\},\{u_3\},\{u_4\},\{u_5\},\{u_6\},\{u_7\},\{u_8\},\{u_9\}, \{u_{10}\},\{u_{11}\},\{u_{12}\}\}$.

PStemp=PStemp∩$_{min}$PS[$<w_{12}, 12>$]={$\{u_0\},\{v_1\},\{v_2\},\{v_3\},\{v_4\},\{v_5\},\{u_6\},\{u_7\},\{w_8\},\{w_9\}, \{w_{10}\},\{w_{11}\},\{w_{12}\}\}$.

(b)$<w_{19}, 19>$ is the rightmost vertex of the valid path fragment $w_{15}$-$w_{16}$-$w_{17}$-$w_{18}$-$w_{19}$ located on the rightmost "parallel path cluster" from which the $P_{del}$=$v_{15}$-$v_{16}$-$v_{17}$-$v_{18}$-$v_{19}$ is deleted. This valid path fragment $w_{15}$-$w_{16}$-$w_{17}$-$w_{18}$-$w_{19}$ belongs to two valid paths $P_3$ and $P_4$. Therefore, we can backward search from $<w_{19}, 19>$ to $<u_0, 0>$ along $P_3$ or $P_4$, respectively. In other words, we can choose the parent vertex $<w_{19}, 19>$ of $<u_{20}, 20>$, the parent vertex $<v_{12}, 12>$ or $<w_{12}, 12>$ of $<u_{13}, 13>$, and the parent vertex $<u_5, 5>$ or $<w_5, 5>$ of $<u_6, 6>$ for backward searching.

That is, PStemp=PStemp∩$_{min}$PS[$<u_{20}, 20>$]={$\{u_0\},\{u_1,v_1,w_1\},\{u_2,v_2,w_2\},\{u_3,v_3,w_3\},\{u_4,v_4,w_4\},\{u_5,v_5,w_5\},\{u_6\},\{u_7\},\{u_8,v_8,w_8\},\{u_9,v_9,w_9\},\{u_{10},v_{10},w_{10}\},\{u_{11},v_{11},w_{11}\},\{u_{12},v_{12},w_{12}\}, \{u_{13}\},\{u_{14}\},\{u_{15},w_{15}\},\{u_{16},w_{16}\},\{u_{17},w_{17}\},\{u_{18},w_{18}\},\{u_{19},w_{19}\},\{u_{20}\}\}$.

PStemp=PStemp∩$_{min}$PS[$<w_{19}, 19>$]={$\{u_0\},\{u_1,w_1\},\{u_2,w_2\},\{u_3,w_3\},\{u_4,w_4\},\{u_5,w_5\},\{u_6\}, \{u_7\},\{v_8,w_8\},\{v_9,w_9\},\{v_{10},w_{10}\},\{v_{11},w_{11}\},\{v_{12},w_{12}\},\{u_{13}\},\{u_{14}\},\{w_{15}\},\{w_{16}\},\{w_{17}\},\{w_{18}\}, \{w_{19}\}\}$.

PStemp=PStemp∩$_{min}$PS[$<u_{13}, 13>$]={$\{u_0\},\{u_1,w_1\},\{u_2,w_2\},\{u_3,w_3\},\{u_4,w_4\},\{u_5,w_5\},\{u_6\}, \{u_7\},\{v_8,w_8\},\{v_9,w_9\},\{v_{10},w_{10}\},\{v_{11},w_{11}\}, \{v_{12},w_{12}\},\{u_{13}\}\}$.

PStemp=PStemp∩$_{min}$PS[$<v_{12}, 12>$]={$\{u_0\},\{u_1\},\{u_2\},\{u_3\},\{u_4\},\{u_5\},\{u_6\},\{u_7\},\{v_8\},\{v_9\}, \{v_{10}\},\{v_{11}\},\{v_{12}\}\}$.

PStemp=PStemp∩$_{min}$PS[$<w_{12}, 12>$]={$\{u_0\},\{w_1\},\{w_2\},\{w_3\},\{w_4\},\{w_5\},\{u_6\},\{u_7\},\{w_8\},\{w_9\}, \{w_{10}\},\{w_{11}\}, \{w_{12}\}\}$.

(c)Since $<v_{19}, 19>$ is not in PStemp=PStemp∩$_{min}$PS[$<u_{20}, 20>$], i.e., $v_{19}\notin$PStemp[19]∩PS$_{<u_{20},20>}$[19], the parent vertex $<v_{19}, 19>$ of $<u_{20}, 20>$, the parent vertex $<u_{12}, 12>$ or $<v_{12}, 12>$ of $<u_{13}, 13>$, and the parent vertex $<v_5, 5>$ or $<w_5, 5>$ of $<u_6, 6>$ along the two invalid basic paths $P_2$ and $P_6$ respectively cannot be visited for backward searching.

In other words, only can the $u_8$-$u_9$-$u_{10}$-$u_{11}$-$u_{12}$ or $v_8$-$v_9$-$v_{10}$-$v_{11}$-$v_{12}$ be visited along the valid basic path $P_1$ or $P_4$ respectively. So is the $v_1$-$v_2$-$v_3$-$v_4$-$v_5$ or $w_1$-$w_2$-$w_3$-$w_4$-$w_5$, for which can be visited only along the valid basic path $P_5$ or $P_3$ respectively.

On the other hand, the $u_8$-$u_9$-$u_{10}$-$u_{11}$-$u_{12}$ or $v_8$-$v_9$-$v_{10}$-$v_{11}$-$v_{12}$ cannot be visited along the invalid basic path $P_6$ or $P_2$ respectively. And the $v_1$-$v_2$-$v_3$-$v_4$-$v_5$ or $w_1$-$w_2$-$w_3$-$w_4$-$w_5$ cannot be visited along the invalid basic path $P_2$ or $P_6$ respectively.

Case 3. There are exactly 27=3×3×3 unique basic paths in *PS*[<$u_{20}$, 20>]. In this case, any a path fragment can be visited for backward searching as the component of the valid basic path after the deleted path $P_{del}$ has been removed from *PS*[<$v$, $k$-1>].

Case 4. If $u$=$u_7$, which is a vertex on a unique "serial path segment" on the basic path from the initial vertex ***S*** to the <$v$, $k$-1>, deleting the $u_7$ from *PS*[<$v$, $k$-1>] will result in $PStemp_{<v,k-1>}[7]$ =∅. So PStemp[<$v$, $k$-1>]={∅}. In this case, there does not exist any basic path from the initial vertex ***S*** through <$v$, $k$-1> to the <$u$, $k$>.

Case 5. If $u$=$u_9$, $u$=$v_{10}$, and $u$=$w_{11}$, i.e., each <$u$, $i$> is located on a distinct parallel path segment of the same unique "parallel path cluster" on the basic path from the initial vertex ***S*** to the <$v$, $k$-1>, PStemp[<$v$, $k$-1>] will contain an empty set ∅ at the same unique cluster after all <$u$, $i$> had been deleted from *PS*[<$v$, $k$-1>]. That is, PStemp[<$v$, 21>]={{$u_0$}, {$u_1$,$v_1$,$w_1$},{$u_2$,$v_2$,$w_2$}, {$u_3$,$v_3$, $w_3$},{$u_4$,$v_4$,$w_4$},{$u_5$,$v_5$,$w_5$},{$u_6$},{$u_7$},{},{},{},{},{$u_{13}$},{$u_{14}$},{$u_{15}$,$v_{15}$,$w_{15}$},{$u_{16}$,$v_{16}$, $w_{16}$},{$u_{17}$,$v_{17}$,$w_{17}$},{$u_{18}$,$v_{18}$,$w_{18}$},{$u_{19}$,$v_{19}$,$w_{19}$},{$u_{20}$},{$v$}}. So PStemp[<$v$, $k$-1>]={∅}. In this case, there does not exist any basic path from the initial vertex ***S*** through <$v$, $k$-1> to <$u$, $k$>.

Consider the situations shown in **Figure 8**.

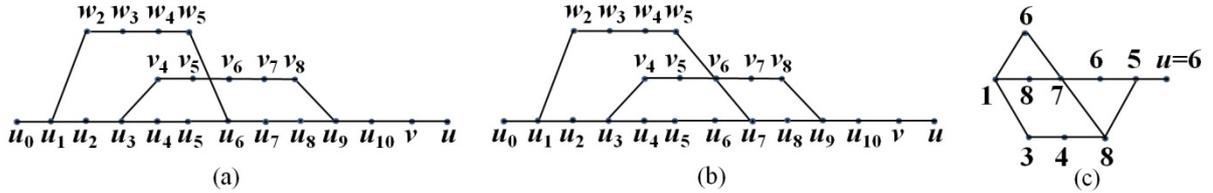

**Figure 8**   Duplicative vertex of <$u$, $k$> on the path set PS[<$v$, $k$-1>]

Without loss of generality assume that all vertices are different except for vertex *u* in **Figure 8(a)** and **8(b)**.

Consider the situations in **Figure** 8(a). Case 1. $u$=$u_{10}$. Since the vertex <$u_{10}$, 10> is located on a unique "serial path segment" on the basic path from the initial vertex ***S*** to the <$v$, $k$-1>, PStemp[<$v$, $k$-1>]={∅} after deleting the $u_{10}$ from *PS*[<$v$, $k$-1>]. Case 2. $u$=$u_8$. Since the vertex <$u_8$, 8> is located on a parallel path segment $u_4$-$u_5$-$u_6$-$u_7$-$u_8$ of a "parallel path cluster" {$u_4$-$u_5$-$u_6$-$u_7$-$u_8$; $v_4$-$v_5$-$v_6$-$v_7$-$v_8$} on the basic path from the initial vertex ***S*** to the <$v$, $k$-1>, the path segment $u_4$-$u_5$-$u_6$-$u_7$-$u_8$ will be deleted from *PS*[<$v$, $k$-1>]. Since the vertex <$u_8$, 8> is located on a serial path segment $u_6$-$u_7$-$u_8$ following a "parallel path cluster" {$u_2$-$u_3$-$u_4$-$u_5$; $w_2$-$w_3$-$w_4$-$w_5$} on the basic path from the initial vertex ***S*** to the <$v$, $k$-1>, the path segment $u_2$-$u_3$-$u_4$-$u_5$ and $w_2$-$w_3$-$w_4$-$w_5$ will be deleted from *PS*[<$v$, $k$-1>]. However, the path segment $u_2$-$u_3$ will be replenished due to the reason of the left action field of <$v_4$, 4>. Therefore the parallel path segment $v_4$-$v_5$-$v_6$-$v_7$-$v_8$ along the valid basic path $u_0$-$u_1$-$u_2$-$u_3$-$v_4$-$v_5$-$v_6$-$v_7$-$v_8$-$u_9$-$u_{10}$-$v$ can be visited in Backward Searches. Case 3. $u$=$u_5$. Since the vertex <$u_5$, 5> is located not only on a parallel path segment $u_4$-$u_5$-$u_6$-$u_7$-$u_8$ of a "parallel path cluster" {$u_4$-$u_5$-$u_6$-$u_7$-$u_8$; $v_4$-$v_5$-$v_6$-$v_7$-$v_8$} but also on a parallel path segment $u_2$-$u_3$-$u_4$-$u_5$ of a "parallel path cluster" {$u_2$-$u_3$-$u_4$-$u_5$; $w_2$-$w_3$-$w_4$-$w_5$}, both the valid path segments $v_4$-$v_5$-$v_6$-$v_7$-$v_8$ and $w_2$-$w_3$-$w_4$-$w_5$ can be visited along distinct basic paths, which are $u_0$-$u_1$-$u_2$-$u_3$-$v_4$-$v_5$-$v_6$-$v_7$-$v_8$-$u_9$-$u_{10}$-$v$ and $u_0$-$u_1$-$w_2$-$w_3$-$w_4$-$w_5$-$u_6$-$u_7$-$u_8$-$u_9$-$u_{10}$-$v$, respectively in Backward Searches. Here, the path segment $u_2$-$u_3$ will be replenished due to the reason of the left action field of $<v_4, 4>$. Case 4. $u=u_2$. Since the vertex $<u_2, 2>$ is located on a parallel path segment $u_2$-$u_3$-$u_4$-$u_5$ of a "parallel path cluster" {$u_2$-$u_3$-$u_4$-$u_5$; $w_2$-$w_3$-$w_4$-$w_5$} on the basic path from the initial vertex $S$ to the $<v, k$-$1>$, the path segment $u_2$-$u_3$-$u_4$-$u_5$ will be deleted from $PS[<v, k$-$1>]$. Since the vertex $<u_2, 2>$ is located on a serial path segment $u_1$-$u_2$-$u_3$ ahead of a "parallel path cluster" {$u_4$-$u_5$-$u_6$-$u_7$-$u_8$; $v_4$-$v_5$-$v_6$-$v_7$-$v_8$} on the basic path from the initial vertex $S$ to the $<v, k$-$1>$, the path segment $u_4$-$u_5$-$u_6$-$u_7$-$u_8$ and $v_4$-$v_5$-$v_6$-$v_7$-$v_8$ will be deleted from $PS[<v, k$-$1>]$. However, the path segment $u_6$-$u_7$-$u_8$ will be replenished due to the reason of the right action field of $<w_5, 5>$. Therefore the parallel path segment $w_2$-$w_3$-$w_4$-$w_5$ along the valid basic path $u_0$-$u_1$-$w_2$-$w_3$-$w_4$-$w_5$ -$u_6$-$u_7$-$u_8$-$u_9$-$u_{10}$-$v$ can be visited in Backward Searches. Case 5. $u=v_6$. Since the vertex $<v_6, 6>$ is located on a parallel path segment $v_4$-$v_5$-$v_6$-$v_7$-$v_8$ of a "parallel path cluster" {$u_4$-$u_5$-$u_6$-$u_7$-$u_8$; $v_4$-$v_5$-$v_6$-$v_7$-$v_8$} on the basic path from the initial vertex $S$ to the $<v, k$-$1>$, the path segment $v_4$-$v_5$-$v_6$-$v_7$-$v_8$ will be deleted from $PS[<v, k$-$1>]$. On the other hand, the valid path segment $u_4$-$u_5$-$u_6$-$u_7$-$u_8$ along its original basic path $u_0$-$u_1$-$u_2$-$u_3$-$u_4$-$u_5$-$u_6$-$u_7$-$u_8$-$u_9$-$u_{10}$-$v$ can be visited in Backward Searches. Further, the valid path segment $w_2$-$w_3$-$w_4$-$w_5$ can be visited along its original basic path $u_0$-$u_1$-$w_2$-$w_3$-$w_4$-$w_5$-$u_6$-$u_7$-$u_8$-$u_9$-$u_{10}$-$v$ in Backward Searches.

    Consider the situations in **Figure** 8(b). Case 1. $u=u_8$. Since the vertex $<u_8, 8>$ is located on a parallel path segment $u_4$-$u_5$-$u_6$-$u_7$-$u_8$ of a "parallel path cluster" {$u_4$-$u_5$-$u_6$-$u_7$-$u_8$; $v_4$-$v_5$-$v_6$-$v_7$-$v_8$} on the basic path from the initial vertex $S$ to the $<v, k$-$1>$, the path segment $u_4$-$u_5$-$u_6$-$u_7$-$u_8$ will be deleted from $PS[<v, k$-$1>]$. Since the vertex $<u_8, 8>$ is located on a serial path segment $u_6$-$u_7$-$u_8$ following a "parallel path cluster" {$u_2$-$u_3$-$u_4$-$u_5$-$u_6$; $w_2$-$w_3$-$w_4$-$w_5$-$v_6$} on the basic path from the initial vertex $S$ to the $<v, k$-$1>$, the path segment $u_2$-$u_3$-$u_4$-$u_5$-$u_6$ and $w_2$-$w_3$-$w_4$-$w_5$-$v_6$ will be deleted from $PS[<v, k$-$1>]$. However, the path segment $u_2$-$u_3$ will be replenished due to the reason of the left action field of $<v_4, 4>$ and the vertex $<v_6, 6>$ will be replenished due to the reason of the left action field of $<v_7, 7>$. Therefore the parallel path segment $v_4$-$v_5$-$v_6$-$v_7$-$v_8$ along the valid basic path $u_0$-$u_1$-$u_2$-$u_3$-$v_4$-$v_5$-$v_6$-$v_7$-$v_8$-$u_9$-$u_{10}$-$v$, and the parallel path segment $w_2$-$w_3$-$w_4$-$w_5$-$v_6$ along the valid basic path $u_0$-$u_1$-$w_2$-$w_3$-$w_4$-$w_5$-$v_6$-$v_7$-$v_8$-$u_9$-$u_{10}$-$v$ can be visited in Backward Searches. Case 2. $u=u_5$. In this case, the path segment $u_4$-$u_5$-$u_6$ will be deleted from $PS[<v, k$-$1>]$. As a result, there are four basic paths $u_0$-$u_1$-$u_2$-$u_3$-$v_4$-$v_5$-$v_6$-$u_7$-$u_8$-$u_9$-$u_{10}$-$v$, $u_0$-$u_1$-$u_2$-$u_3$-$v_4$-$v_5$-$v_6$-$v_7$-$v_8$-$u_9$-$u_{10}$-$v$, $u_0$-$u_1$-$w_2$-$w_3$-$w_4$-$w_5$-$v_6$-$u_7$-$u_8$-$u_9$-$u_{10}$-$v$, and $u_0$-$u_1$-$w_2$-$w_3$-$w_4$-$w_5$-$v_6$-$v_7$-$v_8$-$u_9$-$u_{10}$-$v$ can be visited respectively in Backward Searches. Case 3. $u=u_2$. In this case, the vertex $<u_2, 2>$ will be deleted from $PS[<v, k$-$1>]$, resulting in the path segment $u_3$-$v_4$-$v_5$ and $u_3$-$u_4$-$u_5$-$u_6$ deleted from $PS[<v, k$-$1>]$. As a result, there are two basic paths $u_0$-$u_1$-$w_2$-$w_3$-$w_4$-$w_5$-$v_6$-$u_7$-$u_8$-$u_9$-$u_{10}$-$v$ and $u_0$-$u_1$-$w_2$-$w_3$-$w_4$-$w_5$-$v_6$-$v_7$-$v_8$-$u_9$-$u_{10}$-$v$ can be visited respectively in Backward Searches. Case 4. $u=v_6$. The vertex $<v_6, 6>$ is located not only on a parallel path segment $v_4$-$v_5$-$v_6$-$v_7$-$v_8$ of a "parallel path cluster" {$u_4$-$u_5$-$u_6$-$u_7$-$u_8$; $v_4$-$v_5$-$v_6$-$v_7$-$v_8$} but also on a parallel path segment $w_2$-$w_3$-$w_4$-$w_5$-$v_6$ of a "parallel path cluster" {$u_2$-$u_3$-$u_4$-$u_5$-$u_6$; $w_2$-$w_3$-$w_4$-$w_5$-$v_6$}. Therefore, the vertex $<v_6, 6>$ deleted from $PS[<v, k$-$1>]$ will result in that these path segments $v_4$-$v_5$, $w_2$-$w_3$-$w_4$-$w_5$, and $v_7$-$v_8$ are deleted from $PS[<v, k$-$1>]$. So that the valid path segment $u_2$-$u_3$-$u_4$-$u_5$-$u_6$-$u_7$-$u_8$ can be visited in Backward Searches along its original basic path $u_0$-$u_1$-$u_2$-$u_3$-$u_4$-$u_5$-$u_6$-$u_7$-$u_8$ -$u_9$-$u_{10}$-$v$. In other words, the parent vertex $<v_8, 8>$ of $<u_9, 9>$ and the parent vertex $<v_6, 6>$ of $<u_7, 7>$ respectively cannot be chosen for backward searching due to their absence in PStemp.

Consider the situations in **Figure** 8(c). We have $PS[<1, 0>]=\{\{1\}\}$, $PS[<3, 1>]=\{\{1\},\{3\}\}$, $PS[<6, 1>]=\{\{1\},\{6\}\}$, $PS[<8, 1>]=\{\{1\},\{8\}\}$, $PS[<4, 2>]=\{\{1\},\{3\},\{4\}\}$, $PS[<7, 2>]=\{\{1\}, \{6, 8\},\{7\}\}$, $PS[<6, 3>]=\{\{1\},\{8\},\{7\},\{6\}\}$, $PS[<8, 3>]=\{\{1\},\{3,6\},\{4,7\},\{8\}\}$, $PS[<5, 4>]= \{\{1\},\{3,6,8\},\{4,7\}, \{6,8\},\{5\}\}$. When we consider the vertex $<u, k> = <6, 5>$ joining the $PS[<5, 4>]$, we find that the vertex 6 repeats twice in $PS[<5, 4>]$. After deleting the duplicative vertices of $u=6$, we obtain PStemp$[<5, 4>]=\{\{1\},\{3,8\},\{4,7\},\{8\},\{5\}\}$. If we continue to run the "While (glag2)" loop in the **CM** operator, we will obtain PStemp$[<5, 4>]=\{\{1\},\{3\},\{4\},\{8\},\{5\}\}$. Now, we can modify our **CM** operator so that we don't need to run the "While (glag2)" loop. During the process of backward searching, we have initially PStemp=$PS[<6, 5>]=\{\{1\},\{3,8\},\{4,7\},\{8\},\{5\},\{6\}\}$ at iteration 1, PStemp=PStemp$\cap_{min}PS[<5, 4>]=\{\{1\},\{3,8\},\{4,7\},\{8\},\{5\}\}$ at iteration 2, and PStemp=PStemp $\cap_{min}PS[<8, 3>]=\{\{1\},\{3\}, \{4,7\},\{8\}\}$ at iteration 3. In this case, the parent vertex $<7, 2>$ of $<8, 3>$ will not be chosen to backward search due to $\varnothing \in$ PStemp$\cap_{min}PS[<7, 2>]$, i.e., PStemp=PStemp $\cap_{min}PS[<7, 2>]=\{\{1\}, \{ \}, \{7\}\}$ at iteration 4. On the other hand, the parent vertex $<4, 2>$ of $<8, 3>$ can be chosen to backward search due to $\varnothing \notin$ PStemp$\cap_{min}PS[<4, 2>]$, i.e., PStemp=PStemp$\cap_{min}$ $PS[<4, 2>]=\{\{1\}, \{3\},\{4\}\}$ at iteration 4.

As discussed above, the "broken" invalid path fragment 8 caused by the deletion of the $<6, 3>$, which is located on a unique basic path 1-8-7-6-5, cannot be found to visit in Backward Searches due to the duplicative vertex $<6, 3>$ of $u$ located on the rightmost "parallel path cluster". Further, the "broken" invalid path fragment 8 caused by the deletion of the $<6, 1>$, which is located on a unique basic path 1-6-7-8-5, will not be chosen to visit in Backward Searches due to the duplicative vertex $<6, 1>$ of $u$ located on the leftmost "parallel path cluster". So we must have PStemp$\cap_{min}PS[<7, 2>]=\{\varnothing\}$. Besides, the valid path fragment 8, which is located on a unique basic path 1-3-4-8-5, can be chosen to visit in Backward Searches along the unique valid basic path 1-3-4-8-5. Therefore, when we backward search from $<6, 5>$ to $<1, 0>$, we can choose the parent vertex $<8, 3>$ of $<5, 4>$ for backward searching, but we cannot find the parent vertex $<6, 3>$ of $<5, 4>$ in PStemp$\cap_{min}PS[<6, 3>]$ for backward searching. Further, we can choose the parent vertex $<4, 2>$ of $<8, 3>$ for backward searching, but we will not choose the parent vertex $<7, 2>$ of $<8, 3>$ for backward searching due to PStemp$\cap_{min}PS[<7, 2>]=\{\varnothing\}$. As a result, we don't need to worry about the "broken" invalid path fragments located on some unique "invalid" basic paths from which the path segment $P_{del}$ of the duplicative vertices of $u$ is deleted.

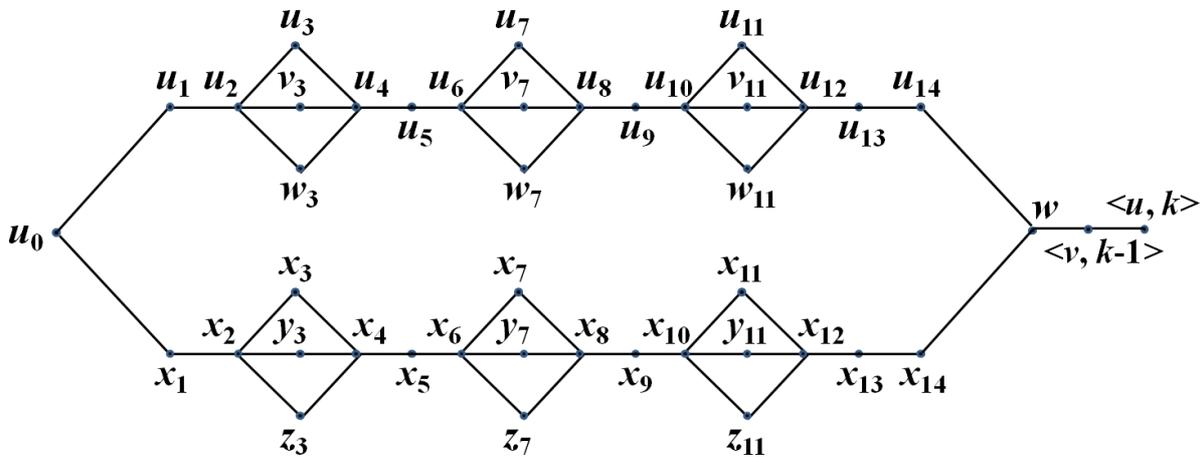

**Figure 9** Duplicative vertex of $<u, k>=<u, 17>$ on the path set $PS[<v, 16>]$

Consider the situations in **Figure** 9. Supposed that there are exactly 6 unique basic paths in $PS[<v, 16>]$. $P_1=u_0$-$u_1$-$u_2$-$u_3$-$u_4$-$u_5$-$u_6$-$u_7$-$u_8$-$u_9$-$u_{10}$-$u_{11}$-$u_{12}$-$u_{13}$-$u_{14}$-$w$-$v$, $P_2=u_0$-$u_1$-$u_2$-$v_3$-$u_4$-$u_5$-$u_6$-$v_7$-$u_8$-$u_9$-$u_{10}$-$v_{11}$-$u_{12}$-$u_{13}$-$u_{14}$-$w$-$v$, $P_3=u_0$-$u_1$-$u_2$-$w_3$-$u_4$-$u_5$-$u_6$-$w_7$-$u_8$-$u_9$-$u_{10}$-$w_{11}$-$u_{12}$-$u_{13}$-$u_{14}$-$w$-$v$, $P_4=u_0$-$x_1$-$x_2$-$x_3$-$x_4$-$x_5$-$x_6$-$x_7$-$x_8$-$x_9$-$x_{10}$-$x_{11}$-$x_{12}$-$x_{13}$-$x_{14}$-$w$-$v$, $P_5=u_0$-$x_1$-$x_2$-$y_3$-$x_4$-$x_5$-$x_6$-$y_7$-$x_8$-$x_9$-$x_{10}$-$y_{11}$-$x_{12}$-$x_{13}$-$x_{14}$-$w$-$v$, $P_6=u_0$-$x_1$-$x_2$-$z_3$-$x_4$-$x_5$-$x_6$-$z_7$-$x_8$-$x_9$-$x_{10}$-$z_{11}$-$x_{12}$-$x_{13}$-$x_{14}$-$w$-$v$.

In this case, we must have $PS[<u_{11}, 11>]=\{\{u_0\},\{u_1\},\{u_2\},\{u_3\},\{u_4\},\{u_5\},\{u_6\},\{u_7\},\{u_8\}, \{u_9\}, \{u_{10}\},\{u_{11}\}\}$, $PS[<v_{11}, 11>]=\{\{u_0\},\{u_1\},\{u_2\},\{v_3\},\{u_4\},\{u_5\},\{u_6\},\{v_7\},\{u_8\},\{u_9\},\{u_{10}\},\{v_{11}\}\}$, $PS[<w_{11}, 11>]=\{\{u_0\},\{u_1\},\{u_2\},\{w_3\},\{u_4\},\{u_5\},\{u_6\},\{w_7\},\{u_8\},\{u_9\},\{u_{10}\},\{w_{11}\}\}$, $PS[<x_{11}, 11>]=\{\{u_0\},\{x_1\},\{x_2\},\{x_3\},\{x_4\},\{x_5\},\{x_6\},\{x_7\},\{x_8\},\{x_9\},\{x_{10}\},\{x_{11}\}\}$, $PS[<y_{11}, 11>]=\{\{u_0\}, \{x_1\},\{x_2\},\{y_3\},\{x_4\},\{x_5\},\{x_6\},\{y_7\},\{x_8\},\{x_9\},\{x_{10}\},\{y_{11}\}\}$, $PS[<z_{11}, 11>]=\{\{u_0\},\{x_1\},\{x_2\}, \{z_3\},\{x_4\},\{x_5\},\{x_6\},\{z_7\},\{x_8\},\{x_9\},\{x_{10}\},\{z_{11}\}\}$.

Therefore, $PS[<u_{12}, 12>]=\{\{u_0\},\{u_1\},\{u_2\},\{u_3,v_3,w_3\},\{u_4\},\{u_5\},\{u_6\},\{u_7,v_7,w_7\},\{u_8\}, \{u_9\},\{u_{10}\},\{u_{11},v_{11},w_{11}\},\{u_{12}\}\}$, $PS[<u_{13}, 13>]=\{\{u_0\},\{u_1\},\{u_2\},\{u_3,v_3,w_3\}, \{u_4\},\{u_5\},\{u_6\}, \{u_7,v_7,w_7\},\{u_8\},\{u_9\},\{u_{10}\},\{u_{11},v_{11},w_{11}\},\{u_{12}\},\{u_{13}\}\}$, $PS[<u_{14}, 14>]=\{\{u_0\},\{u_1\},\{u_2\}, \{u_3,v_3,w_3\}, \{u_4\},\{u_5\},\{u_6\}, \{u_7,v_7,w_7\},\{u_8\},\{u_9\},\{u_{10}\},\{u_{11},v_{11},w_{11}\},\{u_{12}\},\{u_{13}\},\{u_{14}\}\}$, $PS[<x_{12}, 12>]=\{\{u_0\},\{x_1\},\{x_2\},\{x_3,y_3,z_3\},\{x_4\},\{x_5\},\{x_6\},\{x_7,y_7,z_7\},\{x_8\},\{x_9\},\{x_{10}\},\{x_{11}, y_{11},z_{11}\},\{x_{12}\}\}$, $PS[<x_{13}, 13>]=\{\{u_0\},\{x_1\},\{x_2\},\{x_3,y_3,z_3\},\{x_4\},\{x_5\},\{x_6\},\{x_7,y_7,z_7\},\{x_8\}, \{x_9\},\{x_{10}\},\{x_{11},y_{11},z_{11}\},\{x_{12}\},\{x_{13}\}\}$, $PS[<x_{14}, 14>]=\{\{u_0\},\{x_1\},\{x_2\},\{x_3,y_3,z_3\},\{x_4\},\{x_5\}, \{x_6\},\{x_7,y_7,z_7\},\{x_8\}, \{x_9\}, \{x_{10}\}, \{x_{11},y_{11},z_{11}\},\{x_{12}\},\{x_{13}\},\{x_{14}\}\}$.

$PS[<w, 15>]=\{\{u_0\},\{u_1,x_1\},\{u_2,x_2\},\{u_3, v_3, w_3,x_3,y_3,z_3\},\{u_4,x_4\},\{u_5,x_5\},\{u_6,x_6\},\{u_7,v_7, w_7,x_7,y_7,z_7\},\{u_8,x_8\},\{u_9,x_9\},\{u_{10},x_{10}\},\{u_{11}, v_{11}, w_{11},x_{11},y_{11},z_{11}\},\{u_{12},x_{12}\},\{u_{13},x_{13}\},\{u_{14},x_{14}\} ,\{w\}\}$, $PS[<v, k\text{-}1>]=PS[<v, 16>]=\{\{u_0\},\{u_1,x_1\},\{u_2,x_2\},\{u_3, v_3, w_3,x_3,y_3,z_3\},\{u_4,x_4\},\{u_5,x_5\}, \{u_6,x_6\},\{u_7, v_7, w_7,x_7,y_7,z_7\},\{u_8,x_8\},\{u_9,x_9\},\{u_{10},x_{10}\},\{u_{11}, v_{11}, w_{11},x_{11},y_{11},z_{11}\},\{u_{12},x_{12}\}, \{u_{13},x_{13}\},\{u_{14},x_{14}\},\{w\},\{v\}\}$.

Next, we will consider the $<u, k>$ joining $PS[<v, k\text{-}1>]$. Without loss of generality assume that $u=u_3=v_7=w_{11}=x_3=y_7=z_{11}$. So that $PStemp[<v, k\text{-}1>]=\{\{u_0\},\{u_1,x_1\},\{u_2,x_2\},\{v_3, w_3,y_3,z_3\},\{u_4,x_4\}, \{u_5,x_5\},\{u_6,x_6\},\{u_7,w_7,x_7,z_7\},\{u_8,x_8\},\{u_9,x_9\},\{u_{10},x_{10}\},\{u_{11}, v_{11},x_{11},y_{11}\},\{u_{12},x_{12}\},\{u_{13},x_{13}\}, \{u_{14},x_{14}\},\{w\},\{v\}\}$, does not contain $\emptyset$.

Then we call the **CHECK** operator to judge the validity of $PStemp[<v, k\text{-}1>]$. $PStemp[<w, k\text{-}2>]= PS[<w, k\text{-}2>]\cap_{min} PStemp[<v, k\text{-}1>]=\{\{u_0\},\{u_1,x_1\},\{u_2,x_2\},\{v_3, w_3,y_3,z_3\},\{u_4,x_4\},\{u_5,x_5\}, \{u_6,x_6\},\{u_7,w_7,x_7,z_7\},\{u_8,x_8\},\{u_9,x_9\},\{u_{10},x_{10}\},\{u_{11}, v_{11},x_{11},y_{11}\},\{u_{12},x_{12}\},\{u_{13},x_{13}\},\{u_{14},x_{14} \},\{w\}\}$, $PStemp[<u_{14}, 14>]=PS[<u_{14}, 14>]\cap_{min}PStemp[<v, k\text{-}1>]=\{\{u_0\},\{u_1\},\{u_2\},\{v_3,w_3\},\{u_4\}, \{u_5\},\{u_6\}, \{u_7,w_7\},\{u_8\},\{u_9\},\{u_{10}\},\{u_{11},v_{11}\},\{u_{12}\},\{u_{13}\},\{u_{14}\}\}$, $PStemp[<x_{14}, 14>]=PS[<x_{14}, 14>]\cap_{min}PStemp[<v, k\text{-}1>]=\{\{u_0\},\{x_1\},\{x_2\},\{ y_3,z_3\},\{x_4\},\{x_5\},\{x_6\},\{x_7,z_7\},\{x_8\},\{x_9\},\{x_{10}\}, \{x_{11},y_{11}\},\{x_{12}\},\{x_{13}\},\{x_{14}\}\}$, $PStemp[<u_{12}, 12>]=PS[<u_{12}, 12>]\cap_{min}PStemp[<v, k\text{-}1>]=\{\{u_0\}, \{u_1\},\{u_2\},\{v_3,w_3\},\{u_4\},\{u_5\},\{u_6\},\{u_7,w_7\},\{u_8\},\{u_9\},\{u_{10}\}\{u_{11},v_{11}\},\{u_{12}\}\}$, $PStemp[<x_{12}, 12>]=PS[<x_{12}, 12>]\cap_{min}PStemp[<v, k\text{-}1>]=\{\{u_0\},\{x_1\},\{x_2\},\{y_3,z_3\},\{x_4\},\{x_5\},\{x_6\},\{x_7,z_7\},\{x_8\}, \{x_9\},\{x_{10}\},\{x_{11},y_{11}\},\{x_{12}\}\}$. Obviously, these $PStemp[<p, q>]$ do not contain $\emptyset$.

On the other hand, $PStemp[<u_{11}, 11>]=PS[<u_{11}, 11>]\cap_{min}PStemp[<v, k\text{-}1>]$ contains $\emptyset$. So do $PStemp[<v_{11}, 11>]$, $PStemp[<w_{11}, 11>]$, $PStemp[<x_{11}, 11>]$, $PStemp[<y_{11}, 11>]$, and $PStemp[<z_{11}, 11>]$. That is, they all contain $\emptyset$. Therefore, we can infer that $<u_{12}, 12>$ and $<x_{12}, 12>$ cannot be reached for the reason of $<u, k>$. So do $<u_{13}, 13>$, $<x_{13}, 13>$, $<u_{14}, 14>$, $<x_{14}, 14>$, $<w, 15>$, and $<v, k\text{-}1>$. In other words, for all parent vertices $u_{11}$, $v_{11}$, $w_{11}$ of $u_{12}$, we have that $PStemp[<u_{11}, 11>]=$

{∅}, PStemp[<$v_{11}$, 11>]= {∅}, and PStemp[<$w_{11}$, 11>]={∅}, resulting in that PStemp[<$u_{12}$, 12>]= {∅}. Therefore, we should let PStemp[<$u_{12}$, 12>]=PStemp[<$x_{12}$, 12>]=PStemp[<$u_{13}$, 13>]= PStemp[<$x_{13}$, 13>]=PStemp[<$u_{14}$, 14>]=PStemp[<$x_{14}$, 14>]=PStemp[<$w$, 15>]=PStemp[<$v$, $k$-1>]= {∅}. This is the responsibility of the **CHECK** operator.

**Theorem 3.** If there exists a basic path from the initial vertex *S* or a vertex on segment level *j*, where 1≤*j*≤*k*-1, to the vertex <*u*, *k*>, then the **CM** operator, i.e., Consistency Maintain operator, ensures that all the valid basic paths saved in the path set *PS*[<*u*, *k*>] can be found in Backward Searches, while all the invalid path fragments, which are broken and remained in *PS*[<*u*, *k*>], cannot be visited in Backward Searches.

Proof. First of all, any a vertex is located on a "serial path segment", or on a "parallel path cluster", or on both, along a basic path from the initial vertex *S* or a vertex on segment level *j*, where 1≤*j*≤*k*-1, to its parent vertex.

According to the principle of the **CM** operator, for every vertex <*u*, *k*> in the path hologram *H*, when the path set *PS*[<*u*, *k*>] will be generated, each parent vertex <*v*, *k*-1> of <*u*, *k*> is considered to judge whether or not a conflict will occur if the vertex *u* joins the *PS*[<*v*, *k*-1>]. That is whether *u*∈ $PS_{<v,k-1>}[i]$ or not, where 0≤*i*<*k*-1 and $PS_{<v,k-1>}[i]$∈*PS*[<*v*, *k*-1>].

If *u*∉ $PS_{<v,k-1>}[i]$ for each *i*, where 0≤*i*<*k*-1, then the vertex *u* does not appear on any a basic path from the initial vertex *S* or a vertex on segment level *j*, where 1≤*j*≤*k*-1, to the vertex <*v*, *k*-1>. So that, a conflict will not occur and *PS*[<*u*, *k*>] can be obtained by *PS*[<*v*, *k*-1>]⊗{{*u*}}, where ⊗ stands for "join" operation. That is, there exists a basic path from the initial vertex *S* or a vertex on segment level *j*, where 1≤*j*≤*k*-1, to the vertex <*u*, *k*> through the vertex <*v*, *k*-1> in *H*.

If *u*∈ $PS_{<v,k-1>}[i]$ for some *i*, where 0≤*i*<*k*-1, then a conflict will occur. It means that the path, on which *u* reappeared, from the initial vertex *S* or a vertex on segment level *j*, where 1≤*j*≤*k*-1, to the vertex <*u*, *k*> through the vertex <*v*, *k*-1> is not basic and is an "invalid path". So such "invalid path" should be abandoned. In order to correctly abandon this "invalid path", we need to find those "invalid vertices" on this path as much as possible. To achieve this goal, the vertex <*u*, *i*> must be deleted firstly. Then the vertices in the left/right action filed of the vertex <*u*, *i*> should be considered. If they are only located on this "invalid path" shown in **Figure 1**, they should be deleted later. Otherwise they will appear on another valid path and should be retained. For those vertices <*w*, *i*-1> in the left action filed of the vertex <*u*, *i*> or <*w*, *i*+1> in the right action filed of the vertex <*u*, *i*>, if the vertex *w* is deleted but not replenished, i.e., the vertex <*w*, *i*-1> or <*w*, *i*+1> is only located on this "invalid path", those vertices in the left action filed of the vertex <*w*, *i*-1> or in the right action filed of the vertex <*w*, *i*+1> should be considered whether to delete or retain again. Such "consecutive" vertex deleting-replenishing operations should be carried out recursively on the left action field and/or the right action field of "invalid" vertices respectively until another valid path is arrived. That is, no vertices can be deleted. So far, the conflicts caused by <*u*, *i*> have been eliminated as much as possible. For other such *i*1 that *u*∈ $PS_{<v,k-1>}[i1]$, where 0≤*i*1<*i*<*k*-1, we should repeat the above process.

As discussed earlier, the duplicates <*u*, *i*> of the vertex <*u*, *k*>, which will be deleted from *PS*[<*v*, *k*-1>], may be located on a serial path segment and/or on a parallel path segment. In any case, the "broken" invalid path fragments caused by the deletion of <*u*, *i*> will not be visited in future Backward Searches, although these "broken" invalid path fragments are remained in the path set PStemp[<*v*, *k*-1>]. We will prove this conclusion below.

Case 1. If the <*u*, *i*> is located on a unique "serial path segment" on the basic path from the initial vertex $S$ to the $<v, k-1>$, PStemp$[<v, k-1>]$ will contain an empty set $\emptyset$ at the segment level $i$ after the $<u, i>$ had been deleted from $PS[<v, k-1>]$. In this case, there does not exist any a basic path from the initial vertex $S$ to the $<u, k>$ through the $<v, k-1>$. So we let PStemp$[<v, k-1>]=\{\emptyset\}$ and $PS[<u, k>]=\{\{u\}\}$. In this case, $PS[<u, k>]$ does not contain any paths stored in $PS[<v, k-1>]$, resulting in that all "broken" invalid path fragments remained in PStemp$[<v, k-1>]$ cannot be visited in future Backward Searches, because these "broken" invalid path fragments are not in $PS[<u, k>]$ and the parent vertex $<v, k-1>$ is not in $PS[<u, k>]$.

Case 2. If each $<u, i>$ is located on a distinct parallel path segment of the same unique "parallel path cluster" on the basic path from the initial vertex $S$ to the $<v, k-1>$, PStemp$[<v, k-1>]$ will contain an empty set $\emptyset$ at the same unique cluster after all $<u, i>$ had been deleted from $PS[<v, k-1>]$. In this case, there does not exist any a basic path from the initial vertex $S$ to the $<u, k>$ through $<v, k-1>$. So we let PStemp$[<v, k-1>]=\{\emptyset\}$ and $PS[<u, k>]=\{\{u\}\}$. As discussed above, $PS[<u, k>]$ does not contain any paths stored in $PS[<v, k-1>]$, resulting in that all "broken" invalid path fragments remained in PStemp$[<v, k-1>]$ cannot be visited in future Backward Searches.

Case 3. If each $<u, i>$ is located on a parallel path segment of a distinct "parallel path cluster" on the basic path from the initial vertex $S$ to the $<v, k-1>$ and such parallel path segment, on which $<u, i>$ is located, is a unique "serial path segment" on the basic path from the initial vertex $S$ to the ancestral vertex $<w, k-i1>$ of $<v, k-1>$, shown in **Figure** 9, then PStemp$[<w, k-i1>]$ will contain an empty set $\emptyset$ at segment level $i$ after the $<u, i>$ has been deleted from $PS[<v, k-1>]$. In other words, PStemp$[<w, k-i1>]=PS[<w, k-i1>] \cap_{min}$ PStemp$[<v, k-1>]=\{\emptyset\}$. If every PStemp$[<w, k-i1>]=\{\emptyset\}$ for each ancestral vertex $<w, k-i1>$ of $<v, k-1>$, then the vertex $<v, k-1>$ cannot be reached from the initial vertex $S$ through the vertex $<w, k-i1>$ due to the deletion of $<u, i>$, although every segment set in PStemp$[<v, k-1>]$ is not empty in this case. So we let PStemp$[<v, k-1>]=\{\emptyset\}$ and $PS[<u, k>]=\{\{u\}\}$. As discussed above, $PS[<u, k>]$ does not contain any paths stored in $PS[<v, k-1>]$, resulting in that all "broken" invalid path fragments remained in PStemp$[<v, k-1>]$ cannot be visited in future Backward Searches. In this case, PStemp$[<v, k-1>]$ does not inherit from the paths of $PS[<w, k-i1>]$ and the vertex $<v, k-1>$ cannot be reached from the initial vertex $S$ due to the deletion of $<u, i>$.

Case 4. If the $<u, i>$ is located on a parallel path segment of a "parallel path cluster" on the basic path from the initial vertex $S$ to the $<v, k-1>$ and such parallel path segment, on which $<u, i>$ is located, is not a unique "serial path segment" on the basic path from the initial vertex $S$ to the ancestral vertex $<w, k-i1>$ of $<v, k-1>$, then there exist other distinct parallel path segments of the same "parallel path cluster" on which $<u, i>$ is located. Some associated "broken" invalid path fragments will be generated after the $<u, i>$ has been deleted from $PS[<v, k-1>]$. These "broken" invalid path fragments may be located on the left and/or right of $<u, i>$.

Case 4.1. If these "broken" invalid path fragments can also constitute a basic path from the initial vertex $S$ to the $<v, k-1>$ with another parallel path fragment in the same "parallel path cluster" where $<u, i>$ is located, then these "broken" invalid path fragments are all real valid path fragments due to the reason of this "another parallel path fragment" and can be visited in Backward Searches. In other words, these "broken" invalid path fragments are not only on the basic path on which $<u, i>$ is located, but also on other valif basic paths on which $<u, i>$ is not located, from the initial vertex $S$ to the $<v, k-1>$. In this case, we can let PStemp$[<u, k>]=$PStemp$[<v, k-1>] \otimes \{\{u\}\}$, which stores all basic paths from the initial vertex $S$ to $<u, k>$ through $<v, k-1>$ and all the valid basic paths can be visited in future Backward Searches.

Case 4.2. If these "broken" invalid path fragments cannot constitute a basic path from the initial vertex $S$ to the $<v, k\text{-}1>$ with another parallel path fragment in the same "parallel path cluster" where $<u, i>$ is located, then these "broken" invalid path fragments must pass through the unique vertex $<u, i>$, resulting in that there exists the respective rightmost vertex $w_r$ of these "broken" invalid path fragments, as the ancestral vertex of $<v, k\text{-}1>$, such that PStemp[$<w_r, k\text{-}i2>$]=$PS$[$<w_r, k\text{-}i2>$] $\cap_{min}$PStemp[$<v, k\text{-}1>$] contains an empty set $\varnothing$ at segment level $i$. In this case, during the future Backward Searches, the "broken" invalid path fragments located on the left of $<u, i>$ cannot be visited due to the lack of $<u, i>$ in PStemp[$<v, k\text{-}1>$] and the "broken" invalid path fragments located on the right of $<u, i>$ should not be visited due to the reason of PStemp[$<w_r, k\text{-}i2>$]=$PS$[$<w_r, k\text{-}i2>$] $\cap_{min}$PStemp[$<v, k\text{-}1>$]={$\varnothing$}. In other words, we do not choose the $<w_r, k\text{-}i2>$ for backward searching due to PStemp[$<w_r, k\text{-}i2>$]={$\varnothing$}. That is, there does not exist a basic path from the initial vertex $S$ to $<v, k\text{-}1>$ through $<w_r, k\text{-}i2>$ due to the reason of $<u, k>$. So we let PStemp[$<w_r, k\text{-}i2>$] ={$\varnothing$} and PStemp[$<v, k\text{-}1>$] does not inherit from the paths of $PS$[$<w_r, k\text{-}i2>$]. If all the basic paths stroed in PStemp[$<v, k\text{-}1>$] come from PStemp[$<w_r, k\text{-}i2>$], then we let PStemp[$<v, k\text{-}1>$]={$\varnothing$}. Otherwise PStemp[$<v, k\text{-}1>$] saves some basic paths which do not come from PStemp[$<w_r, k\text{-}i2>$]. In this case, all valid path fragments, which do not come from PStemp[$<w_r, k\text{-}i2>$], in PStemp[$<v, k\text{-}1>$] can be visited in Backward Searches, while all "broken" real invalid path fragments, which come from PStemp[$<w_r, k\text{-}i2>$], in PStemp[$<v, k\text{-}1>$] cannot be visited in Backward Searches. Therefore, we can let PStemp[$<u, k>$]=PStemp[$<v, k\text{-}1>$]$\otimes${{$u$}}, which stores all basic paths from the initial vertex $S$ to $<u, k>$ through $<v, k\text{-}1>$.

Since the duplicate of $u$ has at least been deleted from $PS_{<v,k-1>}[i]$, some segment sets may become singletons in PStemp[$<v, k\text{-}1>$] after the "consecutive" deleting-replenishing operations on $<u, i>$ have been accomplished. In this case, the duplicates of the vertex $z$ in a singleton should been handled with the same method as the duplicates of $u$. So that, we should scan PStemp[$<v, k\text{-}1>$] for processing the "duplicates" of singletons from segment level $k$-2 to segment level 0 after handling the duplicative vertex of $u$.

After such two kinds of "duplicative" cases, i.e., the duplicative vertex of $u$ and the duplicative vertex of the singleton, have been handled, if the remainder of $PS$[$<v, k\text{-}1>$]$\neq${$\varnothing$} as the **CHECK** operator has been executed, the remainder of the $PS$[$<v, k\text{-}1>$] contains all valid basic paths which can be found in Backward Searches, while all the "broken" real invalid path fragments cannot be visited in Backward Searches. So that the remainder of $PS$[$<v, k\text{-}1>$] can be used to construct the path from the initial vertex $S$ or a vertex on segment level $j$, where $1\leq j\leq k$-1, to the vertex $<u, k>$ through the vertex $<v, k\text{-}1>$. That is, $PS$[$<u, k>$]=the remainder of $PS$[$<v, k\text{-}1>$]$\otimes${{$u$}} saves all the valid basic paths, which can be found in Backward Searches, from the initial vertex $S$ or a vertex on segment level $j$, where $1\leq j\leq k$-1, to the vertex $<u, k>$ through the vertex $<v, k\text{-}1>$. Further, all the "broken" real invalid path fragments, which is remained in the $PS$[$<u, k>$], cannot be visited in Backward Searches. If the remainder of the $PS$[$<v, k\text{-}1>$] is {$\varnothing$} after calling the **CHECK** operator, then there is not a basic path from the initial vertex $S$ or a vertex on segment level $j$, where $1\leq j\leq k$-1, to the vertex $<u, k>$ through the vertex $<v, k\text{-}1>$.    □

**Theorem 4.** $PS$[$<u, k>$] has no basic paths from the initial vertex $S$ to the vertex $<u, k>$ if and only if some segment set is empty.

Proof. We prove the necessity as follows.

Since some segment set $PS_{<u,k>}[i]$ is empty, all basic paths from the initial vertex $S$ to the vertex $<u, k>$ are broken at the segment level $i$, i.e., all basic paths cannot pass through the segment level $i$, resulting in that $PS$[$<u, k>$] has no basic paths from the initial vertex $S$ to the vertex $<u, k>$.

We prove the sufficiency as follows.

Since $PS[<u, k>]$ has no basic paths from the initial vertex $S$ to the vertex $<u, k>$, for each parent vertex $<v, k-1>$ of $<u, k>$ where $v \in PS_{<u,k>}[k-1]$ we must have the fact that the vertex $u$ appears in all basic paths in $PS[<v, k-1>]$. If not, there must exist a basic path $P=S-\ldots-<v, k-1>$ in $PS[<v, k-1>]$ such that $u$ does not appear in $P$. So that we can obtain a new basic path $P_1=P-<u, k>=S-\ldots-<v, k-1>-<u, k>$, which is a basic path from $S$ to $<u, k>$, contradicting the fact that $PS[<u, k>]$ has no basic paths from the initial vertex $S$ to the vertex $<u, k>$.

For any basic path $P=S-\ldots-<v, k-1>$ in $PS[<v, k-1>]$, after deleting the duplicates of $u$ and signelton as well as executing the "consecutive" deleting-replenishing operations recursively on the left/right action field of invalid vertices respectively, some broken path fragments will be certainly created. If these broken path fragments can also constitute a basic path $P_2$ from the initial vertex $S$ to the $<v, k-1>$ with another parallel path fragment in the same "parallel path cluster" where the duplicates of $u$ is deleted, then $u$ must appear in this basic path $P_2$. If not, we can obtain a new basic path $P_2-<u, k>$, which is a basic path from $S$ to $<u, k>$, contradicting the fact that $PS[<u, k>]$ has no basic paths from the initial vertex $S$ to the vertex $<u, k>$.

Notice that $u$ does not appear in all broken path fragments. Since these broken path fragments cannot constitute a basic path from the initial vertex $S$ to the $<v, k-1>$ with another parallel path fragment $\bar{P}$ in the same "parallel path cluster" where the duplicates of $u$ is deleted, then $u$ must appear in $\bar{P}$, resulting in that the "parallel path cluster" is an empty segment set.

So that some segment set in $PS[<u, k>]$ is empty.

In fact, as the same reason, $PS[<u, k>]$ has no basic paths from the initial vertex $S$ or any vertex on segment level $j$, where $1 \leq j \leq k-1$, to the vertex $<u, k>$ if and only if some segment set is empty.

Of course, in this case, in order to ensure the calculation of the path set of descendant vertices, we let $PS[<u, k>]=\{\{u\}\}$, i.e., the initial value.   □

**Theorem 5.** The **CM** operator is complete.

Proof. For the path set $PS[<v, k-1>]$, when the child vertex $<u, k>$ plans to join $PS[<v, k-1>]$, we will handel the duplicates of $u$ and the duplicates of singletons. If the duplicate belongs to a singleton set, i.e., the duplicate is located on a unique "serial path segment", the empty segment set will occur when the duplicate has been deleted. This is the Case 1 of **Theorem** 3.

If the duplicates belong to each parallel path fragment in the same "parallel path cluster", i.e., each duplicate is located on a distinct parallel path segment of the same unique "parallel path cluster", the empty segment set will occur when these duplicates have been deleted. This is the Case 2 of **Theorem** 3.

Deleting the duplicates will create some broken path fragments. If these broken path fragments are located on a unique basic path passing only through the deleted path fragment $P_{del}$ on which the duplicate is located, this must make the rightmost "parallel path cluster" on this unique basic path empty. This is the Case 3 and 4.2 of **Theorem** 3. If these broken path fragments are also located on another basic path passing through the parallel path fragments of the "parallel path cluster" on which $P_{del}$ is located, then these broken path fragments are valid and can be visited in Backward Searching. This is the Case 4.1 of **Theorem** 3.

We know that $PS[<v, k-1>]$ has saved all the basic paths from $S$ to $<v, k-1>$. If $PStemp[<v, k-1>]$ becomes invalid, i.e., $PStemp[<v, k-1>]$ has not any basic path from $S$ to $<v, k-1>$, due to the limitation of $<u, k>$, then we must have $PStemp_{<v,k-1>}[k-1]=\varnothing$. This means that $PStemp[<w, k-2>]=\{\varnothing\}$ for each parent vertex $<w, k-2>$ of $<v, k-1>$ where $w \in PStemp_{<v,k-1>}[k-2]$. In other words, all valid basic paths in *PS*[<*w, k*-2>] becomes invalid when *PS*[<*w, k*-2>] places restrictions on <*w, k*-2>, <*v, k*-1> and <*u, k*>. At this time, we have PStemp[<*w, k*-2>]=PStemp[<*v, k*-1>] $\cap_{min}$ *PS*[<*w, k*-2>]. In this case, by **Theorem 4**, we will infer PStemp[<*w, k*-2>]={∅} after we have regenerated PStemp[<*w, k*-2>] from segment level 0 to segment level *k*-2 based on PStemp[<*v, k*-1>]$\cap_{min}$*PS*[<*w, k*-2>], as the **Example 4** shown. Therefore, the **CHECK** operator can accomplish this task correctly.

If PStemp[<*v, k*-1>] is still valid due to the limitation of <*u, k*>, we know that PStemp[<*v, k*-1>] has no empty segment sets by **Theorem 4**. In this case, the **CHECK** operator cannot return PStemp[<*v, k*-1>]={∅}.

So far, we have proved the completeness of the **CM** operator.   □

Before proving **Theorem 6**, we analyze the following cases first. As one knows, *PS*[<*u, k*>] saves all basic paths from the initial vertex ***S*** or any vertex on segment level *j*, where 1≤*j*≤*k*-1, to the vertex <*u, k*> in ***H*** if these paths exist. The number of these paths is determined by the size of the segment sets, which is the vertex set consisting of all vertices located on the same segment level. On the other hand, the maximum number of elements in each segment set is *n*-1, where *n* is the number of vertices of the original graph ***G***.

The goal of the "consecutive" deleting-replenishing operations on the duplicative vertex of *u* is to delete the "invalid vertices" on the "invalid" path as much as possible, so as to ensure that the broken invalid path fragments, if they exist, will not be visited in Backward Searches. If the number of the paths stored in *PS*[<*u, k*>] is very large, for example, $d^n$ (*d*≥2) or *n*!, it is very important whether the cost of the "consecutive" deleting-replenishing operations is polynomial time or not. Fortunately, the **CM** operator introduced in this paper can guarantee that the cost is polynomial time.

Consider the situations shown in **Figure 10**.

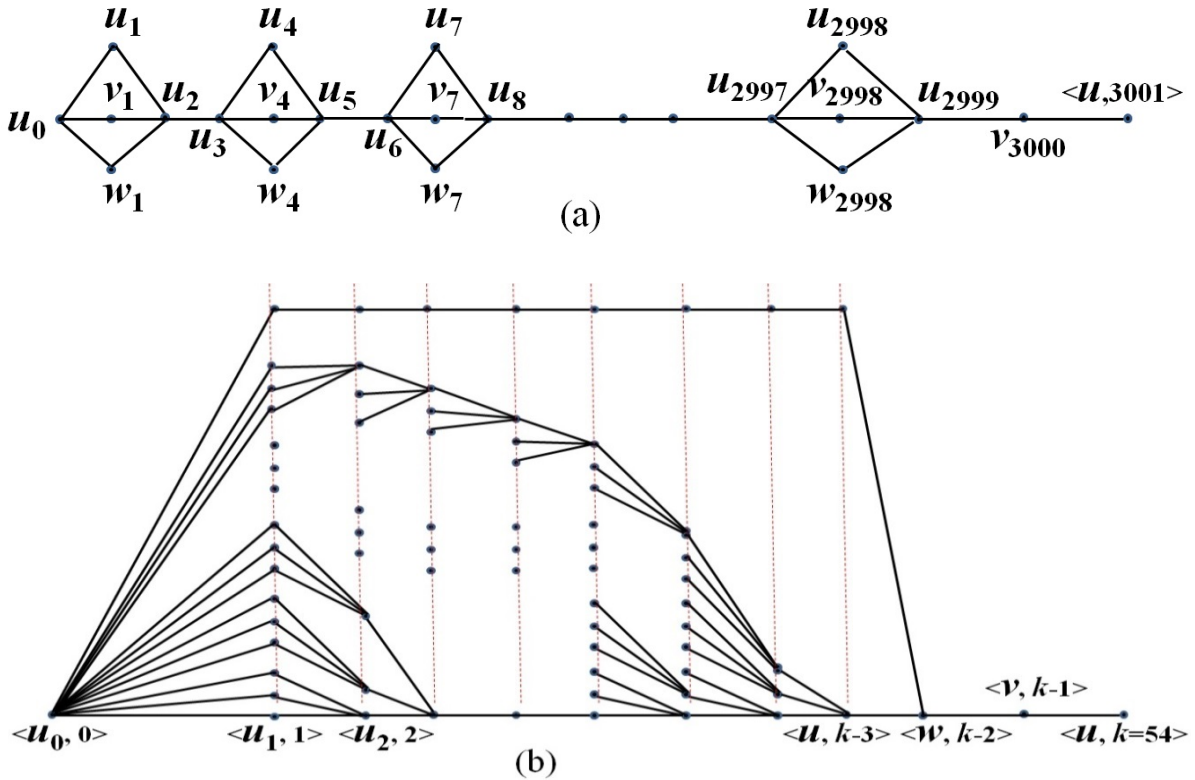

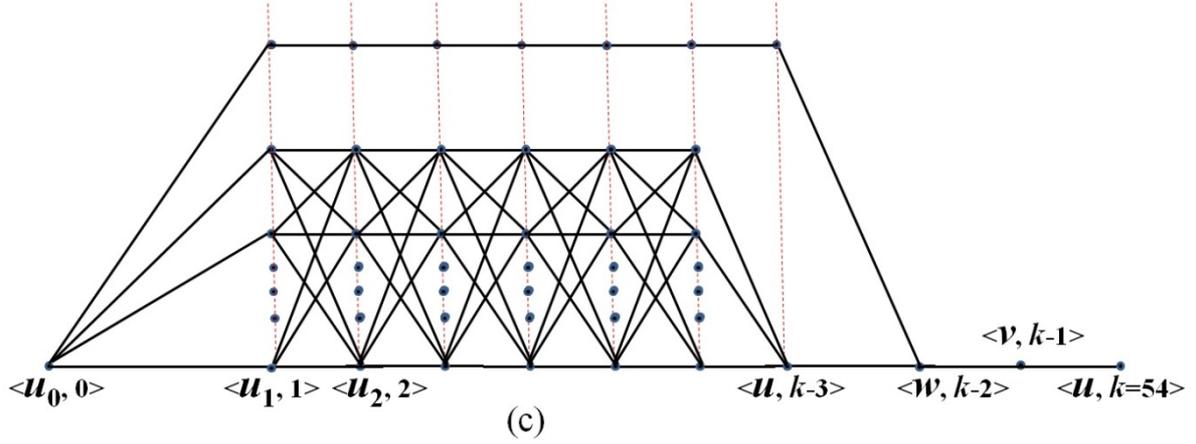

**Figure 10**  Analysis of the cost of "consecutive" vertex deleting-replenishing operations

Case 1. In **Figure 10(a)**, there are exactly $3^{1000}$ unique basic paths in $PS[<v, k-1>]=PS[<v_{3000}, 3000>]$, supposed that every vertex is distinct except for the duplicative vertex of $u$. At this point, the segment set in $PS[<v, k-1>]$ has either 3 elements or 1 element. That is, $PS[<v, k-1>]=\{\{u_0\},\{u_1, v_1, w_1\},\{u_2\},\{u_3\},\{u_4, v_4, w_4\},\{u_5\},…,\{u_{2997}\},\{u_{2998}, v_{2998}, w_{2998}\},\{u_{2999}\},\{v_{3000}\}\}$. In this case, any a path fragment can be visited for backward searching as the component of one valid basic path after the deleted path $P_{del}$, on which $<u, i>$ is located, has been removed from $PS[<v, k-1>]$.

No matter on which "parallel path cluster" the duplicative vertex of $u$ is, only is it deleted. Because each "parallel path cluster" has 5 vertices, i.e., 1 fork vertex $u_i$, 3 parallel vertices $u_{i+1}, v_{i+1}, w_{i+1}$, and 1 meeting vertex $u_{i+2}$, where $i=3(k-1)$ and $1≤k≤1000$, the "consecutive" deleting-replenishing operations on the duplicative vertex of $u$ will be halted after $5×k=5000$ comparisons. Therefore, the time cost of the **CM** operator is $O(5k)=O(n)$, where $k$ is the number of "parallel path cluster" and $5k≤n$. In other words, the time cost of the **CM** operator is not $O(3^{1000})$ although $PS[<v, k-1>]$ saves $3^{1000}$ unique basic paths.

Case 2. In **Figure 10(b)**, supposed that the duplicative vertex of $<u, k>$ is $<u, k-3>$. In the 1st iteration, the $<u, k-3>$ is deleted from $PS_{<v,k-1>}[k-3]$. In the 2nd iteration, $d$ vertices in the left action field set $AF^-(<u, k-3>)$ are deleted from $PS_{<v,k-1>}[k-4]$, where $d≥2$. In the 3rd iteration, $d^2$ vertices in the left action field set of the vertices in $AF^-(<u, k-3>)$ are deleted from $PS_{<v,k-1>}[k-5]$, and so on. Finally, in the 51th iteration, $d^{50}$ vertices are deleted from $PS_{<v,k-1>}[1]$.

However, we must remember that each segment set can contain up to $n$-1 elements. That is, we have $d^{50}≤n$-1. On the other hand, the "consecutive" deleting-replenishing operations on the left action field of the "invalid" vertices are performed only on each segment set. So at most $n$-1 elements will be deleted during each iteration. This will result in a total of $51×(n-1)$ deletion operations and of $51×d×(n-1)$ comparisons after the 51th iteration.

Case 3. In **Figure 10(c)**, supposed that the duplicative vertex of $<u, k>$ is $<u, k-3>$ and the deleted "invalid" vertices come from a complete subgraph $K_{50}$. Since only are $d=50$ "invalid" vertices deleted from each segment set during each iteration, this will result in a total of $51×d$ deletion operations and of $51×d×d$ comparisons after the 51th iteration.

**Theorem 6.** The **CM** operator can be computed in polynomial time.
Proof. First, the **LAFDR** operator and the **RAFDR** operator can be performed in $O(n×d)$ time respectively, where $d$ is the maximum degree of vertex. The purpose of **LAFDR** and **RAFDR** are just to accomplish the "consecutive" deleting-replenishing operations on invalid vertices. Although $d^{k-1}$ vertices may be deleted at the *k*-th iteration, where *d* is the maximum vertex degree, we must remember that each segment set has a maximum of *n*-1 elements. Thus, $d^{k-1}$≤*n*-1 no matter how big *k* is. So that after *k* iterations, we can delete *k*(*n*-1) elements at most, resulting in that this operation of the "consecutive" deleting-replenishing can be accomplished in *O*(*k*(*n*-1)) time in the worst case.

The **CHECK** operator can be performed in O($n^7$×$d^2$) time in the worst case. The purpose of **CHECK** is just to decide finally PStemp[<*v*, *k*-1>]={∅} or not.

So we pay close attention to the cost of the "While (flag2)" loop, i.e., the number of iterations. The purpose of the "While (flag2)" loop is just to handle the duplicates of the singletons.

Case 1. There are not singletons or the duplicates of singletons. In this case, the "While (flag2)" loop runs only once. So that the "While (flag2)" loop can halt in *O*(1) time.

Case 2. There exist the duplicates of singletons. Without loss of generality, suppose that the singleton is {*z*}. After deleting the duplicate of vertex *z* and the vertices in its left/right action field recursively, if the replenishing operations are executed on those vertices in the left/right action fields, the size of all segment sets is either decreased or unchanged.

In other words, the size of the segment set which contains the duplicative vertex of *z* must decrease by 1, while the size of the other segment sets is either decreased or unchanged. Once some vertices have been really deleted from some segment sets, i.e., these vertices have been deleted but not replenished, the "While (Flag2)" loop continues within the next iteration. Once there do not exist singletons or the duplicates of singletons in the next iteration, or the case of "unchanged" occurs in the next iteration, the "While (Flag2)" loop halts immediately.

We can notice that the size of each segment set is limited, i.e., *n*-1 vertices at most, and the length of a path set is also limited, i.e., *n*-1 segment sets at most. If the duplicates of the singletons exist and only is one element deleted from one segment set during the "While (Flag2)" loop each iteration, then in the worst case the "While (Flag2)" loop must halt after having iterated $(n-1)^2$ times due to the empty segment set arisen.

In other words, during each iteration of the "While (Flag2)" loop, (1) if no vertex has been deleted, then the "While (flag2)" loop will halt in 1 step; (2) if only has one vertex been deleted, then the "While (flag2)" loop will halt up to $(n-1)^2$ steps; (3) if many vertices have been deleted, then the "While (flag2)" loop will halt far less than $(n-1)^2$ steps.

Therefore, in the worst case the "While (Flag2)" loop iterates $(n-1)^2$ times, while the body of the "While (Flag2)" loop is performed in O($n^5$×*d*) time. As a result, the operation on the "While (Flag2)" loop can be performed in polynomial time in each case.    □

**Theorem 7.** The **LPM** operator, i.e., Longest Path Merging operator, satisfies the optimal principle of greedy strategy.

Proof. According to the principle of the **LPM** operator, all the same longest paths from the initial vertex ***S*** or any vertex on segment level *j*, where 1≤*j*≤*k*-1, to the vertex <*u*, *k*> through its each parent vertex <*v*, *k*-1> have been stored in *PS*[<*u*, *k*>] after calling **LPM**(*PS*[<*u*, *k*>], PStemp[<*u*, *k*>]).

We know that *PS*[<*v*, *k*-1>] stores all the longest paths from the initial vertex ***S*** or any vertex on segment level *j*, where 1≤*j*≤*k*-2, to the vertex <*v*, *k*-1>. If we obtain PStemp[<*v*, *k*-1>]={∅} after calling **CM**(*PS*[<*v*, *k*-1>], <*u*, *k*>), then there does not exist any basic path from the initial vertex ***S*** or any vertex on segment level *j*, where 1≤*j*≤*k*-1, to the vertex <*u*, *k*> through its parent vertex <*v*, *k*-1>. In this case, *PS*[<*u*, *k*>]={{*u*}} has the longest path.

Therefore, we consider the case of PStemp[<*v*, *k*-1>]≠{∅} below. When we call the **LPM** operator,

we will obtain $PS[<u, k>]=\cup_{max}\{PStemp[<v, k-1>]\otimes\{\{u\}\} \mid \forall (<v, k-1>, <u, k>)\in E_H$ and $PStemp[<v, k-1>]$ is valid for the vertex $<u, k>\}$. So that the length of the paths stored in $PS[<u, k>]$ is either $k+1$ or $k-j+1$ due to the path from the initial vertex $S$ or from any vertex on segment level $j$, where $1\leq j\leq k-1$. If the length is $k+1$, then the longest path is the one from the initial vertex $S$ to the vertex $<u, k>$ through its parent vertex $<v, k-1>$ and stored in $PS[<u, k>]$ after calling **LPM**($PS[<u, k>]$, $PStemp[<u, k>]$). If the length is $k-j+1$, then all valid paths through its parent vertex $<v, k-1>$ come from the vertices on segment level $j$, where $1\leq j\leq k-1$. In this case, all the paths from the initial vertex $S$ or from any vertex on segment level $j-1$, where $1\leq j\leq k-1$, to the vertex $<v, k-1>$ are invalid for the vertex $<u, k>$. So we cannot obtain such a path, which comes from these invalid paths to the vertex $<u, k>$ through its parent vertex $<v, k-1>$. In other words, we cannot obtain such a path of length greater than $k-j+1$. It is the fact that every path from the initial vertex $S$ or any vertex on segment level $j$, where $1\leq j\leq k-1$, to the vertex $<u, k>$ should pass through its parent vertex $<v, k-1>$ unless $PStemp[<v, k-1>]=\{\varnothing\}$. So that after the **LPM** operator has been performed, $PS[<u, k>]$ stores all the same longest paths from the initial vertex $S$ or any vertex on segment level $j$, where $1\leq j\leq k-1$, to the vertex $<u, k>$. So the optimality of the $PS[<u, k>]$ is guaranteed.    □

**Theorem 8.** If the original graph $G$ is Hamiltonian, then any a Hamiltonian cycle can be found by the **FHC** operator in deterministic polynomial time.

Proof. The principle of the **FHC** operator is that it searches backward for a basic path from the final vertex $D$ to the initial vertex $S$ in the path hologram $H$ by selecting a non-redundant, valid parent vertex based on the condition of $\varnothing\notin$**CHECK**($PStemp\cap_{min}PS[<parent\ vertex>]$) iteratively at each step.

By the **Theorems 3~7**, we know that $PS[<u, k>]$ saves all the longest valid basic paths from the initial vertex $S$ or any vertex on segment level $j$, where $1\leq j\leq k-1$, to the vertex $<u, k>$. Each vertex of every segment set in $PS[<u, k>]$, which is located on the valid path fragment, can be visited in Backward Searches, but those vertices $<w_{lx}, lx>$ located on the "broken" invalid path fragments cannot be visited in Backward Searches due to the reason of $\varnothing\in$**CHECK**($PStemp\cap_{min}PS[<w_{lx}, lx>]$) or $w_{lx}\notin PStemp[lx]$. Since the original graph $G$ is Hamiltonian, there must exist a basic path from the initial vertex $S$ to the final vertex $D$ which is saved in $PS[D]$ by the **Theorem 2**.

According to **Theorem 4**, $PS[D]$ has a basic path from $S$ to $D$ if and only if each segment set of $PS[D]$ is not $\{\varnothing\}$. Thus, there must exist a parent vertex $<v, n-1>$ of $D$, whose $PS[<v, n-1>]$ satisfies $\varnothing\notin$**CHECK**($PS[<v, n-1>]\cap_{min}PS[<D>]$). Further, we can determine the parent vertex $<v, n-1>$ of $D$ in the $deg(D)$ steps, i.e., in a deterministic steps. Similarly, $PS[<v, n-1>]\cap_{min}PS[<D>]\neq\{\varnothing\}$ will inevitably infer that there must exist a parent vertex $<w, n-2>$ of $<v, n-1>$, whose $PS[<w, n-2>]$ satisfies $\varnothing\notin$**CHECK**($PS[<w, n-2>]\cap_{min}PS[<v, n-1>]\cap_{min}PS[<D>]$), and so on until we reach the $S$ by Backward Searching.

At each step of the Backward Searching, we only consider $deg(v)$ parent vertices of $<v, i>$ and we must find such a parent vertex $<w, i-1>$ satisfying that $\varnothing\notin$**CHECK**($PS[<w, i-1>]\cap_{min}PStemp$). In other words, a parent vertex can be selected deterministically at each step. On the other hand, $PS[D]$ has only $n+1$ segment sets, this means that the Backward Searching will stop after $n+1$ steps. Thus, a basic path $P=S-\ldots-<v, n-2>-<u, n-1>-D$ from the initial vertex $S$ to the final vertex $D$ can be found by the **FHC** operator in deterministic polynomial time.

Since we can select any a deterministic valid parent vertex in Backward Searching at each step, we can find any a Hamiltonian cycle by the **FHC** operator in deterministic polynomial time.    □